%% file: paper1.tex
\documentclass[longauth]{aa}  
\usepackage{graphicx}
\usepackage{txfonts}
\usepackage{natbib}
\bibpunct{(}{)}{;}{a}{}{,} 
\usepackage{graphicx}
\usepackage[colorlinks=true,allcolors=blue]{hyperref}
\usepackage{lscape}
\usepackage{xcolor}
\usepackage{multicol}
\usepackage{multirow}
\usepackage{placeins}

\newcommand{\ms}{$M_{\odot}$}

\newcommand{\Ni}{$^{56}$Ni}
\newcommand{\mni}{$M_{\rm Ni}$}
\newcommand{\sneii}{SNe~II}
\newcommand{\snii}{SN~II}
\newcommand{\halpha}{H$_{\alpha}$}
\newcommand{\cd}{\emph{Cd}}
\newcommand{\pd}{\emph{pd}}
\newcommand{\optd}{\emph{optd}}
\newcommand{\sone}{$s_{\rm 1}$}
\newcommand{\stwo}{$s_{\rm 2}$}
\newcommand{\sthr}{$s_{\rm 3}$}
\newcommand{\ttrans}{$t_{\rm trans}$}

\newcommand{\tpt}{$t_{\rm PT}$}
\newcommand{\mbolend}{$M_{\rm bol,end}$}
\newcommand{\mboltail}{$M_{\rm bol,tail}$}

\defcitealias{paper2_submitted}{Paper~II}
\defcitealias{paper3_submitted}{Paper~III}

\begin{document}

	\title{Type II supernovae from the \textit{Carnegie Supernova Project-I}}
	\subtitle{I. Bolometric light curves of 74 SNe~II using $uBgVriYJH$ photometry}
	
	\author{L. Martinez \inst{1,2,3}
		\and M.~C. Bersten \inst{1,2,4}
        \and J.~P. Anderson \inst{5}
        \and M. Hamuy \inst{6,7}
        \and S. González-Gaitán \inst{8}
        \and M. Stritzinger \inst{9}
        \and \\ M.~M. Phillips \inst{10}
        \and C.~P. Guti\'errez \inst{11,12}
        \and C. Burns \inst{13}
        \and C. Contreras \inst{10}
        \and T. de Jaeger \inst{14,15}
        \and K. Ertini \inst{1,2}
        \and G. Folatelli \inst{1,2,4}
        \and \\ F. Förster \inst{16,17,18,19}
        \and L. Galbany \inst{20}
        \and P. Hoeflich \inst{21}
        \and E.~Y. Hsiao \inst{21}
        \and N. Morrell \inst{10}
        \and M. Orellana \inst{3,22}
        \and P.~J. Pessi \inst{2,5}
        \and \\ N.~B. Suntzeff \inst{23}
        }

	\institute{Instituto de Astrofísica de La Plata (IALP), CCT-CONICET-UNLP. 
			Paseo del Bosque S/N, B1900FWA, La Plata, Argentina\\
			\email{laureano@carina.fcaglp.unlp.edu.ar}
		\and
			Facultad de Ciencias Astronómicas y Geofísicas,
			Universidad Nacional de La Plata, Paseo del Bosque S/N, B1900FWA, La Plata, Argentina
		\and
            Universidad Nacional de Río Negro. Sede Andina, Mitre 630 (8400) Bariloche, Argentina
        \and
            Kavli Institute for the Physics and Mathematics of the Universe (WPI), The University of Tokyo, 5-1-5 Kashiwanoha, Kashiwa, Chiba 277-8583, Japan
        \and
            European Southern Observatory, Alonso de Córdova 3107, Casilla 19, Santiago, Chile
        \and
            Vice President and Head of Mission of AURA-O in Chile, Avda. Presidente Riesco 5335 Suite 507, Santiago, Chile
        \and
            Hagler Institute for Advanced Studies, Texas A\&M University, College Station, TX 77843, USA
        \and
            CENTRA-Centro de Astrofísica e Gravitaçäo and Departamento de Física, Instituto Superio Técnico, Universidade de Lisboa, Avenida Rovisco Pais, 1049-001 Lisboa, Portugal
        \and
            Department of Physics and Astronomy, Aarhus University, Ny Munkegade 120, DK-8000 Aarhus C, Denmark
        \and
            Carnegie Observatories, Las Campanas Observatory, Casilla 601, La Serena, Chile
        \and 
            Finnish Centre for Astronomy with ESO (FINCA), FI-20014 University of Turku, Finland
        \and
            Tuorla Observatory, Department of Physics and Astronomy, FI-20014 University of Turku, Finland
        \and 
            Observatories of the Carnegie Institution for Science, 813 Santa Barbara St., Pasadena, CA 91101, USA
        \and
            Institute for Astronomy, University of Hawaii, 2680 Woodlawn Drive, Honolulu, HI 96822, USA
        \and
            Department of Astronomy, University of California, 501 Campbell Hall, Berkeley, CA 94720-3411, USA.
        \and
            Data and Artificial Intelligence Initiative, Faculty of Physical and Mathematical Sciences, University of Chile
        \and
            Centre for Mathematical Modelling, Faculty of Physical and Mathematical Sciences, University of Chile
        \and
            Millennium Institute of Astrophysics, Chile.
        \and
            Department of Astronomy, Faculty of Physical and Mathematical Sciences, University of Chile
        \and
            Institute of Space Sciences (ICE, CSIC), Campus UAB, Carrer de Can Magrans, s/n, E-08193 Barcelona, Spain.
        \and
            Department of Physics, Florida State University, 77 Chieftan Way, Tallahassee, FL 32306, USA
        \and
            Consejo Nacional de Investigaciones Científicas y Tećnicas (CONICET), Argentina.
        \and
            George P. and Cynthia Woods Mitchell Institute for Fundamental Physics and Astronomy, Department of Physics and Astronomy, Texas A\&M University, College Station, TX 77843
            }

\titlerunning{Bolometric light curves for 74 SNe~II from the CSP-I}

\date{Received XXX; accepted XXX}
 

\abstract 
{The present study is the first of a series of three papers where we characterise the type II supernovae (\sneii) from the \textit{Carnegie Supernova Project-I} to understand their diversity in terms of progenitor and explosion properties.
In this first paper, we present bolometric light curves of 74 \sneii.
We outline our methodology to calculate the bolometric luminosity, which consists of the integration of the observed fluxes in numerous photometric bands ($uBgVriYJH$) and black-body (BB) extrapolations to account for the unobserved flux at shorter and longer wavelengths.
BB fits were performed using all available broadband data except when line blanketing effects appeared.
Photometric bands bluer than $r$ that are affected by line blanketing were removed from the fit, which makes near-infrared (NIR) observations highly important to estimate reliable BB extrapolations to the infrared.
BB fits without NIR data produce notably different bolometric light curves, and therefore different estimates of \snii\ progenitor and explosion properties when data are modelled.
We present two methods to address the absence of NIR observations: (a) colour-colour relationships from which NIR magnitudes can be estimated using optical colours, and (b) new prescriptions for bolometric corrections as a function of observed \snii\ colours.
Using our 74 \snii\ bolometric light curves, we provide a full characterisation of their properties based on several observed parameters. We measured magnitudes at different epochs, as well as durations and decline rates of different phases of the evolution.
An analysis of the light-curve parameter distributions was performed, finding a wide range and a continuous sequence of observed parameters which is consistent with previous analyses using optical light curves. 
}


\maketitle

\section{Introduction} 
\label{sec:intro} 

The \textit{Carnegie Supernova Project-I} \citep[CSP-I,][]{hamuy+06} was a five-year programme to obtain well-calibrated high-cadence optical and near-infrared (NIR) light curves (LCs) and optical spectra of every type of supernova \citep[see e.g.][]{contreras+10,stritzinger+11,taddia+13,krisciunas+17,stritzinger+18}.
The current work is the first in a series of three papers where the main goal is to analyse a large sample of type II supernovae (\sneii) observed by the CSP-I in order to understand \snii\ diversity in terms of progenitor and explosion properties. 
In the present study, we calculate and present the most homogeneous sample of 74 \snii\ bolometric LCs to date.
In the second and third papers in this series, we fit these bolometric LCs to \snii\ explosion models. This affords a characterisation of the progenitor and explosion properties of \sneii\ \citep[][hereafter Paper~II]{paper2_submitted}, and an analysis of the underlying physical parameters defining \snii\ diversity \citep[][hereafter Paper~III]{paper3_submitted}.

\sneii\ are the explosions of massive stars ($\gtrsim$8$-$10~\ms) at the end of their evolution showing prevalent hydrogen lines in their spectra \citep{minkowski41}. \sneii\ were initially divided into sub-classes according to the shape of their LC. \sneii\ with almost constant luminosity for a few months are classified as SNe~IIP, and those with fast-declining LCs were historically classified as SNe~IIL \citep{barbon+79}. However, it is now generally accepted that they belong to a single family as they exhibit a continuous sequence in their LC decline rates \citep{anderson+14_lc,sanders+15,galbany+16,valenti+16,rubin+16,dejaeger+19}.

Over the years, many subgroups of \sneii\ have been introduced based on their spectral and photometric characteristics. SNe~IIb show hydrogen lines at early times that weaken and disappear with time, and seem to be transitional events between \sneii\ and SNe~Ib \citep[][although see \citealt{pessi+19}]{filippenko+93}. SNe~IIn show narrow emission lines in their spectra and a diversity of photometric characteristics \citep[e.g.][]{schlegel90,taddia+13}, possibly due to an interaction between the ejecta and a dense circumstellar material (CSM). Meanwhile, the 1987A-like events display typical characteristics of \snii\ spectra, but long-rising LCs that resemble that of SN~1987A \citep[e.g.][]{arnett+89,taddia+12}.
The properties of these subgroups are distinct from the \sneii\ we focus on in this paper.
Therefore, we exclude SNe~IIb, IIn, and 1987A-like from all analyses and use `SNe~II' to refer to SNe that would historically have been classified as SNe~IIP or SNe~IIL together.

The strong and enduring hydrogen lines in \snii\ spectra indicate that their progenitors have retained a significant fraction of their hydrogen-rich envelope prior to explosion. Indeed, the first models of \sneii\ \citep{grassberg+71,falk+77} demonstrate that the LC morphology of slow-declining \sneii\ (SNe~IIP) can be reproduced assuming a red supergiant (RSG) progenitor with an extensive hydrogen envelope.
More recently, the direct detections of the progenitor stars of nearby \sneii\ confirm RSG stars as their progenitors \citep[e.g.][]{vandyk+03,smartt09}.

\sneii\ are the most common type of SNe in nature \citep{li+11a,shivvers+17}, and their study is closely related to stellar evolution, nucleosynthesis, and chemical enrichment of the interstellar medium, among other astrophysical processes. In addition, \sneii\ have been proposed as distance indicators \citep[e.g.][]{kirshner+74,hamuy+02,rodriguez+14,dejaeger+15} and metallicity tracers \citep{dessart+14,anderson+16,gutierrez+18} which further motivates their study.

\sneii\ show a significant diversity in their observed properties.
Studies of large samples have defined several parameters to examine this diversity finding a continuous sequence within a large range of values for most of the parameters analysed, including magnitudes, decline rates and durations of different LC phases \citep[e.g.][]{anderson+14_lc}; velocities and pseudo-equivalent widths of \halpha\ profiles \citep[e.g.][]{schlegel96,anderson+14_ha,gutierrez+14} in addition to many other spectral lines \citep[e.g.][]{gutierrez+17II,gutierrez+17I}; and \snii\ colours \citep[e.g.][see also \citealt{patat+94,hamuy03,faran+14a,faran+14b,gonzalez+15,anderson+16,gutierrez+18,davis+19} for more general studies of \snii\ diversity]{dejaeger+18}.

Efforts have already been made to characterise and analyse the spectral and photometric properties of the CSP-I \snii\ sample.
\citet{anderson+14_ha} and \citet{gutierrez+14} presented a detailed spectroscopic analysis of the H$_{\rm \alpha}$ profiles, particularly of the blueshifted emission, velocity, and ratio of absorption to emission. Moreover, both works searched for correlations between these spectral parameters and observed parameters of the \emph{V}-band LCs (magnitudes, decline rates, and time durations) measured by \citet{anderson+14_lc}.
A subsequent study was presented in \citet{gutierrez+17II,gutierrez+17I}. These authors measured expansion velocities and pseudo-equivalent widths for several spectral lines and analysed correlations between these spectral properties and the spectral and photometric parameters defined in previous works.
Additionally, \citet{dejaeger+18} analysed the properties of the colour curves.
These previous works attempted to understand (mostly qualitatively) observed \snii\ diversity and correlations through differences in physical parameters of the explosions. In the present series of papers we attempt to go a step further and quantitatively derive physical parameters of \sneii\ from robustly modelling their data.

This first paper follows a number of previous investigations in the literature concerning \snii\ bolometric LCs and their properties \citep[e.g.][]{bersten+09,lyman+14,pejcha+15b,lusk+17,faran+18}.
In the current study, we present bolometric LCs for 74 \sneii\ observed by the CSP-I. Additionally, we outline our bolometric LC creation methodology and discuss important sources of uncertainty when estimating the bolometric flux of a SN~II.
The CSP-I sample comprises high-quality optical ($uBgVri$) and NIR ($YJH$) photometry homogeneously obtained and processed.
The release of the final CSP-I \snii\ photometry is presented in Anderson et al. (in prep.).
In the present work, we find that NIR observations are essential for a reliable calculation of the bolometric luminosity at times later than $\sim$20$-$30~days after explosion. Therefore, the high-quality LCs and the considerable amount of NIR data in the CSP-I sample present the opportunity to calculate the most homogeneous and largest sample of \snii\ bolometric LCs with high observational cadence to date.
The 74 bolometric LCs from the CSP-I sample enable new calibrations for bolometric corrections (BCs) as a function of observed \snii\ colours from a statistically significant sample. Previous studies using an approach similar to that used in this paper only employed a small number of \snii\ events \citep{bersten+09,lyman+14}.
Furthermore, such a large sample allows a full characterisation of \snii\ bolometric LC properties.

The current paper is organised as follows. We give a brief description of the data sample in Sect.~\ref{sec:sample}. In Sect.~\ref{sec:blcs} we describe the methodology to calculate bolometric LCs and present them for the entire CSP-I \snii\ sample.
Section~\ref{sec:results} presents an analysis of the bolometric LC parameter distributions. New prescriptions for BCs versus optical colours are presented in Sect.~\ref{sec:bc}.
Finally, some concluding remarks are made in Sect.~\ref{sec:conclusions}.

\section{Supernova sample}
\label{sec:sample}

The data analysed in this study includes 74 \sneii\ from the CSP-I \citep[][PIs: Phillips and Hamuy; 2004$-$2009]{hamuy+06}. The sample is characterised by a high cadence and quality of photometric and spectroscopic observations, and a 
photometric coverage over a wide wavelength range for most \sneii.
The data set comprises LCs covering optical ($uBgVri$) and NIR ($YJH$) bands, and optical spectral sequences of nearby \sneii\ ($z$~$<$~0.05). A detailed description of the data reduction and photometric calibration is found in \citet{hamuy+06}, \citet{contreras+10}, \citet{stritzinger+11}, \citet{folatelli+13}, and \citet{krisciunas+17}.
Release of the final photometry for the SN~II sample is presented in Anderson et al. (in prep.).
Previously, CSP-I \snii\ \emph{V-}band photometry was published by \citet{anderson+14_lc}. CSP-I optical spectra were published by \citet{gutierrez+17II,gutierrez+17I}.
LCs, spectra, and colours of these \sneii\ were already analysed in several previous papers \citep{gutierrez+14,anderson+14_ha,anderson+14_lc,anderson+16,gutierrez+17II,gutierrez+17I,dejaeger+18,pessi+19}. 

The distance of each \snii\ was taken from \citet{anderson+14_lc} except for SN~2005kh, SN~2009A, and SN~2009aj since these \sneii\ were not included in that study. 
The weighted mean and standard deviation of redshift-independent distances taken from NED\footnote{\url{http://ned.ipac.caltech.edu/}} were employed for SN~2005kh and SN~2009A, where only Tully-Fisher estimates are available.
In the case of SN~2009aj, we used the distance from \citet{rodriguez+20}.
In \citet{anderson+14_lc} distances were estimated using cosmic-microwave background corrected recession velocities if this value is higher than 2000~km~s$^{-1}$, together with an $H_{\rm 0}$~=~73~km~s$^{-1}$~Mpc$^{-1}$, $\Omega_{m}$~=~0.27, and $\Omega_{\rm \Lambda}$~=~0.73.
Redshift-independent distances taken from NED were used when recession velocities are lower than 2000~km~s$^{-1}$.

Explosion epochs were taken from \citet{gutierrez+17I} with one exception. In the case of SN~2008bm, the new estimate by \citet{rodriguez+20} was used, who include pre-explosion non-detections closer to the SN discovery. 
It should be noted that SN~2005gk, SN~2005hd, and SN~2005kh do not allow for reliable explosion epoch estimates because of the lack of pre-discovery non-detections and/or spectra. 
Table~\ref{table:sn_info} lists the sample of \sneii\ analysed in this work, their distance modulus, explosion epochs, and Milky Way reddening values.

All magnitudes were first corrected for Milky Way extinction using the values from \citet{anderson+14_lc} and assuming a standard Galactic extinction law of 3.1 \citep{cardelli+89}.
While there are a number of methods in the literature to estimate the host-galaxy reddening, such as the \ion{Na}{i}~D equivalent width, the Balmer decrement, the strength of the diffuse interstellar bands, and the $V-I$ colour excess at the end of the plateau phase \citep[e.g.][]{hobbs74,munari+97,olivares+10}, the accuracy of these methods is unsatisfactory \citep[][among others]{phillips+13,galbany+16} and any attempt to correct for host-galaxy extinction is uncertain.
Therefore, we did not correct for host-galaxy extinction.
In addition, \citet{faran+14a}, \citet{gutierrez+17II}, and \citet{dejaeger+18} show that several correlations between observed parameters are stronger when host-galaxy extinction correction is not performed, suggesting that such corrections are simply adding noise to parameter estimations.
\citet{gutierrez+17II} and \citet{dejaeger+18} also suggest that host-galaxy extinction is relatively small for most \sneii\ in the current sample.
However, some \sneii\ within our sample do show particularly red colours \citep[SNe 2005gk, 2005lw, 2007ab, 2007sq, and 2009ao,][]{dejaeger+18} and/or high \ion{Na}{I}~D absorption equivalent widths \citep{anderson+14_lc} that may imply significant host-galaxy extinction in those specific cases.
In \citetalias{paper2_submitted}, we derive progenitor and explosion properties of the \sneii\ in CSP-I sample via hydrodynamical modelling of their bolometric LCs and expansion velocities. 
Given our ignorance on the degree of host-galaxy reddening for each \snii, to account for such effects we included a scale factor to our fitting procedure through the definition of our priors. This allows for more luminous bolometric LCs (through the effects of unaccounted host extinction) during the fitting \citepalias[see][for details]{paper2_submitted}.
Following the above, it is important that any future user of our published bolometric LCs considers these issues during their analysis.

\section{Bolometric light curves}
\label{sec:blcs}

\subsection{Light-curve phases}
\label{sec:lc_phases}

Throughout this article we refer to different phases of \snii\ LCs, so it is important to provide a clear definition of each phase and a brief discussion of the underlying physical processes involved. We do so in this section. 

When the core of a massive star collapses, a large amount of energy is deposited in the inner layers of the progenitor and a powerful shock wave starts to propagate outwards through the star's envelope.
The shock arrives to the stellar surface and photons begin to diffuse outwards providing the first electromagnetic signal of the SN. The shock emergence is referred to as the `shock breakout' and it is characterised by the fast increase of the bolometric luminosity.
Although this increase of the flux is expected in all photometric bands, it is much more significant at short wavelengths -- due to the high temperature of about 10$^{5}$\,K -- of the outermost ejecta during the shock-breakout phase \citep{grassberg+71,bersten+11}. As a consequence, the shock-breakout signal is expected to be detected more easily in X-ray or ultraviolet (UV) regimes. 
The combination between the non-predictability of the explosion and the short duration of the shock-breakout phase makes its detection extremely difficult with only few claimed cases \citep{campana+06,soderberg+08,garnavich+16,bersten+18}.

The shock breakout is followed by the fast expansion and rapid cooling of the outermost layers of the ejecta. At these times the bolometric LC declines rapidly, which we refer to as the cooling phase.
During the cooling phase the bulk of the emission shifts to longer wavelengths increasing the flux in the optical bands.
The duration of this phase is mostly sensitive to the progenitor size.
The increase of early time data has enabled recent studies to note that the cooling phase is sometimes larger than model predictions \citep[and rise times to maximum light of optical LCs are shorter, e.g.][]{gonzalez+15}.
The presence of additional material close to the progenitor is a possible solution to this issue. The interaction between the ejecta and a CSM lost by the progenitor star shortly before core collapse can boost the bolometric luminosity through the conversion of kinetic energy into radiation and produce longer cooling phases and shorter rise times in optical LCs \citep{morozova+18}. 
Additionally, the occurrence of such interaction in many \sneii\ is supported by the detection of narrow emission lines in very early spectra suggesting a confined CSM \citep[e.g.][]{khazov+16,yaron+17,bruch+21}.

Once the temperature drops to $\sim$6000~K, the LC settles on a phase commonly known as the `plateau' although the luminosity may not necessarily be constant. However, in the current work we decide to continue using this term because it is frequently used in the literature.
During the plateau, hydrogen recombination occurs at different layers of the ejecta as a recombination wave moves inwards in mass coordinates \citep{grassberg+71,bersten+11}. Hydrogen recombination reduces the opacity (dominated by electron scattering during this phase) allowing the radiation to escape. In addition, the plateau phase is also partially powered by the energy released from the radioactive decay chain of \Ni~$\rightarrow$~$^{56}$Co~$\rightarrow$~$^{56}$Fe \citep{kasen+09}.
The end of the plateau phase is marked by a sharp decline in luminosity known as the transitional phase.
The combination of the above mentioned phases is also known as optically-thick phase or photospheric phase.
After the transition from the plateau the luminosity is mostly powered by $^{56}$Co decay. This last phase is commonly referred to in the literature as the radioactive tail.

\subsection{Bolometric luminosity calculations}
\label{sec:blc_calc}

Bolometric LCs are extremely useful to constrain models of SN explosions. One of the goals of this series of papers is to derive physical properties for \sneii\ in the CSP-I sample from the modelling of their bolometric LC and expansion velocities.
Therefore, the first step is to calculate reliable bolometric luminosities for the entire sample. 

Two methodologies are often used in the literature to obtain the bolometric luminosity of SNe. 
Either the observed flux is integrated over all available photometric bands in addition to some corrections for the missing flux at shorter and longer wavelengths, or BCs are used to convert magnitudes in a certain photometric band (usually $V$- or $g$-band magnitudes) into bolometric magnitudes.
\citet{bersten+09} and \citet{lyman+14} provide relations of the BC as a function of optical colours during the photospheric phase of \sneii\ \citep[see also][]{pejcha+15b}. These allow one to easily calculate bolometric LCs when only optical magnitudes are available.
In our present study, we have a large set of LCs covering both optical and NIR bands (from $u$ to $H$). 
Having such a wide wavelength coverage justifies the calculation of the bolometric luminosity via full integration.
In this context, SN~1987A has one of the most accurate bolometric LCs calculated to date \citep{suntzeff+90} since it was observed from $U$ to far-infrared bands.
These observations allow the estimation of the fraction of the flux emitted outside the optical bands during the photospheric and radioactive tail phases, which were used as bolometric corrections to calculate the bolometric luminosity for other \sneii\ \citep[e.g.][]{schmidt+93,clocchiatti+96}.

As a first step, we calculated the pseudo-bolometric flux by direct integration of the broadband data. The pseudo-bolometric flux represents only the optical and NIR regime of the spectral energy distribution (SED) of the SN and does not consider the unobserved flux that falls at redder or bluer wavelengths. 
Although the radiation that falls beyond the NIR regime is not expected to contribute significantly to the bolometric flux (at least during the photospheric phase), it is necessary to take it into account to avoid a systematic underestimation. Something similar happens on the UV part of the SED. This region has its major contribution at early times ($\lesssim$\,20~days after explosion). However, the UV contribution drops as the SN cools and expands, and peak emission is set into longer wavelengths. 
The bolometric flux is then reached by summing the three flux components of the \snii: UV extrapolation plus optical--NIR integration plus IR extrapolation. 
We describe each method in detail in the following subsections. Some of these techniques were used in \citet{bersten+09}, \citet{lyman+14}, \citet{lusk+17}, and \citet{faran+18}.
Finally, the flux was transformed into luminosity using the distances given in Table~\ref{table:sn_info}. 

\subsubsection{Pseudo-bolometric flux estimation}
\label{sec:pseudobol}

\input{colour_calibration.tab}

\begin{figure}
\centering
\includegraphics[width=0.49\textwidth]{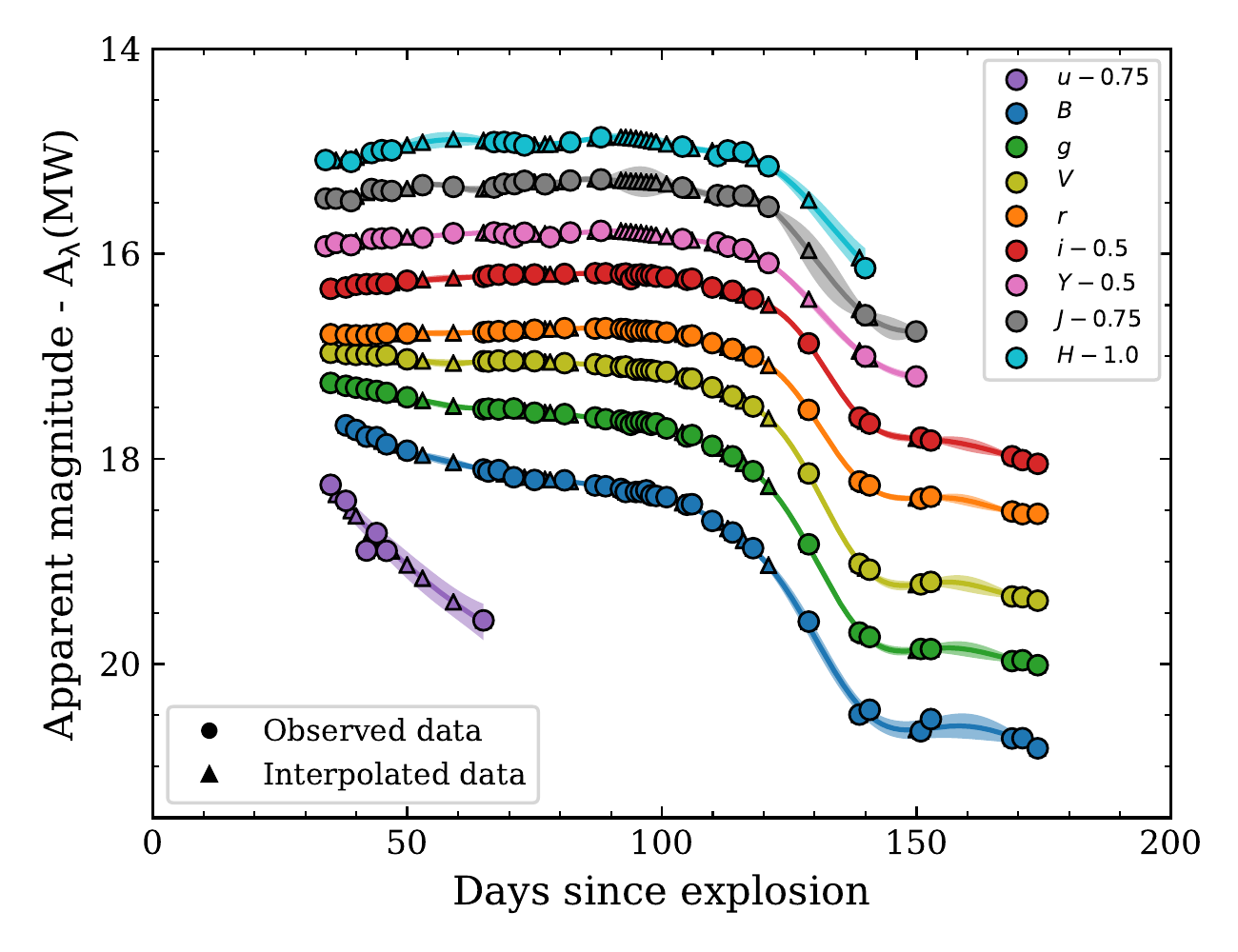}
\caption{Optical and NIR LCs of SN~2008ag corrected by Milky Way extinction. The dots are the observed data and the triangles are the interpolated magnitudes. Solid lines are the re-sampled LCs via Gaussian processes and the shaded regions represent the 95\% confidence intervals of the interpolation. The error bars of the observed magnitudes are smaller than the dot size.}
\label{fig:gp}
\end{figure}

\begin{figure}
\centering
\includegraphics[width=0.49\textwidth]{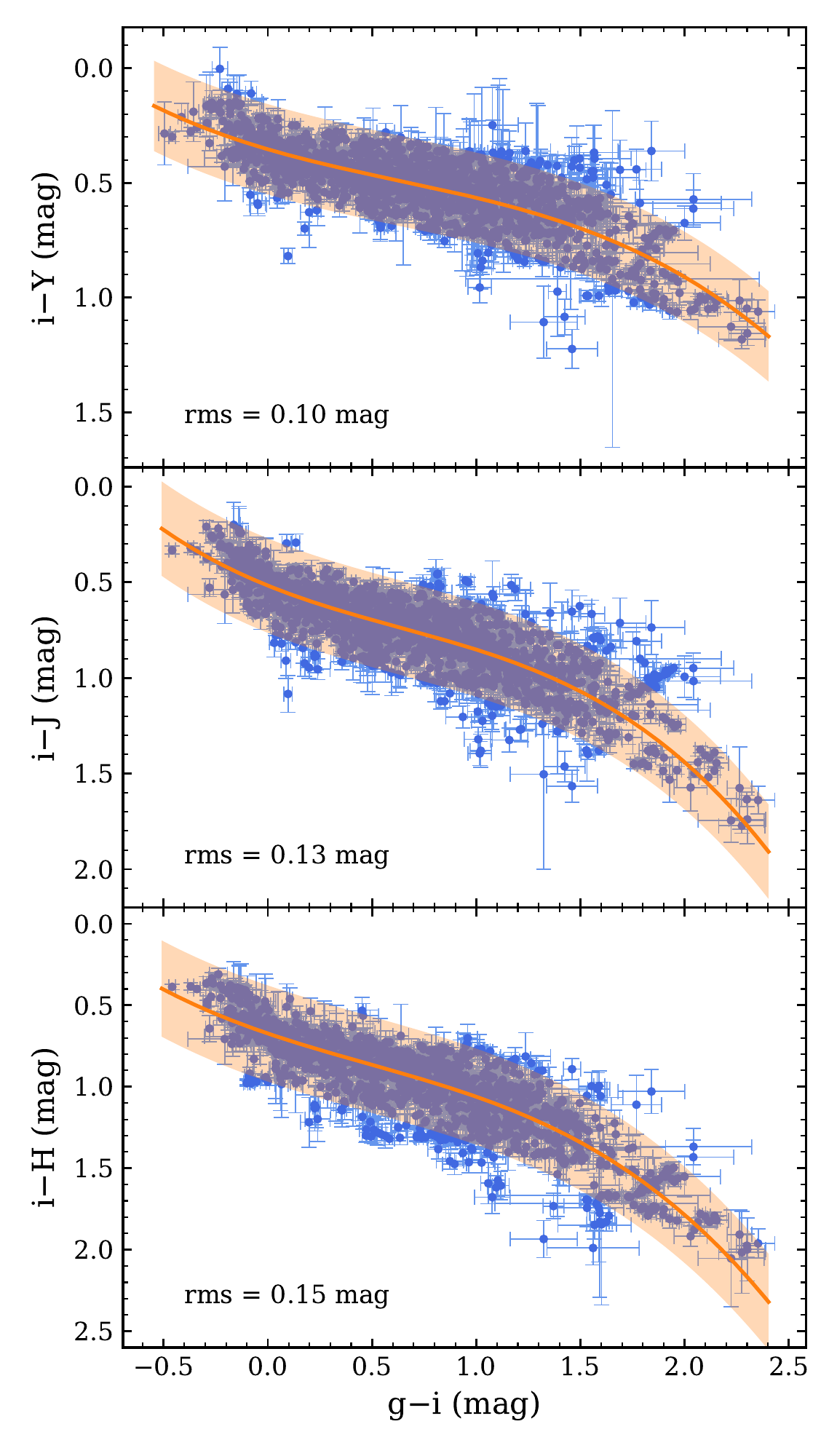} \\
\caption{Colour-colour diagrams for all objects. All measurements were first corrected for Milky Way extinction. Blue dots indicate different colour measurements. The solid orange line shows a polynomial fit to the data. Shaded regions represent the 95\% confidence intervals.}
\label{fig:colour_calibration}
\end{figure}

CSP-I photometry was obtained in optical ($uBgVri$) and NIR ($YJH$) bands for most objects. Although the CSP-I sample is characterised by high-cadence photometry, not all bands ($u$ to $H$) are always available at a given epoch. To obtain the full set of magnitudes at each epoch of observation we interpolated the LCs using Gaussian processes.
Gaussian processes were implemented using the \texttt{python} library \texttt{scikit-learn} \citep{scikit-learn}. 
Interpolated values are confident when observational gaps are small. A good representation of the missing part of the LC is easier to achieve when the observed LC is well sampled. In our sample, the largest gap during the photospheric phase is of the order of 40~days.
Something similar occurs during the radioactive tail phase. However, such large gaps are not common in the CSP-I data sample as it is characterised by a high cadence of observations. The average cadence in optical and NIR bands during the photospheric phase is 4.9 and 8.8~days, respectively. These values increase to 16.2 and 12.8~days during the radioactive tail phase.
Given that during the plateau and radioactive phases the LC changes smoothly, the interpolation was always assumed to be robust.
The most difficult part for the interpolation is the transition from the plateau to the radioactive tail. During the transition, the interpolation resembles the behaviour of the LC if observations exist near both sides of the transition (e.g. Fig.~\ref{fig:gp}).
If such observations were not available we did not interpolate the LCs during the transition to the radioactive tail. 
Extrapolations were only used when the re-sampled LCs were well-behaved and for times less than two days from the first or last epoch of observation.

CSP-I built a large sample of 74 \sneii\ where 47 have optical and NIR LCs with good temporal coverage, which allows a robust interpolation in all bands.
However, 10 out of the 74~\sneii\ present well-sampled optical LCs, while NIR LCs are partially sampled (e.g. the temporal coverage of the NIR data is around half that of the optical LCs), and 17 have scarce or no NIR data at all.
Thus, despite using interpolation, sometimes it is not possible to obtain a complete set of magnitudes at each epoch for the last two groups, particularly due to the scarce NIR band coverage.
Therefore, it is desirable to have a method to complete the NIR LCs of those 27~\sneii\ where NIR data availability is limited or non-existent (this is shown to be of significant importance in Sect.~\ref{sec:bb}).

To obtain NIR magnitudes for \sneii\ where no such data exist, we developed a method based on intrinsic optical colours using our whole sample of \sneii.
Once such relationships are established (see below for details), we are able to predict NIR photometry for events that lack these data, from their optical colours.
For those limiting cases where NIR LCs are partially sampled, we used the observed and interpolated magnitudes at all possible epochs and the predicted NIR magnitudes where these data were missing.
We used the $g-i$ colour, and related this to $i-Y$, $i-J$, and $i-H$ colours. These colours were chosen because: (a) \emph{g-} and \emph{i-}band LCs are always well sampled during the entire evolution of the \sneii\ in CSP-I sample, which allows the use of these relations in a wide range of $g-i$ colours; and (b) we produced polynomial fits for several combinations of colours and found that these sets have the smallest dispersions.
Third-order polynomial fits were found to produce satisfactory results. While higher-order polynomials might represent the bluest part better, these polynomials miss the reddest end of the distribution, where the latter is more important for our aims. As we show in Sect.~\ref{sec:bb}, NIR observations are crucial at times later than $\sim$30~days after explosion when \sneii\ are intrinsically redder than at earlier times.
Polynomial fits were computed using Markov chain Monte Carlo (MCMC) methods with the \texttt{python} \texttt{emcee} package \citep{emcee}.
Intrinsic colours together with the polynomial fits are shown in Fig.~\ref{fig:colour_calibration}. The coefficients of the polynomial shown correspond to the median of the marginal distributions, and are presented in Table~\ref{table:colour_calibration}. 
The relationship between $g-i$ and $i-Y$ colours presents the lowest dispersion about the polynomial fit with a root mean square (rms) of 0.10~mag. Dispersions of 0.13 and 0.15~mag are found for $i-J$ and $i-H$, respectively.

The above relations are based on observations that may correspond to different phases of the \snii\ evolution. Thus, we also tested the colour calibrations but separating the colours by different phases of the LC evolution: cooling phase (t~<~\ttrans), plateau phase (\ttrans~<~t~<~\tpt), and radioactive tail phase (t~>~\tpt), where \ttrans\ corresponds to the time of the transition between the initial decline and the decline of the plateau, and \tpt\ is the mid-point of the transition between the optically-thick phase and the radioactive tail \citep[see][and Sect.~\ref{sec:distributions} for details]{anderson+14_lc}. The values of \ttrans\ and \tpt\ were taken from \citet{anderson+14_lc}. These calibrations are shown in Fig.~\ref{fig:ccd_phases}.
We find that the dispersions of these relations are usually larger than when all colours are taken together. In addition, the continuous behaviour between different epochs tend to favour the utilisation of the combined data set. 
A large fraction of the observations during the tail phase ($\sim$30\%) corresponds to only one \snii\ (SN~2008bk) which has a peculiar behaviour in the colour curves during this phase evolving in time to bluer $g-i$ colours and redder ($i-NIR$~magnitude) colours. 
This is because the $i$-band LC declines faster (0.014~mag per day) than the $g$-, $Y$-, $J$-, and $H$-band LCs ($\sim$0.010~mag per day).
We explore whether this behaviour is intrinsic of low-luminosity \sneii, but we do not have data during the tail phase for other low-luminosity events to generalise such a tendency. No other \sneii\ in the sample display such behaviour. However, we remark that our observations during the radioactive tail are limited and no strong conclusions can be drawn.
The large amount of observations of SN~2008bk and its peculiar behaviour during the radioactive tail may bias the colour-colour relations at this phase. Therefore, we choose not to separate our sample in different phases until more data are obtained in the radioactive tail phase and we continue to use the complete colour curves without any separation respect to their LC evolutionary phases (Fig.~\ref{fig:colour_calibration}).

We developed a method based on intrinsic optical colours in which we use polynomial relations to obtain $i-Y$, $i-J$, and $i-H$ colour measurements from $g-i$ colours for the \sneii\ in our sample with limited (or no) NIR data. Therefore, we can estimate NIR magnitudes when photometry in these bands is not available. The errors of each estimate were obtained using the rms of the residuals. 
As outlined in the following subsection, NIR data are crucial for reliable BB fits and estimation of the missing flux at longer wavelengths. These simple relations (see Table~\ref{table:colour_calibration}) allow one to estimate NIR magnitudes from optical colours and can be used in future works\footnote{The validity of using these NIR estimations for the use of bolometric LC production is checked in the analysis in Sect.~\ref{sec:comp_blcs}.}.

Armed with photometric measurements or estimations in all $u$ to $H$ bands at all epochs, pseudo-bolometric LCs were computed. The magnitudes were converted to monochromatic fluxes at the mean wavelength of the filter using the transmission functions and zero-points of the photometric system available on the CSP website\footnote{\url{https://csp.obs.carnegiescience.edu/data/filters}}. The monochromatic fluxes were then integrated using the trapezoidal method, and the pseudo-bolometric flux was obtained. The flux bluewards and redwards of the mean filter wavelength range was set to zero.

\subsubsection{Extrapolation to redder wavelengths}
\label{sec:bb}

\begin{figure}
\centering
\includegraphics[width=0.49\textwidth]{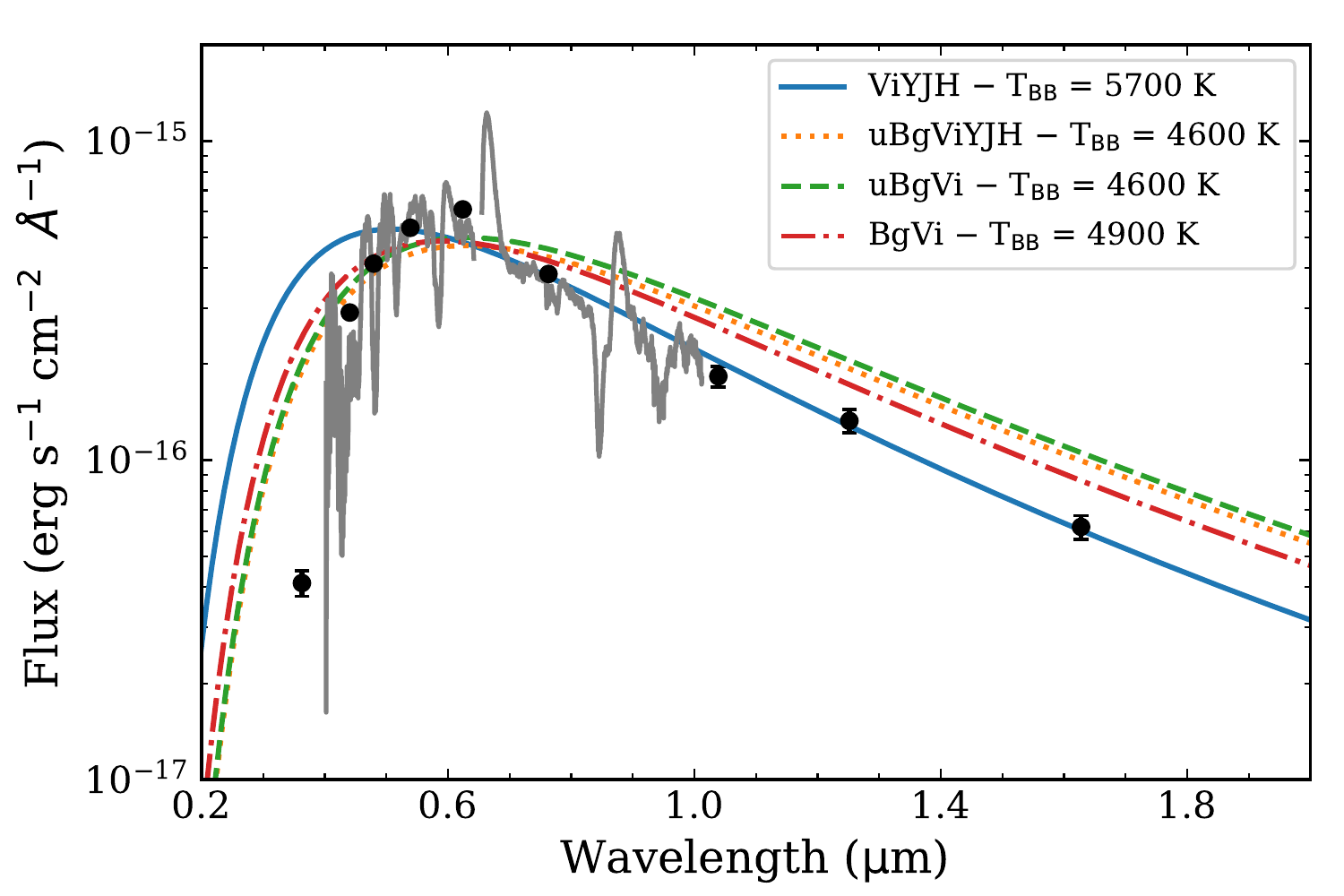}
\caption{BB fits for SN~2008if at 72 days after explosion. Different linestyles and colours correspond to different set of bands used for the fitting process. The blue solid line is the optimal model since data affected by line blanketing ($u$, $B$, and $g$ bands) are removed from the fit. The best-fitting BB temperatures (T$_{\rm BB}$) are indicated in the legend. An optical spectrum taken at 70~days after explosion is presented in grey. It shows that \halpha\ line profile produces an increment of the $r$-band flux above the BB model.}
\label{fig:bbfits}
\end{figure}

\begin{figure}
\centering
\includegraphics[width=0.5\textwidth]{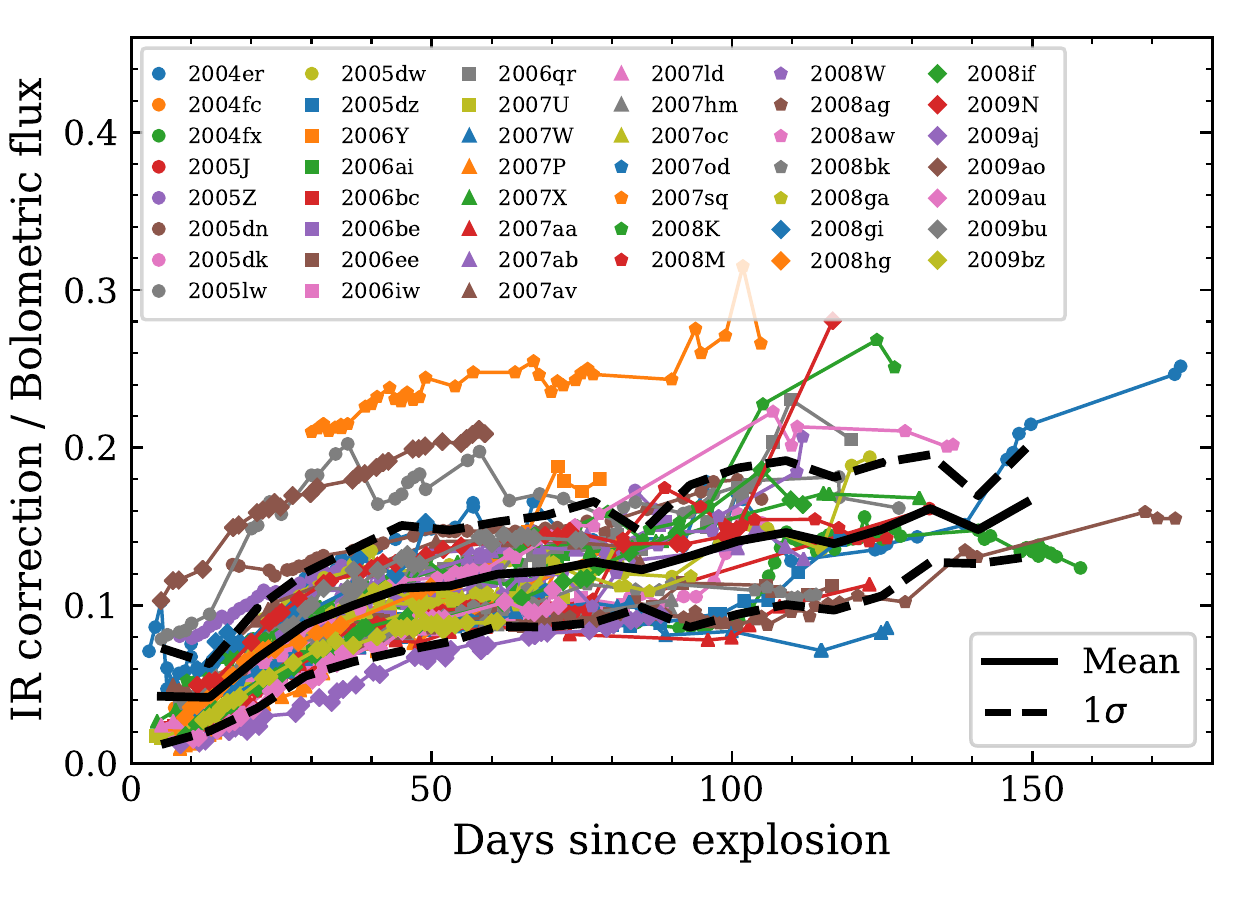}
\caption{Fraction of the IR correction over the bolometric flux as a function of time. The black solid line indicates the mean values within each time bin, while the dashed lines represent the standard deviation. We only include \sneii\ with observed $YJH$ photometry. Three \sneii\ (SNe 2005lw, 2007sq and 2009ao) display unusual values beyond 2$\sigma$ from the mean. This may be due to a significant uncorrected reddening in the host galaxy (see text).}
\label{fig:ircorr}
\end{figure}

One of the most frequently used techniques to correct for the missing flux in the IR is to assume that the SN emission in this region is well described by a BB function. At early epochs, the BB model provides a good representation of the SN flux in all optical and NIR bands. As the SN cools and expands, the SN emission at short wavelengths starts to depart from the BB model because of the increasing line blanketing produced by iron-group elements. 
Figure~\ref{fig:bbfits} shows the effect of considering the bands affected by line blanketing in the fitting process. In this example, we take the fluxes for SN~2008if at 72~days after explosion. At this time, the $u$, $B$, and $g$ bands are already affected by line-blanketing effects and are removed for the fitting (blue solid line).
Figure~\ref{fig:bbfits} clearly shows that when these bands are taken into account for the fitting, the BB model tends to peak at redder wavelengths resulting in an overestimation of the extrapolated IR flux (from $H$ band to longer wavelengths), and therefore in an overestimation of the bolometric luminosity (see Sect.~\ref{sec:comp_blcs}).
To solve this issue, we followed a procedure similar to that of \citet{faran+18}. When the \emph{u-}band flux dropped 1$\sigma$ below the BB model, we ran the BB fitting process again but this time excluding the \emph{u} band from the fitting. The same was done for the $B$, $g$, and $V$ bands.
The SN emission also departs from the BB model at long wavelengths since the flux past $\sim$2~$\mu$m is dominated by free-free emission \citep{davis+19}. However, this should not affect our analysis.

Additionally, during the fitting process, we noted that the $r$-band flux increases with time with respect to the BB model and is located above the BB for the vast majority of the \sneii\ in our sample. 
This is because \halpha\ increases in strength over time. Figure~20 from \citet{gutierrez+17I} shows that the equivalent widths of both absorption and emission components of \halpha\ increase with time. However, the ratio of absorption to emission of the \halpha\ profile is almost constant at $\sim$0.3 after 40~days from explosion \citep[see][their Fig.~22]{gutierrez+17I}. This means that the \halpha\ feature becomes a stronger (positive) contribution to the $r$-band flux with time with respect to the continuum.
In Fig.~\ref{fig:bbfits}, we include a spectrum taken 70 days after explosion showing the strength of \halpha\ profile at that time.
Thus, when the $r$ band departed 1$\sigma$ from the BB fits, we removed this band in the same way we did with the bluer bands.

The above indicates that $uBgVr$ bands are omitted from the fitting around the middle of the photospheric phase. We conclude that NIR data are crucial for the BB fitting since the $i$ band is the only optical band that is not removed from the fits.
In this context, the method presented in Sect.~\ref{sec:pseudobol} is necessary to obtain NIR magnitudes for \sneii\ from observed optical colours.

Once we found a BB model for the observed SED (excluding line blanketing and \halpha\ effects), the missing flux in the IR was estimated by the integral of the BB emission between the reddest observed band and infinity, and is referred to as the `IR correction' (see Fig.~\ref{fig:ircorr}). This procedure was repeated at every epoch of observation.
After the recombination phase the ejecta turns optically thin and line emission dominates the radiation of the \snii. During this phase, BB fits are not appropriate to extrapolate the SN radiation in the IR regime from the physical point of view.
However, we used them to estimate IR corrections given that in Appendix~\ref{app:extrapolations} we showed that they can represent most of the flux at long wavelengths.

Figure~\ref{fig:ircorr} shows the contribution of our IR correction with respect to the total flux as a function of time, for those \sneii\ with observed $YJH$ photometry.
As expected, the contribution of the IR at early times is small ($\sim$5\% of the total flux). The IR correction increases with time taking a mean value of $\sim$10\% during the plateau phase, and $\sim$16\% in the radioactive tail phase.
We note that three \sneii\ are notably different during the photospheric phase: SN~2005lw, SN~2007sq, and SN~2009ao. It is not surprising that these three \sneii\ are within the 10\% reddest of the sample \citep{dejaeger+18}, which may imply significant host-galaxy extinction. If this is the case, a significant uncorrected reddening in the host galaxy would produce larger fluxes in the bluest bands with respect to the NIR bands yielding smaller contributions to the IR, and thus eliminating these objects as outliers.

\subsubsection{Extrapolation to UV wavelengths}
\label{sec:uvcorr}

UV flux contributes significantly at early epochs when the \snii\ emission peaks in these wavelengths. 
\citet{bersten+09} used synthetic \snii\ spectra and noted that UV corrections can be as large as $\sim$50$-$80\% of the total flux at early times when the \snii\ is very blue, but becoming almost negligible when the plateau phase is settled.
This means that the contribution of the UV regime is significant, at least at very early epochs.
Therefore, it is certainly important to have UV data during the first weeks after explosion to account for the real UV flux. However, high-cadence observations from UV to NIR bands for a large sample of \sneii\ are difficult to obtain. 
Consequently, a method is required to estimate the UV contribution for our sample.

In the current work, the UV correction was extrapolated from the \emph{u} band to $\lambda$~=~0 using the BB model on all epochs except when the \emph{u} band was removed from the fitting process (see Sect.~\ref{sec:bb}). In the latter cases, the UV correction was defined as the flux below the straight line from the bluest-band flux to zero flux at $\lambda$~=~2000~\AA, as proposed by \citet{bersten+09}. These authors chose $\lambda$~=~2000~\AA\ based on synthetic \snii\ spectra showing that the flux bluewards of 2000~\AA\ is negligible in comparison with the bolometric flux.
After the transition to the radioactive tail phase, we did not consider any UV correction since it becomes unimportant at these epochs \citep[see][their Fig.~1]{bersten+09}.

\citet{lyman+14} utilised \emph{Swift}-UV data for six stripped-envelope SNe and two \sneii\ to test the above-mentioned approaches and found generally good agreement. The linear extrapolation shows a discrepancy of less than 5\% of the bolometric luminosity for most of the observations, but unfortunately, most of them correspond to stripped-envelope SNe, with only one \snii\ being used (showing good agreement).
In Appendix~\ref{app:extrapolations}, we used seven \sneii\ in the CSP-I sample with available \emph{Swift}-UV data to test our extrapolation method to the UV regime. The details of this test are outlined and these give confidence to our bolometric calculations for times $\gtrsim$\,20~days from explosion.
At the same time, our results involving bolometric luminosities at times $\lesssim$\,20~days should be taken with caution as our calculation method may not predict the actual UV contribution at those epochs.

\subsubsection{CSP-I SN~II bolometric light curves}
\label{sec:csp_snii_blcs}

\begin{figure}
\centering
\includegraphics[width=0.5\textwidth]{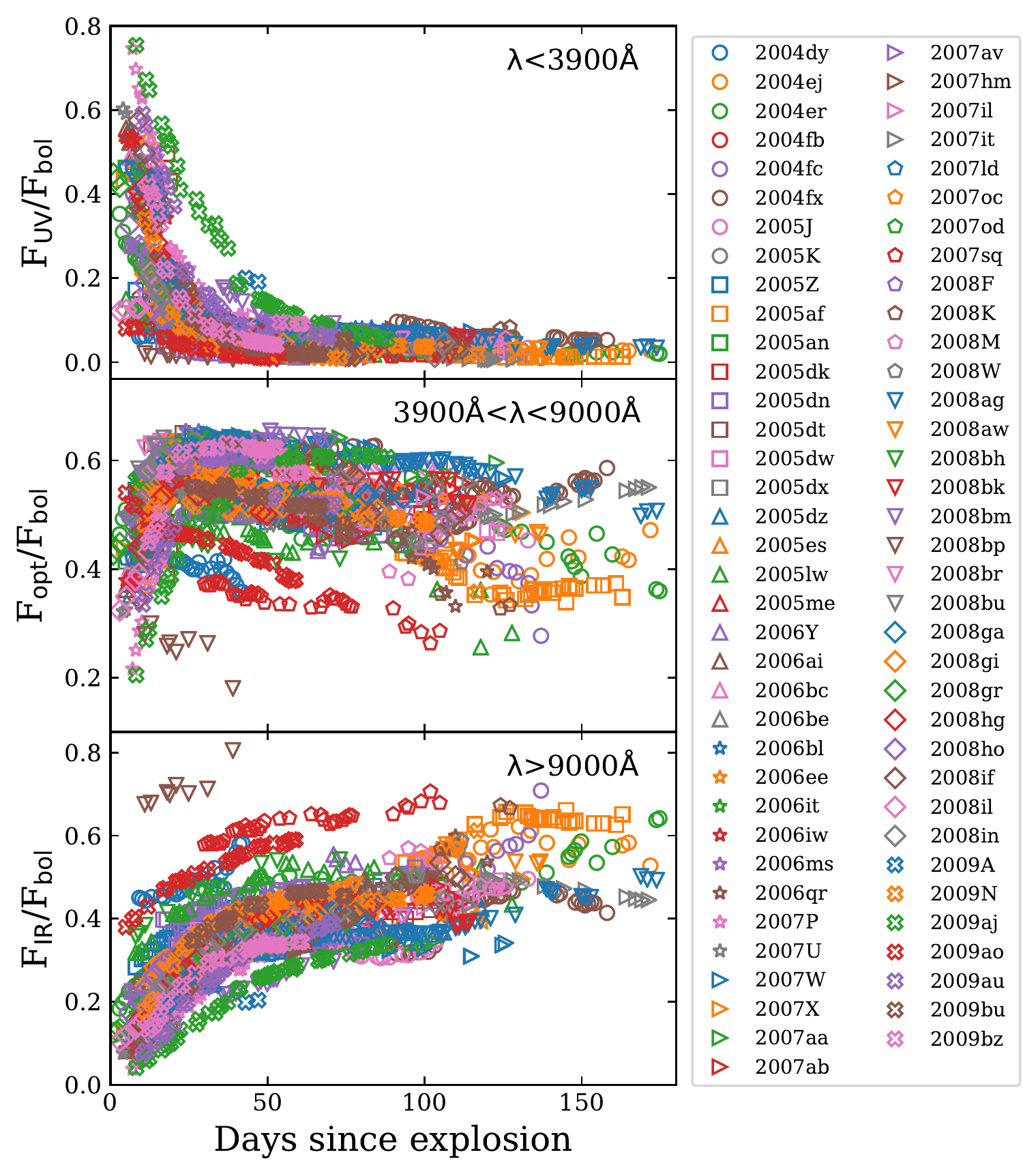}
\caption{Contribution of the UV, optical, and IR fluxes to the bolometric flux of CSP-I \sneii\ as a function of time. UV and IR fluxes include extrapolations. Only \sneii\ with explosion epoch estimates are shown.}
\label{fig:frac_flux}
\end{figure}

\begin{figure}
\centering
\includegraphics[width=0.49\textwidth]{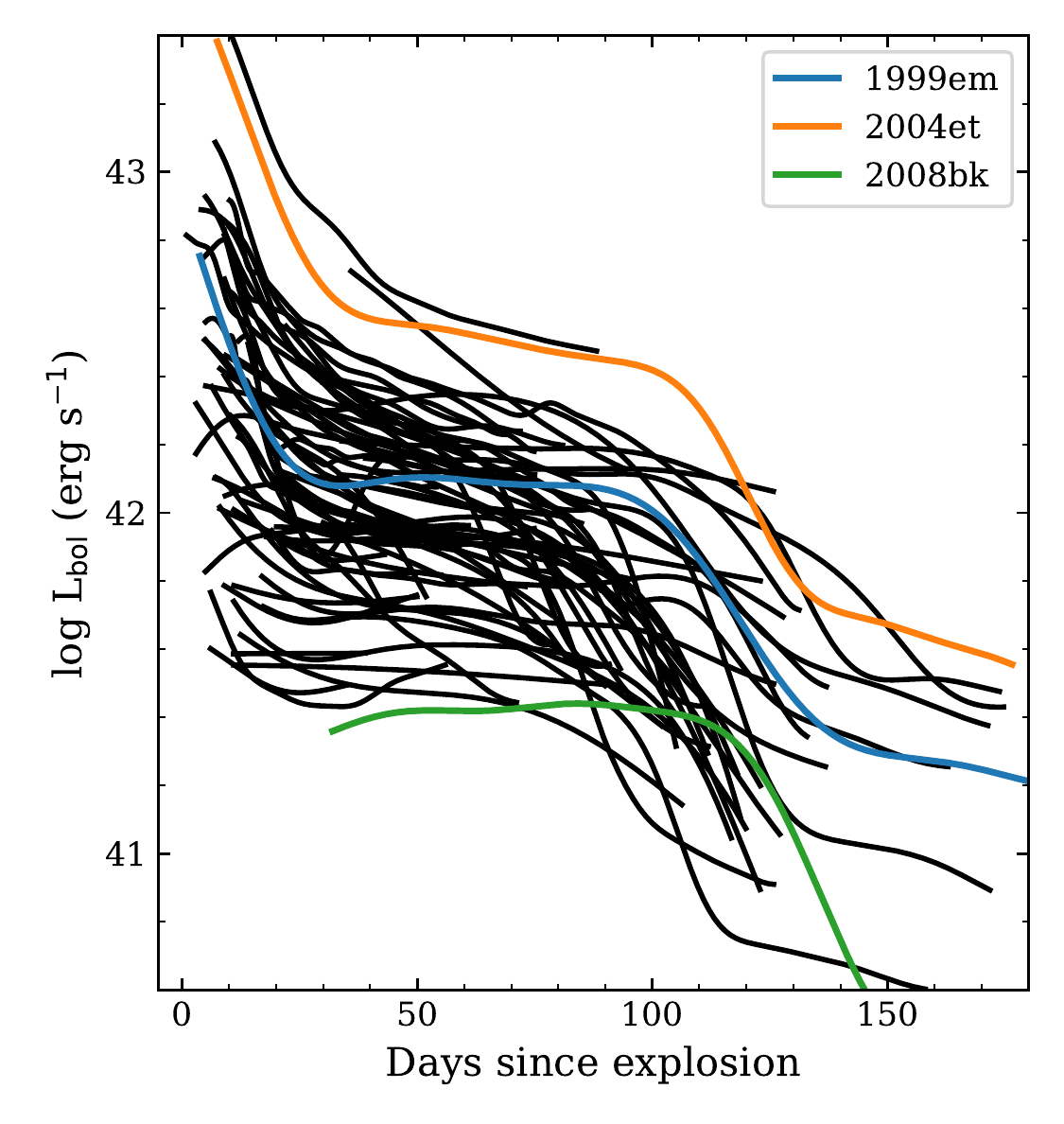}
\caption{Bolometric LCs of the CSP-I \snii\ sample. LCs are re-sampled via Gaussian processes for better visualisation, and displayed by black lines. For comparison, we show in colours three well-studied \sneii: 1999em \citep{bersten+09}, 2004et \citep{martinez+20}, and 2008bk (this work). Only \sneii\ with explosion epochs estimates are shown.}
\label{fig:blcs}
\end{figure}

The sum of the pseudo-bolometric flux and the missing flux falling at unobserved wavelengths equates to the bolometric flux. 
The calculation of each contribution of the bolometric flux was performed through a simple Monte Carlo procedure. 
For each of the thousand Monte Carlo simulations we randomly sampled the broadband magnitudes within their uncertainties and integrated the pseudo-bolometric flux, found the best-fitting BB model, and calculated the IR and UV corrections. 

Fig.~\ref{fig:frac_flux} shows the observed plus extrapolated flux in the UV ($\lambda$~$\leq$~3900~$\AA$), optical (3900~$\leq$~$\lambda$~$\leq$~9000~$\AA$), and IR ($\lambda$~$\geq$~9000~$\AA$) with respect to the bolometric flux as a function of time.
In general, optical fluxes dominate the SN emission representing between 50 and 70\% of the bolometric flux for most \sneii\ in the sample during the entire evolution.
NIR fluxes contribute $\sim$10\% of the total flux at early epochs, increasing with time until values of $\sim$50\%.
The UV regime provides a large contribution at early times reaching values of about 80\% for a few \sneii, but after 30~days from explosion its contribution drops to less than $\sim$10\% for most \sneii\ in the CSP-I sample.
We note that three \sneii\ (SN~2007sq, SN~2009ao, and SN~2008bp) show markedly lower (larger) optical (NIR) fractions than the general trend. The first two objects are within the 10\% reddest of the CSP-I \snii\ sample \citep{dejaeger+18}, while SN~2008bp shows particularly red colours and a high pseudo-equivalent width of \ion{Na}{i}~D, possibly implying significant uncorrected host-galaxy extinction (see also Sect.~\ref{sec:bb}).

The mean bolometric flux of the thousand simulations was calculated and taken as the final bolometric flux. We took the standard deviation as its error. This procedure was employed for each epoch of observation.
Then, we transformed flux into luminosity using the distance to each object. The distance errors were propagated to the final luminosity.
In the calculation of the bolometric luminosity we deal with two major sources of systematic uncertainties: host-galaxy extinction and distance.
Previously, we stated that we do not correct for host-galaxy extinction, which may produce systematically dimmer bolometric LCs.
Distance is the largest source of uncertainty, sometimes producing notably different bolometric luminosity LCs for distance values within error bars.
Therefore, we publish bolometric flux LCs instead of luminosities. In this way, any future user of our bolometric LCs can choose different distance estimations for their analysis.
Bolometric fluxes for the full sample can be found at the CSP website\footnote{\url{https://csp.obs.carnegiescience.edu/data}}.
Figure~\ref{fig:blcs} shows the bolometric LCs for our full SN~II sample. LCs were re-sampled via Gaussian processes for better visualisation. Figure~\ref{fig:blcs} shows a large range in luminosities and LC morphologies, the analysis of which is presented in Sect.~\ref{sec:distributions}.

\subsection{Comparison of bolometric light curves from different calculation methods}
\label{sec:comp_blcs}

\input{physical_pars.tab}

\begin{figure*}
\centering
\includegraphics[width=0.45\textwidth]{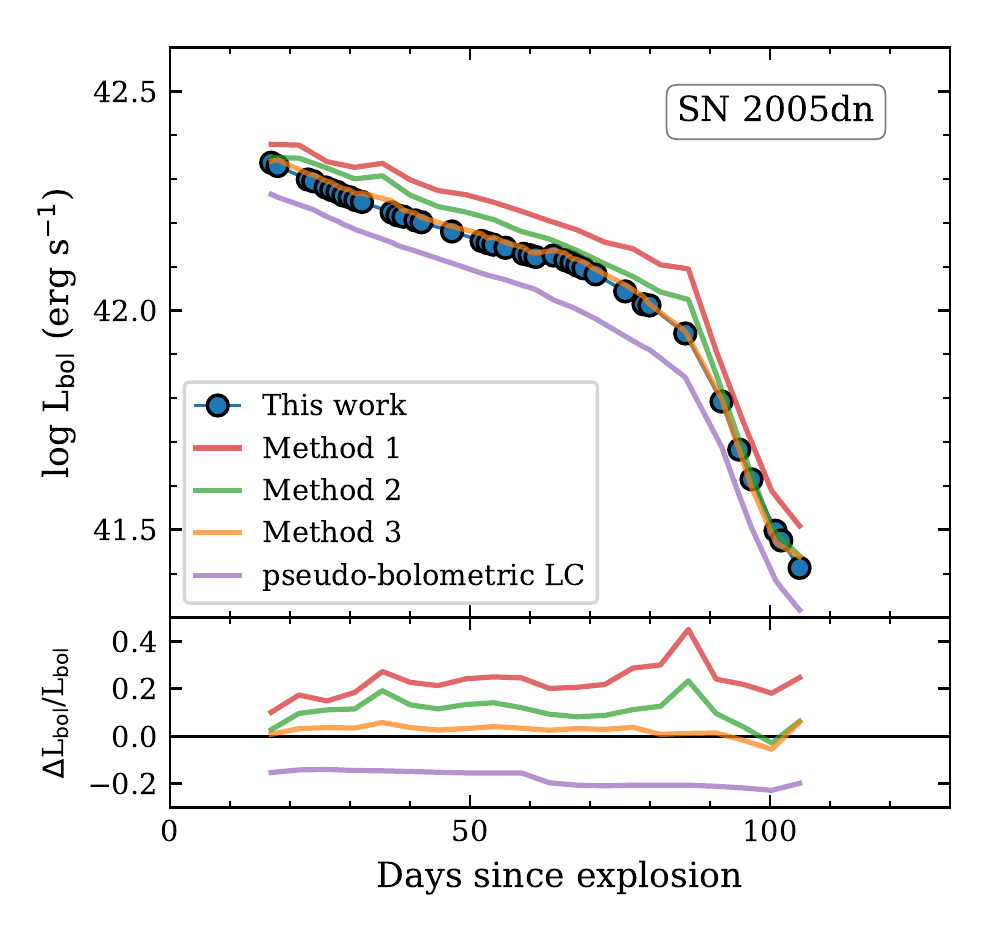}
\includegraphics[width=0.45\textwidth]{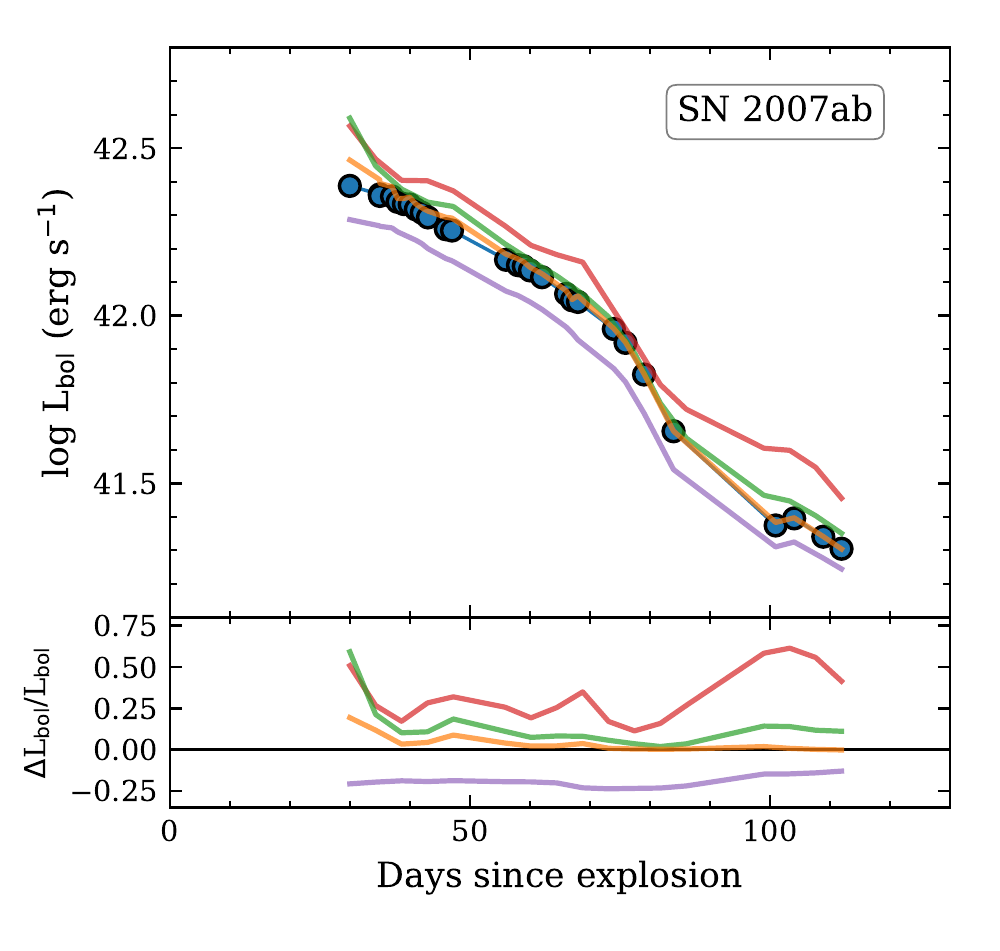}
\includegraphics[width=0.45\textwidth]{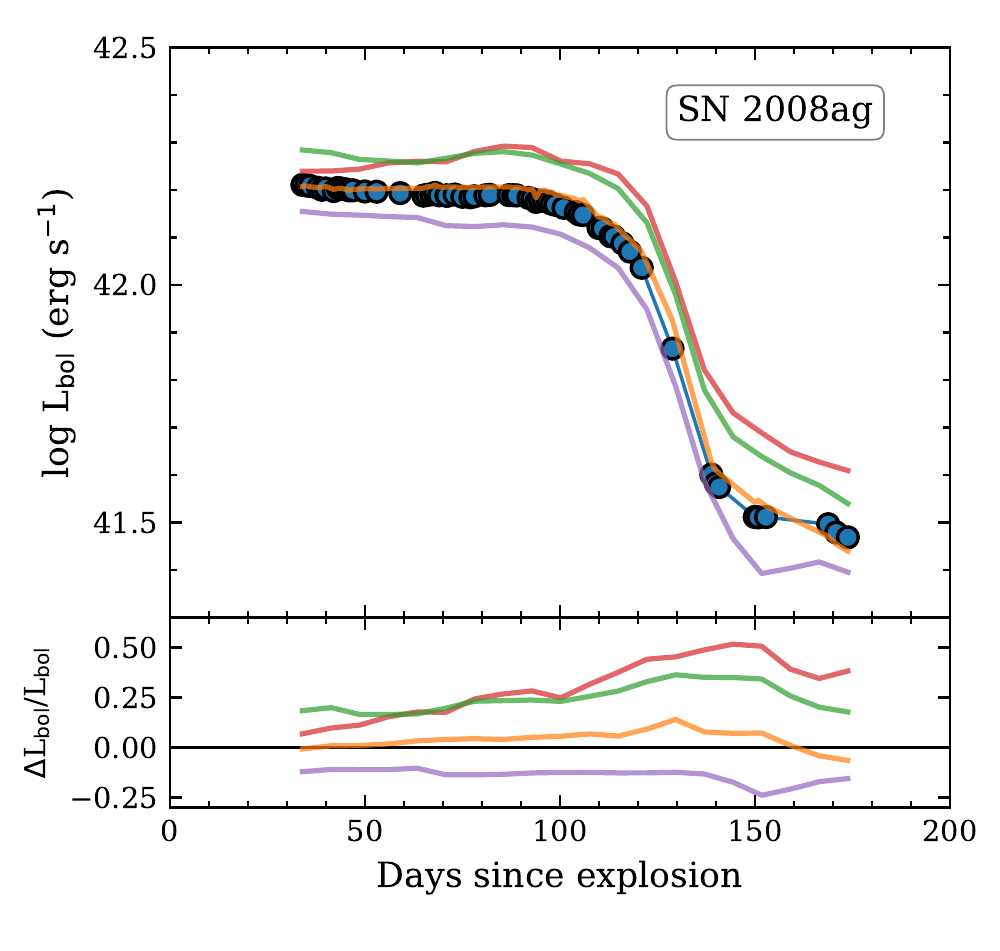}
\includegraphics[width=0.45\textwidth]{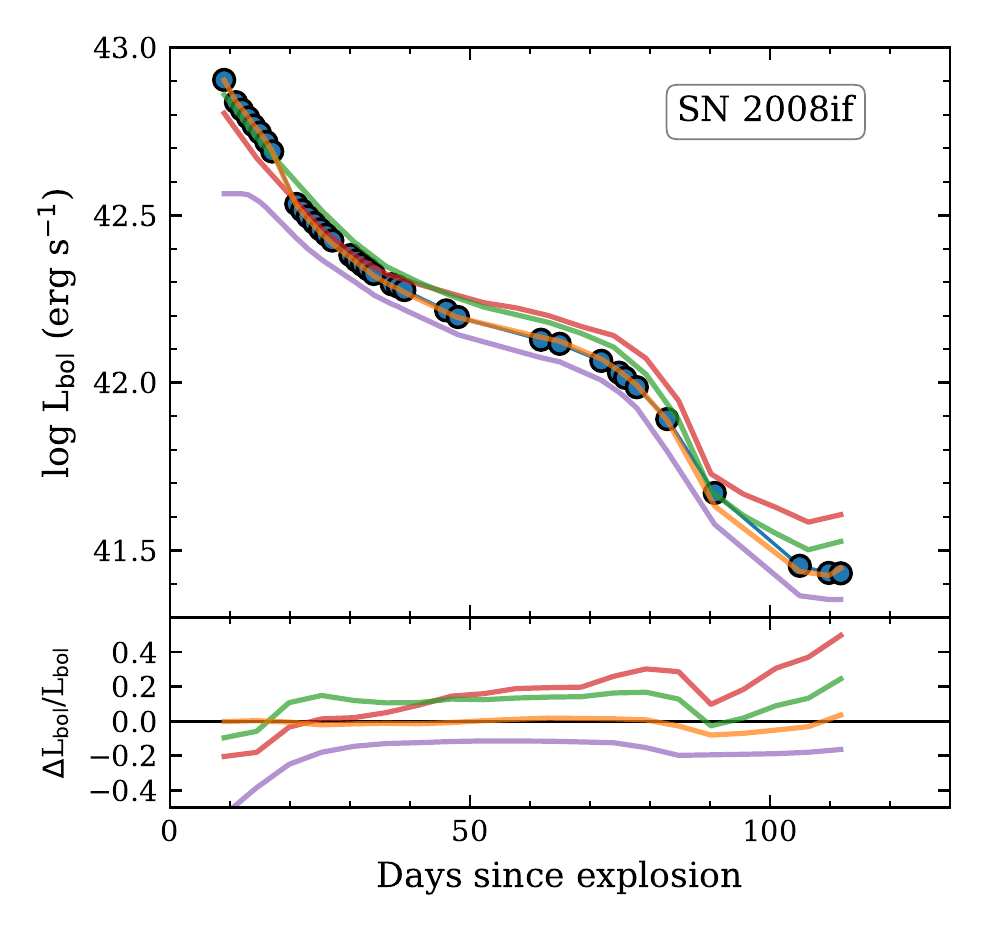}
\caption{Comparison of bolometric LCs for SN2005dn (top left), SN~2007ab (top right), SN~2008ag (bottom left), and SN~2008if (bottom right) using different methods for their calculation (see text). The relative differences are with respect to the bolometric luminosities calculated using our method.}
\label{fig:comp_blcs}
\end{figure*}

In the present study, we construct the largest sample of \snii\ bolometric LCs to-date employing a consistent method for all the objects using high-quality and high-cadence observations characterised by a wide wavelength coverage including optical and NIR photometry.
However, such high-quality data are often not common in samples of \sneii. \snii\ LCs evolve for hundreds of days, which make follow-up observations in several optical and NIR bands time consuming. For this reason, a large number of \sneii\ in the literature are only observed in optical bands, sometimes with very few NIR observations.

Having only optical observations it is possible to estimate a bolometric LC by means of BCs \citep[][see also Sect.~\ref{sec:bc}]{bersten+09,lyman+14,pejcha+15b}, or a pseudo-bolometric LC, that is, through integration of the flux over the optical range \citep[e.g.][]{valenti+16}.
However, many studies have estimated bolometric luminosities from BB fits using only optical data, which may cause large systematic errors. In the current section, we analyse these issues.

To test these systematics, we calculated the bolometric LCs for the CSP-I \snii\ sample again but using only optical data through the following two methods commonly used in the literature.
The first method (`Method~1') integrates the flux from the optical bands and then performs BB fits from optical photometry without excluding any of the bluest bands affected by line blanketing effects. UV and IR corrections are estimated by BB extrapolations at each epoch.
The second method (`Method~2') finds the best-fitting BB parameters from optical photometry but excluding the band which includes the \halpha\ line. Then, the luminosity is calculated from the Stefan–Boltzmann law by means of the temperature and radius derived from the fits.
Additionally, we included another method (`Method~3') to analyse the effect of using the colour-colour relations presented in Sect.~\ref{sec:pseudobol} in the bolometric luminosities (see details below).

We start by comparing the bolometric LCs resulting from the first two methods against those resulting from our method (Sect.~\ref{sec:blc_calc}). Methods 1 and 2 almost always produce higher luminosities than our approach, with only a few events with similar bolometric LCs. In addition, it is generally found that bolometric LCs from Method 1 are more luminous than those from Method 2.
Figure~\ref{fig:comp_blcs} compares bolometric LCs for the four \sneii\ with the largest differences. Additionally, the relative differences are displayed at the bottom of each subplot with respect to the bolometric luminosity calculated using our method.
SN~2005dn (top-left panel) and SN~2007ab (top-right panel) show the behaviours discussed above. Bolometric LCs from Methods 1 and 2 are more luminous than those from our calculation method, with those from Method 1 being the most luminous. In the case of SN~2005dn, the relative differences are almost constant during the \snii\ evolution being around 10\% and 20\% for Methods 2 and 1, respectively.
Something similar happens for SN2007ab. For Method 2 the differences are between 10$-$15\%, except for the first value that shows a discrepancy of about 60\% in luminosity. Method 1 gives a bolometric LC that is $\sim$25\% more luminous during plateau, increasing to $\sim$50$-$60\% at the radioactive tail phase.
For SN~2008ag and SN~2008if, the bolometric LCs from Methods 1 and 2 are similar but both showing large differences compared to our method.

For the comparison with Method~3, we used the same four \sneii\ as above. In this case, we omitted observed NIR magnitudes and instead, we predicted them using the colour calibrations from which NIR magnitudes can be estimated using $g-i$ colours.
For consistency, we re-calculated the polynomial coefficients of these relations for each of these four cases, every time removing the object under analysis.
We note excellent agreement between our bolometric LCs and those calculated with predicted NIR photometry using our colour-colour relations (Fig.~\ref{fig:comp_blcs}). 
This validates the use of our method to estimate NIR magnitudes in the absence of such observations.
For completeness, in Fig.~\ref{fig:comp_blcs} we also include the pseudo-bolometric LCs, that is the integration of the observed fluxes extending from $u$/$B$ to $H$ band. These pseudo-bolometric LCs are $\sim$20\% less luminous than their bolometric counterparts.

In all the examples shown in this section, it is clearly seen that the bolometric luminosities from Methods 1 and 2 (i.e. BB fits without NIR data) highly overestimate the luminosity with respect to our more robust methodology. 
Figure~\ref{fig:bbfits} shows the reason for this discrepancy. Clearly, BB fits without NIR data (green dashed line and red dash-dot line) are above the observed NIR photometry. Therefore, the flux redwards of the $i$ band is overestimated.
We conclude that NIR observations are crucial to better reproduce the behaviour of the BB fits at longer wavelengths, and therefore for reliable BB extrapolations to the IR.
In this sense, we encourage future works to use the relations presented in Table~\ref{table:colour_calibration} to estimate NIR magnitudes from optical colours when NIR observations are not available. Alternatively, one may use the calibrations of BC against colours \citep{bersten+09,lyman+14,pejcha+15b}, as we discuss further in Sect.~\ref{sec:bc}.

The bolometric LCs from this study are used to derive \snii\ progenitor and explosion properties through hydrodynamical modelling in \citetalias{paper2_submitted}. Therefore, we also quantify the differences in the physical parameters derived via hydrodynamical modelling that occur when the bolometric LCs are calculated from different methods. This highlights the need for accurate bolometric LC construction. 

For this test, we used three of the \sneii\ previously mentioned (SN~2005dn, SN~2007ab, and SN~2008if). The case of SN~2008ag is analysed later. In all cases, we used our bolometric LCs and those from Method 1, as the latter is the method producing the greatest differences.
\snii\ physical properties using our bolometric LCs are presented and analysed in \citetalias{paper2_submitted}. Here, we briefly discuss our modelling procedure and then focus on the analysis of the discrepancies. 
The reader is referred to \citet{martinez+20} and \citetalias{paper2_submitted} for a detailed description.
We used the stellar evolution code \texttt{MESA} \citep{paxton+11,paxton+13,paxton+15,paxton+18} to obtain RSG-star models at time of collapse for initial masses between 9 and 25~\ms\ in steps of 1~\ms. 
For each progenitor model, we calculated bolometric LC and photospheric velocity models for a large range of explosion parameters: explosion energy ($E$), \Ni\ mass (\mni), and its spatial distribution within the ejecta, \Ni\ mixing. We used a one-dimensional code \citep{bersten+11} that simulates the explosion of the progenitor star.
This grid of hydrodynamical models was previously presented in \citet{martinez+20}. Finally, we employed MCMC methods to find the posterior probability of the model parameters of each observed \snii.

Table~\ref{table:physical_pars} compares the progenitor and explosion properties using the bolometric LCs presented in the current study and the bolometric LCs from Method~1.
The largest differences are found for $E$ and \mni. Discrepancies of the order of 0.3~foe (1~foe~$\equiv$~10$^{51}$~erg) are found for the explosion energies of SN~2007ab and SN~2008if. A smaller difference of $\sim$0.2~foe is found for the explosion energy of SN~2005dn.
Differences in \mni\ are even more notable. The more luminous tail phases derived from Method 1 (Fig.~\ref{fig:comp_blcs}) produce significantly larger \Ni\ masses, with differences of 0.018 and 0.023~\ms\ for SN~2007ab and SN~2008if, respectively.
This analysis shows that for these three \sneii, \mni\ estimations using bolometric LCs from Method~1 might be overestimated between $\sim$20 and 55\%.
This may have important implications for the previously analysed  distributions of \snii\ \Ni\ masses as compared to stripped-envelope SNe \citep{kushnir15,anderson19}.
We emphasise that differences are always offset in the same direction, that is, larger $E$ and \mni\ are found for the LCs calculated using Method~1. This is because the LCs from Method~1 are always more luminous during plateau and radioactive phases.
With respect to progenitor and ejecta mass estimates, no significant differences are found.

SN~2008ag presents large discrepancies between our bolometric LC and that from Method 1, with the latter being significantly more luminous.
Unfortunately, we cannot model the LC of SN~2008ag constructed from Method 1 for the following reason. 
The grid of explosion models used covers a wide parameter space. Specifically, \mni\ ranges from 0.0001~\ms\ to 0.08~\ms\ in different intervals \citep[see][and \citetalias{paper2_submitted}]{martinez+20}.
In \citetalias{paper2_submitted}, we derived a \Ni\ mass near the upper limit of our range for SN~2008ag, \mni~=~0.065~\ms.
When trying to model the LC from Method 1, we realised that it requires larger \mni\ values than those within our parameter space. Therefore, we cannot derive all physical parameters simultaneously for the LC from Method 1.
However, due to the large discrepancy found during the radioactive tail phase it may be enlightening to estimate the difference in the \Ni\ yields between both bolometric LC calculation methods. Thus, to estimate \mni, the procedure presented in \citet{hamuy03} is followed. We found \mni\ $\sim$0.14~\ms, that is, a factor of 2.2 larger than with our method of bolometric luminosity calculation.
\Ni\ mass should be the most reliable parameter to determine since it is directly associated with the luminosity of the tail phase \citep{hamuy03}.
However, an inadequate method to construct bolometric LCs can highly overestimate \Ni\ mass yields.

We conclude this analysis emphasising that bolometric luminosities should not be calculated from BB fits to the SED without NIR observations. Otherwise, significant discrepancies up to $\sim$30\% in the plateau and $\sim$60\% in the radioactive tail are found in the bolometric LCs, and therefore, discrepancies also appear in the physical properties derived through bolometric LC modelling.
If NIR data are missing, we recommend to use the relations presented in Table~\ref{table:colour_calibration} to predict NIR magnitudes from optical colours, or the prescriptions of BCs against colours (Sect.~\ref{sec:bc}) to estimate bolometric luminosities.

\section{\snii\ properties}
\label{sec:results}

In the following sections, we first analyse the distributions of the bolometric parameters (Sect.~\ref{sec:distributions}), and then the temperature evolution of the sample as a function of time (Sect.~\ref{sec:temp}).

\subsection{\snii\ bolometric parameter distributions}
\label{sec:distributions}

\begin{figure}
\centering
\includegraphics[width=0.47\textwidth]{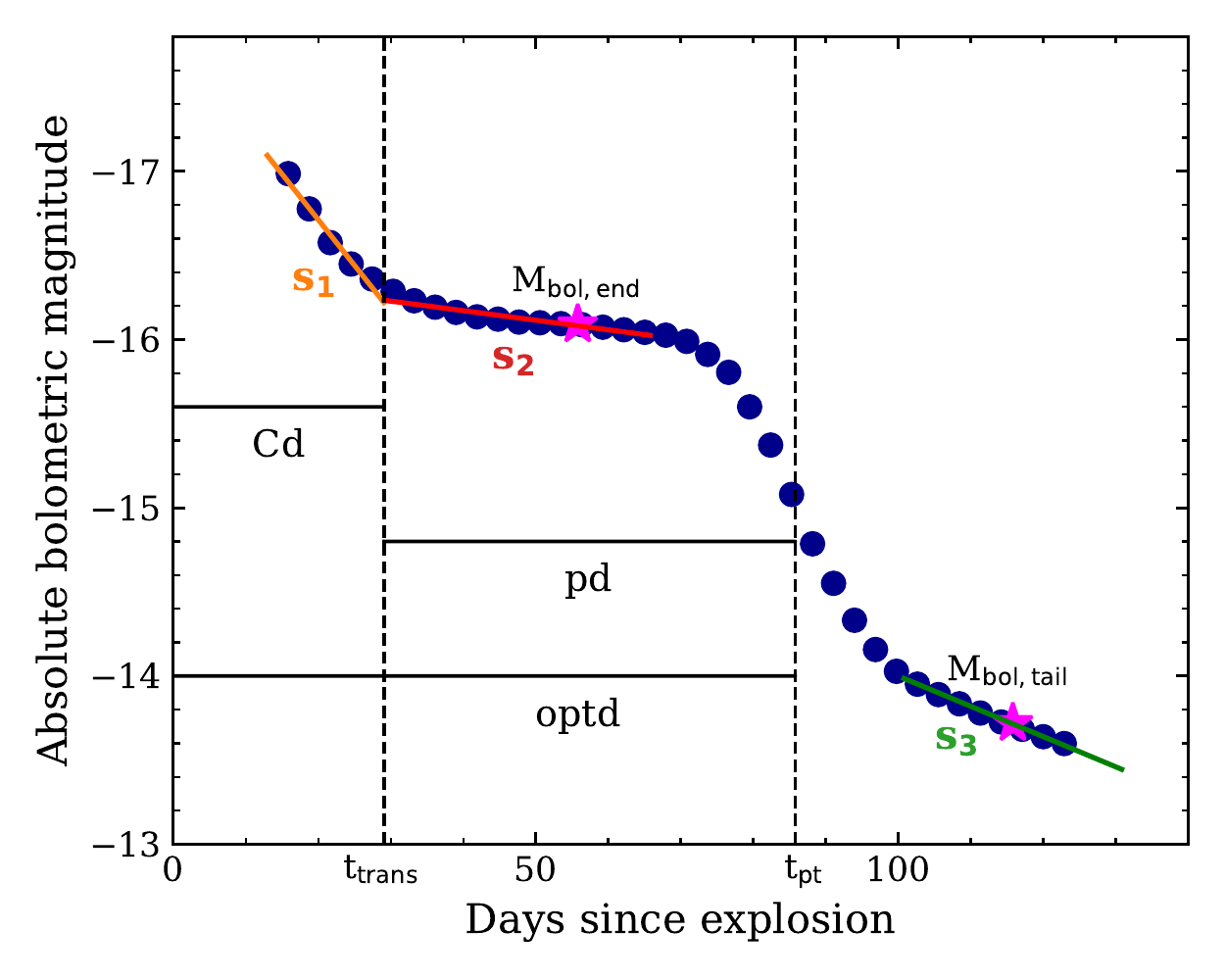}
\caption{Definition of the bolometric LC parameters measured for each SN. We use the interpolated bolometric LC of SN~2008M as an example case.}
\label{fig:blc_parameters}
\end{figure}

\begin{figure}
\centering
\includegraphics[width=0.47\textwidth]{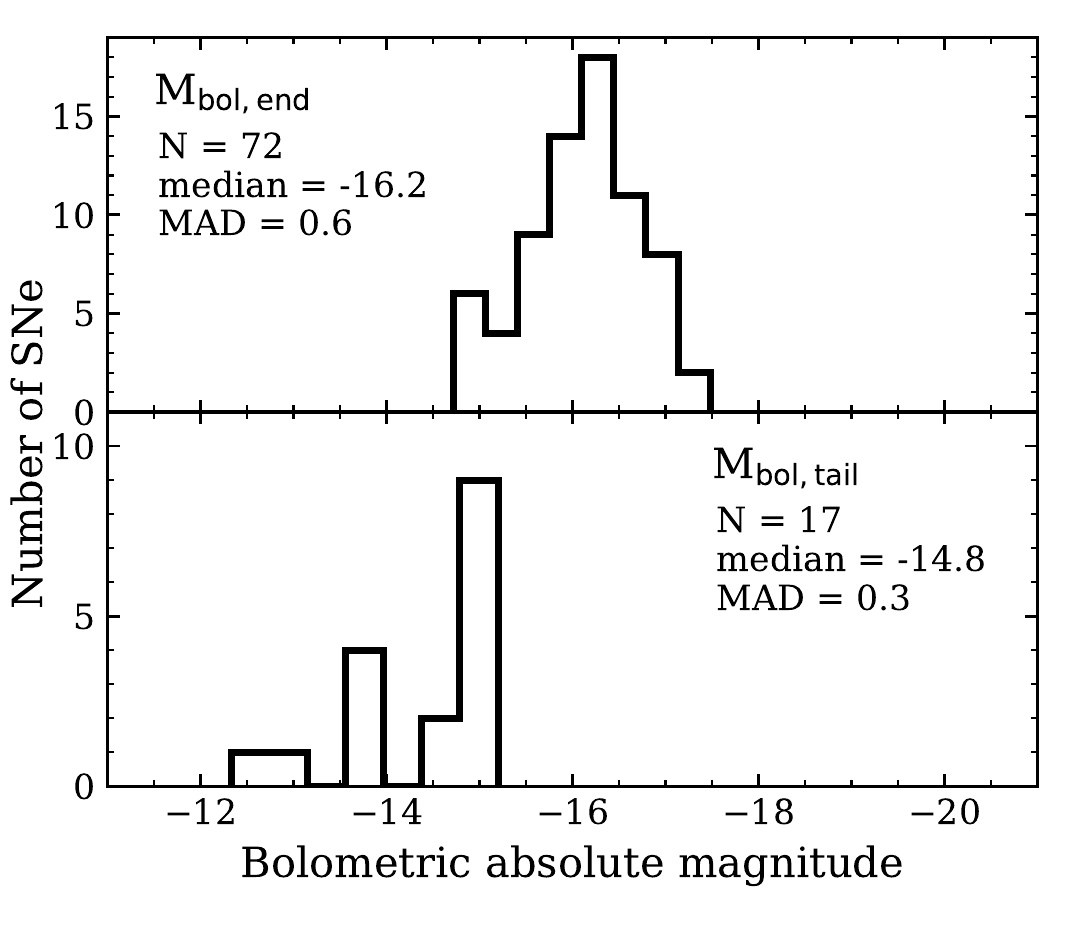}
\caption{Histograms of the measured bolometric magnitudes. Top panel: magnitudes at the end of the optically-thick phase (\mbolend). Bottom panel: magnitudes at the radioactive tail phase (\mboltail). In each panel, the number of \sneii\ (N) is listed, together with the median and the median absolute deviation (MAD).}
\label{fig:hist_mags}
\end{figure}

\begin{figure}
\centering
\includegraphics[width=0.47\textwidth]{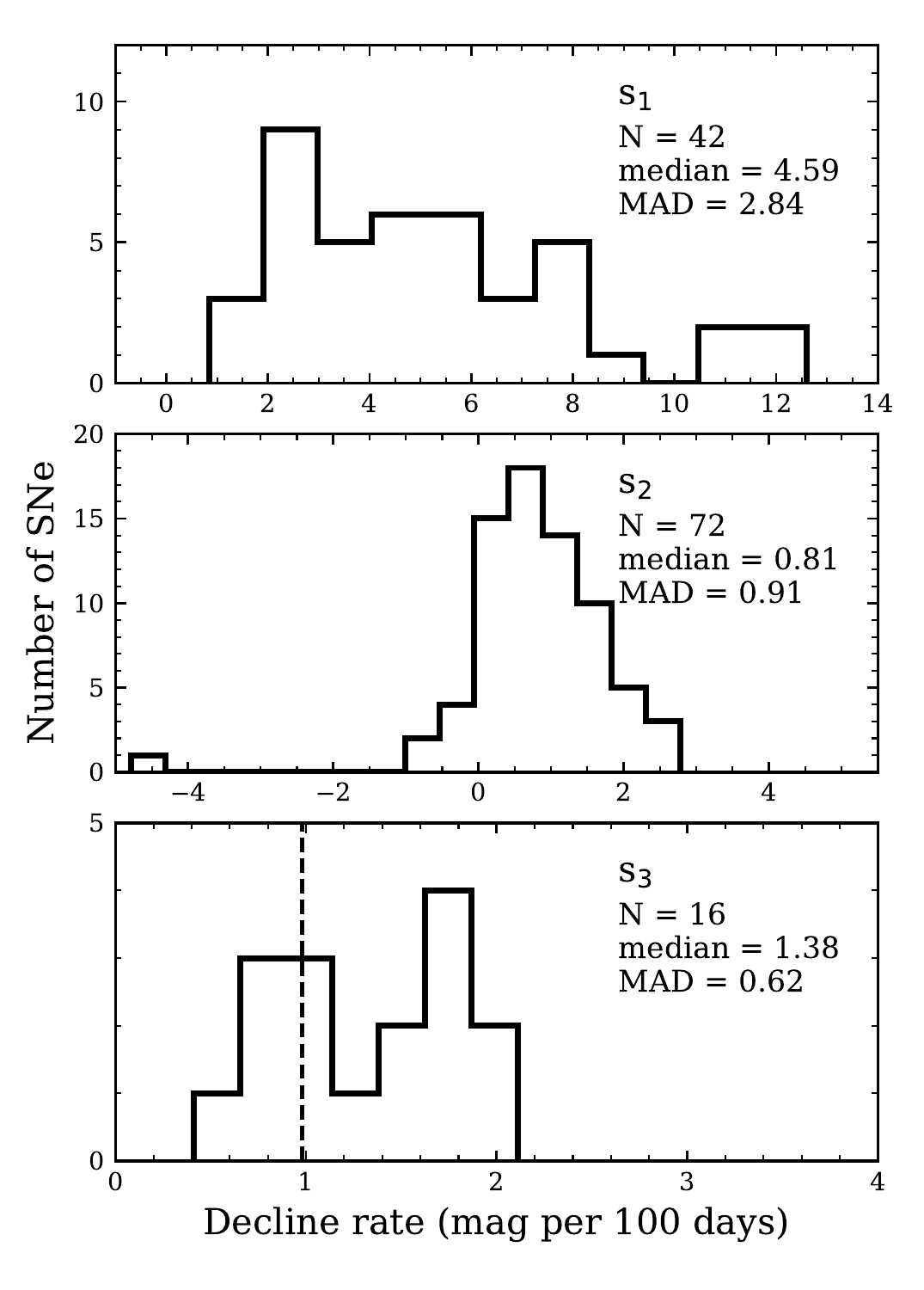}
\caption{Histograms of the three measured decline rates. Top panel: decline rates of the cooling phase (\sone). Middle panel: plateau decline rates (\stwo). Bottom panel: decline rates on the radioactive tail (\sthr). In this last plot, the dashed line indicates the expected decline rate for full trapping of $^{56}$Co decay. In each panel, the number of \sneii\ (N) is listed, together with the median and the median absolute deviation (MAD).}
\label{fig:hist_slopes}
\end{figure}

\begin{figure}
\centering
\includegraphics[width=0.47\textwidth]{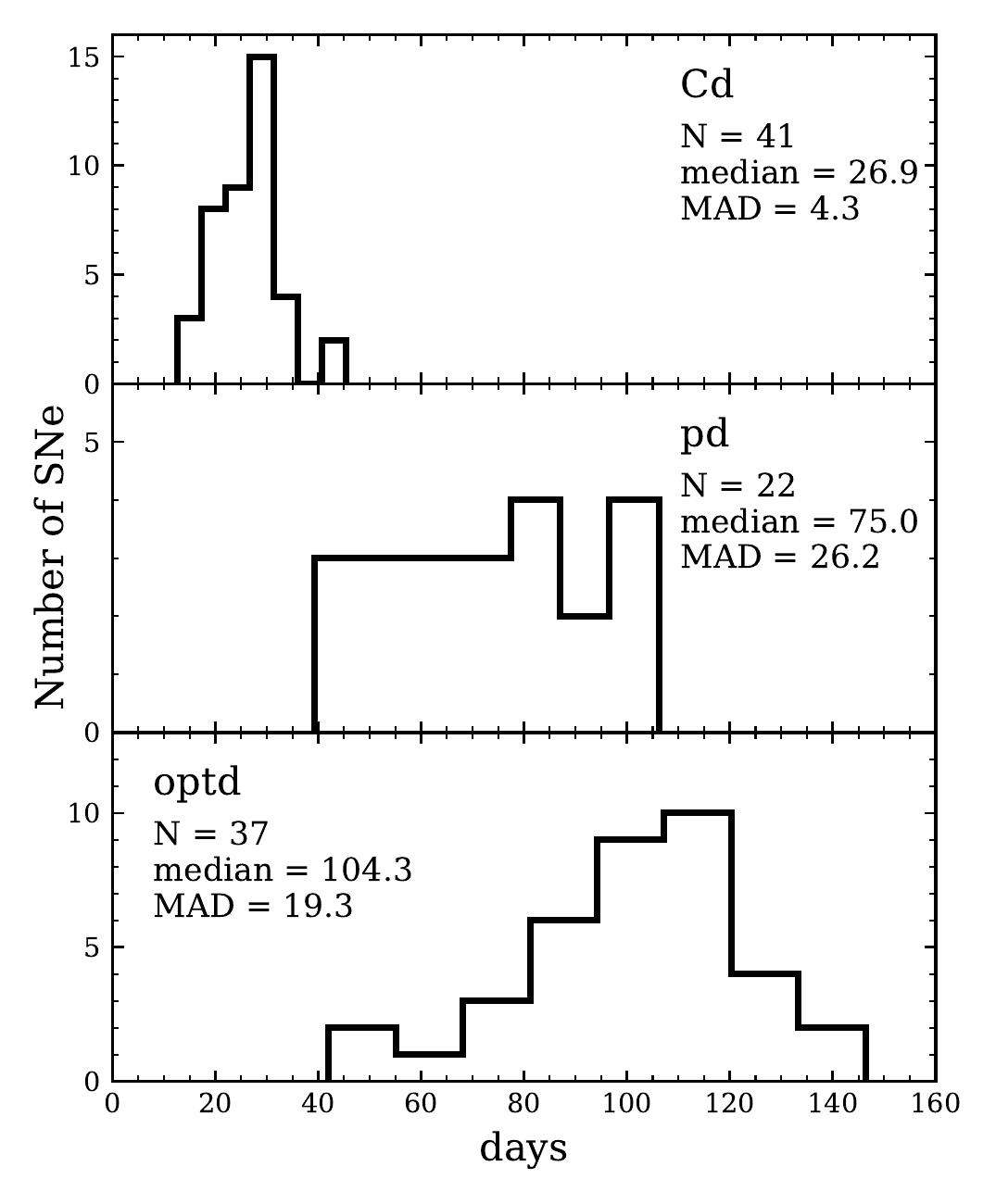}
\caption{Histograms of the three measured time durations of \sneii. Top panel: duration of the cooling phase (\cd). Middle panel: duration of the plateau phase (\pd). Bottom panel: duration of the optically-thick phase (\optd). In each panel, the number of \sneii\ (N) is listed, together with the median and the median absolute deviation (MAD).}
\label{fig:hist_durations}
\end{figure}

With bolometric LCs in hand, we characterised them by measuring several parameters. We started by measuring the mid-point of the transition from plateau to radioactive tail (\tpt) by fitting the LC around the transition with the function reported in Eq.~(1) from \citet{valenti+16}. A complementary description of this function is found in \citet{olivares+10}. The fit was computed using MCMC methods via the \texttt{emcee} package \citep{emcee}.

The parameters studied in the present work were already defined by \citet{anderson+14_lc} and \citet{gutierrez+17II} for \emph{V-}band LCs. 
Figure~\ref{fig:blc_parameters} presents an example LC indicating the bolometric parameters. The observables are:
\begin{enumerate}
    \item \mbolend: defined as the bolometric magnitude measured 30~days before \tpt. If \tpt\ cannot be estimated, \mbolend\ corresponds to the magnitude of the last observation in the optically-thick phase.
    \item \mboltail: defined as the bolometric magnitude measured 30~days after \tpt. If \tpt\ cannot be estimated but the radioactive tail phase is clearly observed, \mboltail\ is the magnitude at the nearest point after transition.
    \item \sone: defined as the decline rate in magnitudes per 100~days of the cooling phase.
    \item \stwo: defined as the decline rate in magnitudes per 100~days of the plateau phase. We reiterate that `plateau' does not necessarily refer to a phase of constant luminosity.
    \item \sthr: defined as the decline rate in magnitudes per 100~days of the slope in the radioactive tail phase.
    \item \ttrans: corresponds to the epoch of transition between the cooling decline (\sone) and the plateau decline (\stwo).
    \item \optd: is the duration of the optically-thick phase and is equal to \tpt. If \tpt\ cannot be estimated but it is clear that the \snii\ has observations during the transition, we take the time of the last observation as the \optd.
    \item \pd: corresponds to the duration of the plateau phase. It is equal to \tpt~$-$~\ttrans. 
    \item \cd: corresponds to the duration of the cooling phase, defined between the explosion epoch and \ttrans.
\end{enumerate}

\mbolend\ and \mboltail\ were interpolated to the chosen epochs when \tpt\ is defined, and their errors were estimated by propagation of the uncertainties in the implicated magnitudes.
The decline rates were measured by fitting a straight line to each of the three phases. \sone\ and \stwo\ (and \ttrans) were measured in the same way as in \citet[][see also \citealt{anderson+14_lc} for more details]{gutierrez+17II}. For \sthr\ we simply fitted a straight line to the radioactive tail with at least three data points.

Finally, we point out that \pd\ and \optd\ were defined differently than in \citet{anderson+14_lc}. For this reason, we named our parameters using lowercase acronyms (\pd, \optd) instead of $Pd$ and $OPTd$ as in \citet{anderson+14_lc}.
These authors defined $Pd$ ($OPTd$) as the time between \ttrans\ (explosion epoch) and the end of the plateau, defined as when the extrapolation of \stwo\ becomes 0.1~mag more luminous than the $V$-band LC. In the present work, we use \tpt\ as a measure of the length of the optically-thick phase (\optd) and, therefore, \pd~=~\tpt~$-$~\ttrans.
In Table~\ref{table:blc_parameters}, we present the measured bolometric parameters as defined above. Additionally, Figs.~\ref{fig:fits1}, \ref{fig:fits2} and \ref{fig:fits3} show the bolometric LC for each \snii\ in the CSP-I sample individually, together with their measured bolometric parameters.

In Figs.~\ref{fig:fits1} and \ref{fig:fits3}, we note that SN~2004er and SN~2009ao display an unusual behaviour in their early bolometric LCs. Both \sneii\ show early rising LCs until the `peak' is reached.
While it is now more common to observe the early rise of \snii\ LCs, the peak is observed in optical and NIR bands. This is due to the shift of the SN spectrum to redder bands as the ejecta cools.
An initial peak in the bolometric luminosity is expected but in this case it is due to the arrival of the shock wave to the stellar surface.
The shock-breakout detection is extremely difficult to observe due to its short duration. This situation may change if the SN explodes inside a dense environment. We further discuss possible explanations for the observed behaviour.

The early bolometric LCs for SN~2004er and SN~2009ao resemble the behaviour seen in some early pseudo-bolometric LCs \citep[see e.g. Fig.~10 from][]{valenti+16}. 
Therefore, the problem may arise from the extrapolated UV flux. The underestimation of the UV correction causes lower luminosities at early times, when the UV flux contributes significantly.
Another possibility is that this behaviour is produced by significant uncorrected host-galaxy extinction. 
In that direction, SN~2004er and SN~2009ao are both within the 20\% reddest of the CSP-I sample \citep{dejaeger+18}. 
Re-emission of energy by dust takes place at longer wavelengths than those observed by the CSP-I and it is not taken into account in our BB models.
Thus, host-galaxy extinction that has been unaccounted for will increase the flux in the bluest bands producing a larger contribution in the UV, while the flux in the reddest bands is less affected. 
This effect will be larger at early times, when the SN is intrinsically blue, which means that extinction will affect the bolometric LC as a function of time, that is, it will change the shape of the bolometric LC. 
We tested this hypothesis by correcting the photometry of SN~2004er for additional extinction. We found that an extra $A_{V}^{\rm host}$ of 0.8~mag changes the initial rising LC to a typical declining LC at early epochs.
Therefore, the unusual rising behaviour of SN~2004er and SN~2009ao might be caused by missing UV flux, probably due to uncorrected host-galaxy reddening\footnote{However, the effect of uncorrected host-galaxy extinction on the shape of bolometric LCs will not affect the overall analysis. This effect becomes smaller as \sneii\ evolve to redder colours. For this reason, this is only significant at early times ($\lesssim$~30~days) and for large $A_{V}^{\rm host}$ values. We emphasise that the first 30~days of evolution are removed from the fitting when determining progenitor and explosion properties of \sneii\ \citepalias[see][]{paper2_submitted}}.
Additionally, we note that the presence of dense CSM surrounding the progenitor star may also explain this behaviour. 
Recent photometric and early spectroscopic analyses suggest that such dense CSM may be present in many \sneii\ \citep[e.g.][]{gonzalez+15,yaron+17,forster+18,morozova+18,bruch+21}.
Under this hypothesis, the observed peak in the bolometric LCs of SN~2004er and SN~2009ao may alternatively be caused by the delay of shock breakout due to ejecta-CSM interaction\footnote{Although the actual properties of the CSM are unclear and beyond the scope of this work.} \citep[e.g.][]{moriya+18,haynie+20}.

Figure~\ref{fig:hist_mags} shows the distributions of the two measured bolometric magnitudes: \mbolend\ and \mboltail. Our sample is characterised by the following median values and median absolute deviations (MAD): \mbolend~=~$-$16.2~mag (MAD\,=\,0.6, 72~\sneii) and \mboltail~=~$-$14.8~mag (MAD\,=\,0.3, 17~\sneii).
As expected, the distributions peak at lower luminosities as time evolves.
\mbolend\ cover a range of $\sim$2.8~mag from $-$14.7~mag (SN~2008bk) to $-$17.5~mag (SN~2009aj).
SN~2008bk is a well-studied \snii\ \citep{pignata13,vandyk+12a,vandyk13,lisakov+17,maund17,jerkstrand+18,eldridge+19b,martinez+19,martinez+20} that belongs to a class of sub-luminous \sneii\ \citep{spiro+14}. \citet{spiro+14} proposes that these faint events represent the lowest-luminosity part of a continuous distribution of \sneii. This was later verified by a number of studies \citep[e.g.][]{anderson+14_lc,sanders+15,gutierrez+17II}.
In the current paper, we find the same continuous distributions in the two measured bolometric magnitudes.
SN~2009aj is found in the upper limit of the \mbolend\ distribution. This SN is more luminous than typical \sneii\ probably due to CSM interaction \citep{rodriguez+20}.
At the radioactive tail phase, the sample ranges from $-$12.3~mag (SN~2008bk) to $-$15.2~mag (SN~2007X). SN~2008bk is again the lowest-luminosity event. This is not surprising as low-luminosity \sneii\ are characterised, among other properties, by the very-low luminosity during the radioactive tail which indicates that the mass of \Ni\ ejected during the explosion is considerably low \citep{pastorello+04,spiro+14,lisakov+18}.

Figure~\ref{fig:hist_slopes} presents histograms of the  distributions of the decline rates (\sone, \stwo, and \sthr). The median values of the decline rates are: \sone~=~4.59~mag per 100 days (MAD\,=\,2.84, 42~\sneii), \stwo~=~0.81~mag per 100 days (MAD\,=\,0.91, 72~\sneii) and \sthr~=~1.38~mag per 100 days (MAD\,=\,0.62, 16~\sneii).
The bolometric decline rates show a continuum in their distributions with some minor exceptions: the \sone\ distribution shows four \sneii\ (SNe~2005dk, 2006bl, 2008bu, and 2009A) with rates higher than 10~mag per 100 days. 
In the case of \stwo, the fastest decliner is SN~2009au with a decline rate of 2.78~$\pm$~0.17~mag per 100 days, while SN~2009A has the lowest \stwo\ value with a rate of $-$4.79~$\pm$~0.47~mag per 100 days, that is, this \snii\ shows a rising plateau phase. SN~2009A has a peculiar behaviour as it shows double-peaked optical and bolometric LCs (Anderson et al. in prep.). However, the lack of data after 50~days from explosion prevents us to reveal their properties.
SN~2008hg has the second lowest \stwo\ with $-$0.96~$\pm$~0.35~mag per 100 days.
In total, eight \sneii\ have rising plateau phase, but two of them have insufficient data during the plateau phase for reliable estimates.

Unfortunately, the decline rate of the radioactive tail phase (\sthr) can only be measured for 16 objects due to insufficient data during this phase.
A significant number of events (11 of 16) decline quicker than expected if gamma-ray photons from the radioactive decay of $^{56}$Co are fully trapped in the ejecta \citep[0.98~mag per 100 days;][]{woosley+89} implying some gamma-ray loss. This confirms the results from \citet{anderson+14_lc} where \emph{V}-band data were used\footnote{However the fraction showing a steeper decline is significantly higher here. This is discussed further in \citetalias{paper3_submitted}.}.

In Fig.~\ref{fig:hist_durations} the distributions of the three measured time durations (\cd, \pd, and \optd) are displayed.
A large range of \pd\ and \optd\ values are observed. 
\optd\ is characterised by an average of 104~days (MAD\,=\,19, 37~\sneii). This value is very similar to the historical 100-day \optd\ of SNe~IIP LCs \citep{barbon+79}.
The shortest and longest \optd\ values are 42\,$\pm$\,2~days (SN~2004dy) and 146\,$\pm$\,2~days (SN~2004er) respectively.
The average \pd\ is 75~days (MAD\,=\,26, 22~\sneii) and ranges from 39~$\pm$~2~days for SN~2008bu to 106~$\pm$~4~days for SN~2004fc. Only 22 \pd\ values are available since it is the most difficult time duration -- of the three analysed in the present work -- to be measured. Given its definition, \pd\ requires the estimation of the \optd, for which the explosion epoch and the transition to the radioactive phase is needed, as well as \ttrans.
\cd\ shows a narrower distribution with a median of 27~days (MAD\,=\,4, 41~\sneii).
All measured time durations display a continuum in their distribution with no signs of bi-modality indicating the existence of different sub-classes.

\subsection{Temperature evolution}
\label{sec:temp}

\input{temperature.tab}

\begin{figure}
\centering
\includegraphics[width=0.47\textwidth]{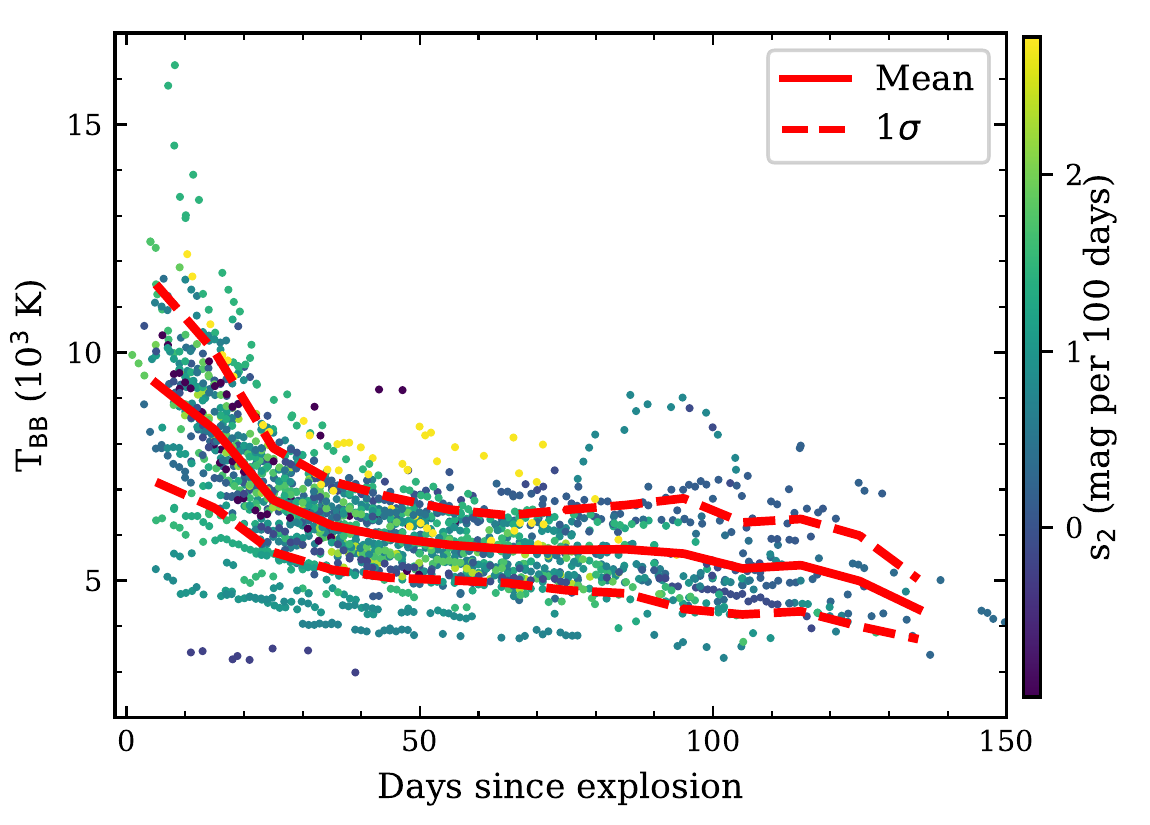}
\includegraphics[width=0.47\textwidth]{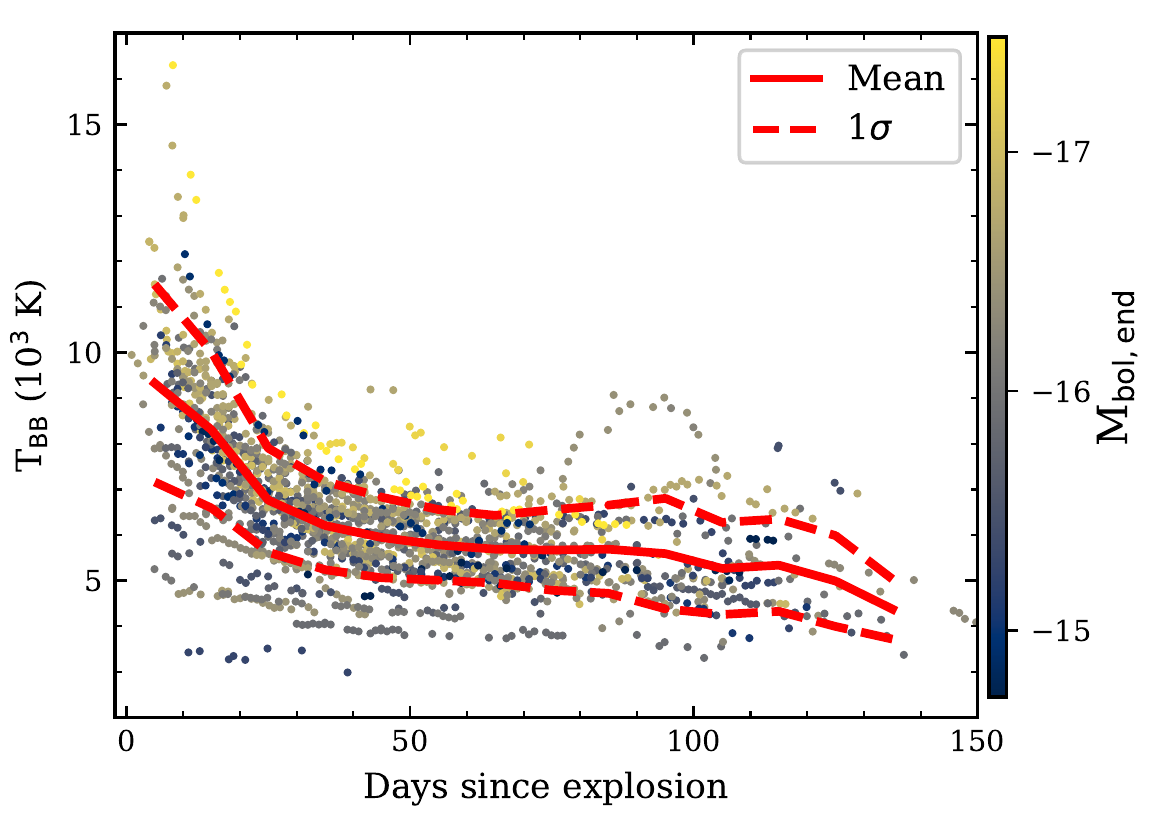}
\caption{BB temperature evolution for \sneii\ in the CSP-I sample with explosion epochs defined. The red solid line indicates the mean temperature within each time bin, while the dashed red lines represent the standard deviation. These values are available in Table~\ref{table:temp}. \sneii\ are colour-coded with the plateau decline rate (\stwo, top panel) and \mbolend\ (bottom panel).}
\label{fig:temp}
\end{figure}

The temperature evolution with time for the \sneii\ in the sample is obtained from the BB fits and is presented in Fig.~\ref{fig:temp}. This figure also includes the mean temperature and the standard deviation within each time bin of 10~days. These values are presented in Table~\ref{table:temp}.

Shock breakout models predict temperatures of about 10$^{5}$\,K or higher \citep{grassberg+71,kozyreva+20}. Indeed, the modelling of spectra of SN~2013fs taken six hours after explosion revealed a temperature of around 5$-$6\,$\times$\,10$^{4}$\,K \citep{yaron+17}. After shock breakout, a rapid cooling of the outermost layers of the ejecta is expected. 
During the first ten days, the mean temperature in our sample is about 9300\,K but with a large standard deviation of 2200\,K. The hottest \snii\ at this stage is SN~2009aj presenting a temperature of 16300~K.
As time progresses, the temperature starts evolving more slowly compared to earlier phases. Between 30 and 40 days after explosion, the temperature enters a phase where it remains nearly constant. This is consistent with hydrogen recombination that occurs when the hydrogen-rich envelope reaches $\sim$6000~K. At this stage a recombination wave appears moving inwards in mass coordinate until the entire hydrogen envelope is recombined and the recombination front arrives to the innermost parts of the ejecta.

We find six \sneii\ that enter the plateau phase with temperatures below 5000~K. In particular, SN~2008bp has estimated temperatures around 3500~K. One may assume that such low temperatures could be obtained if (uncorrected) host-galaxy extinction is considerably high. 
However, for low temperature values (as those of SN~2008bp), the temperature after correcting the photometry by host-galaxy extinction remains almost the same \citep[see][their Fig.~6]{faran+18}. Therefore, the underlying reason for these low temperatures is unclear.
On the other hand, we find SN~2009A entering the plateau phase with temperatures of the order of 9000~K. However, this is a very peculiar \snii\ given that it shows double-peaked optical LCs (see Anderson et al. in prep. for details).

\citet{valenti+16} found that fast-declining \sneii\ display systematically lower temperatures at 50 days after explosion and argue that this is to be expected since fast decliners evolve more quickly and hence are close to the end of the recombination phase at similar (to slow decliners) epochs. 
In the top panel of Fig.~\ref{fig:temp}, the temperature curves are colour-coded with their \stwo\ values. Six objects present temperatures lower than 5000~K and only one, SN~2005lw, is a fast decliner with a \stwo\ value of 1.53~$\pm$~0.08~mag per 100 days, while the others (SNe~2004dy, 2007sq, 2008bh, 2008bp, and 2009ao) have \stwo\ values lower than 1~mag per 100~days.
Moreover, one of the hottest \snii\ at 50~days, SN~2008bm, has a decline rate \stwo~=~2.74~$\pm$~0.12~mag per 100 days.
This is in the opposite direction of what \citet{valenti+16} found.
SN~2008bm belongs to a subgroup of \sneii\ with peculiar characteristics: blue colours, low expansion velocities, and more luminous than `normal' \sneii\ \citep{rodriguez+20}. In addition, \citet{rodriguez+20} find that these properties can be reproduced by the ejecta-CSM interaction model, which may also explain the high temperature. The other two \sneii\ presented in \citet{rodriguez+20}, SN~2009aj and SN~2009au, also display high temperatures falling above the standard deviation. 
Indeed, SN~2009aj is the hottest \sneii\ in our sample until $\sim$30~days after explosion.
CSM interaction causes a boost in the SN luminosity and greater photospheric temperatures because of the additional energy source \citep{hillier+19}.
Leaving aside these outliers, we do not find a clear trend of the temperature with \stwo\ as found by \citet{valenti+16}.

\snii\ spectra show that some iron-group absorption lines appear earlier in low-luminosity events, which may be associated with differences in temperatures and/or metallicity \citep{gutierrez+17I}.
That is to say, low-luminosity \snii\ line forming regions cool faster allowing metal lines to appear sooner.
Therefore, in the bottom panel of Fig.~\ref{fig:temp}, we show the temperature curves for all \sneii\ in the CSP-I sample colour-coded according to \mbolend.
We find a remarkable trend with low-luminosity (high-) \sneii\ displaying low (high) temperatures.

\section{Bolometric corrections}
\label{sec:bc}

In Sect.~\ref{sec:blc_calc}, we determined bolometric luminosities from the estimation of the bolometric flux, that is, from the integration of the observed flux in several photometric bands covering the major part of the \snii\ emission and the modelling of the unobserved flux using different techniques for the UV and IR regimes.
This is the most accurate technique in order to calculate bolometric fluxes, but only in the case where extensive photometric coverage is available. When this requirement is not fulfilled, the use of BCs might be more appropriate.

The BC enables a transformation of magnitudes in certain bands into bolometric magnitudes. By definition, 
\begin{equation}
\label{eq:bc}
    \mathrm{BC}_{j} = m_{\rm bol} - m_{j} \,,
\end{equation}
where $m_{j}$ is the magnitude of the SN in the band $j$ that has been corrected for extinction, and $m_{\rm bol}$ is the bolometric magnitude. The BC is independent of the distance since it is defined as a magnitude difference. Therefore, Eq.~\ref{eq:bc} can also be defined in absolute magnitudes (BC$_{j}$~=~$M_{\rm bol}-M_{j}$).
During the SN evolution, the luminosity changes and the SED rapidly evolves to redder colours. This means that the BC is not constant in time, which complicates its calculation and further implementation.

\citet{bersten+09} calculated BCs for three well-observed \sneii\ (SNe~1987A, 1999em, and 2003hn) and derived calibrations of BCs against optical colours. This is an easy and quick technique to implement as it allows the calculation of bolometric luminosities using only two optical Johnson-Cousins filters.
The same procedure was implemented in \citet{lyman+14}. In the latter study, the authors analysed a larger sample of six \sneii, three of which being those previously studied in \citet{bersten+09}, and three others: SN~2004et, SN~2005cs, and SN~2012A. In addition, \citet{lyman+14} also presented calibrations for BCs from optical colours in Sloan bands.
However, \citet{lyman+14} obtained Sloan magnitudes through the interpolation of the SED constructed using Johnson filters. Although they corrected their magnitudes for the mean offset found by means of the comparison with synthetic magnitudes from spectra, this technique might add systematic errors to the BC calibrations against colours.
For this reason, \citet{lyman+14} point to a re-evaluation of their calibrations once an appreciable data set of \sneii\ observed directly in Sloan bands with good NIR coverage exists, which is what we present in the current study.
More recently, \citet{pejcha+15b} utilised 26 well-observed \sneii\ and a model that disentangles the observed multi-band LCs and expansion velocities into radius and temperature changes by simultaneously fitting the entire set of observations. With this model, \citet{pejcha+15b} were able to provide polynomial fits to the BCs as a function of every colour defined within the large set of filters used. The latter study includes observations in the \emph{Swift}-UV bands which is the main difference with the previous works.

All the previously mentioned studies \citep{bersten+09,lyman+14,pejcha+15b} constructed calibrations for the BC as a function of colours. \citet{maguire+10} presented BCs against time for four \sneii\ and found a large scatter in the BC relative to the $V$ band, although the BC to the $R$ band is more homogeneous but only after 50~days since explosion. This may indicate that colour is a better indicator of the BC than time for \sneii.

\subsection{Calibrations of bolometric corrections versus colours}

\input{bc.tab}

\begin{figure}
\centering
\includegraphics[width=0.49\textwidth]{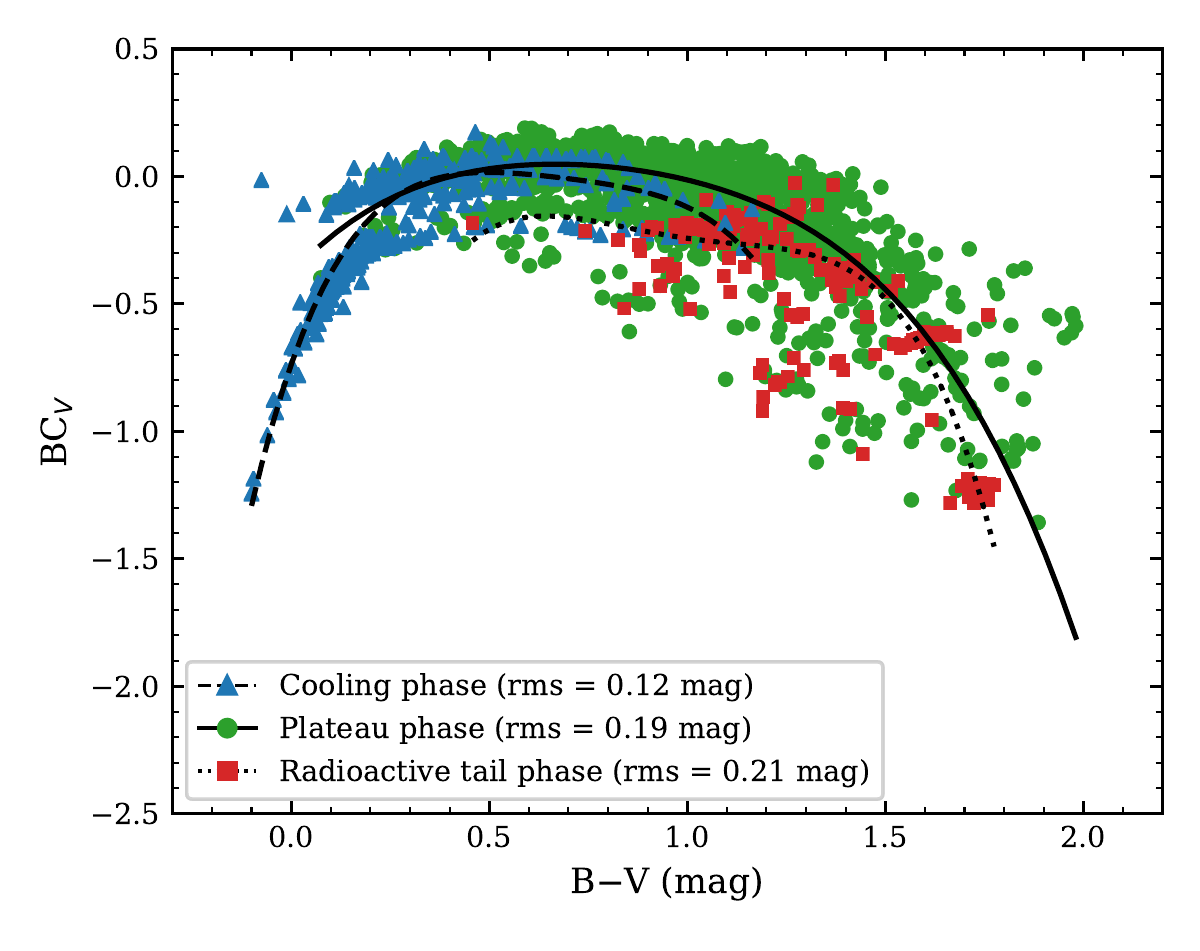}
\includegraphics[width=0.49\textwidth]{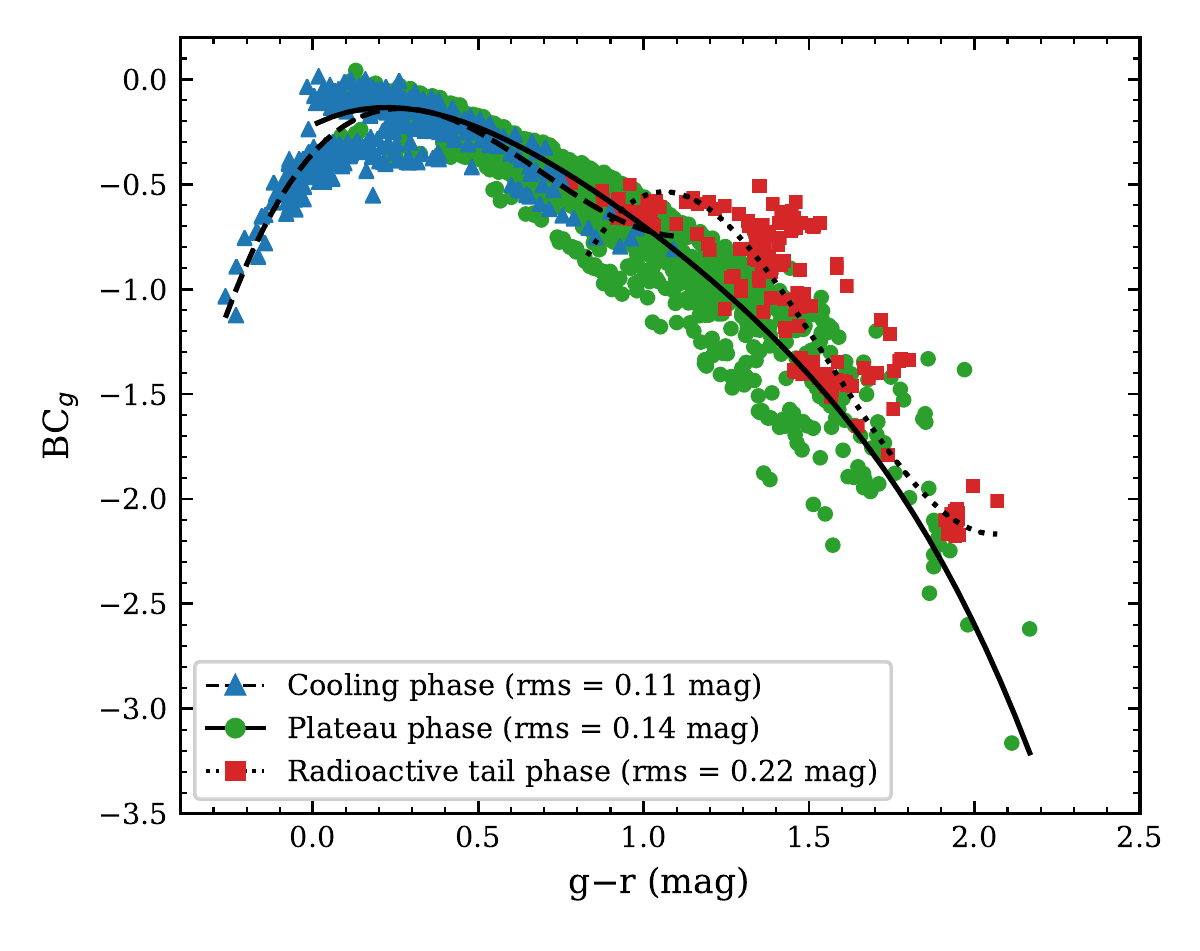}
\includegraphics[width=0.49\textwidth]{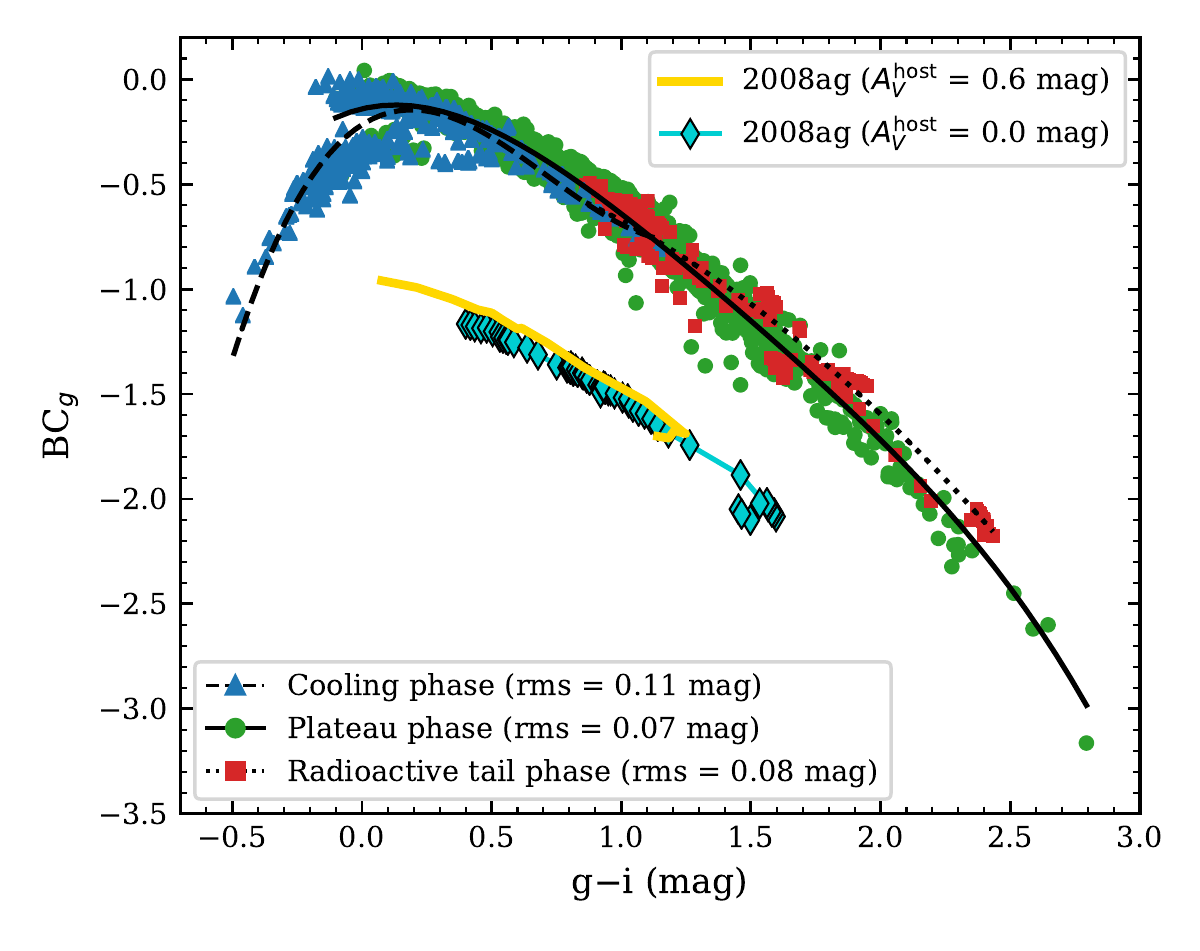}
\caption{Bolometric corrections to the $V$ band as a function of $B-V$ colour (top panel), and to the $g$ band as a function of $g-r$ colour (middle panel) and $g-i$ colour (bottom panel) for all \sneii\ in the sample. The data and fits are separated into three epochs: cooling phase (blue triangles, dashed line), plateau phase (green dots, solid line), and radioactive tail phase (red squared, dotted line). The bottom panel also shows the effect of reddening in this plot by re-analysing SN~2008ag with $A_{V}^{\rm host}$~=~0.6~mag. Both curves for SN~2008ag are offset from the data for visualisation purposes.}
\label{fig:bc}
\end{figure}

In this section, we calculate BCs for our entire sample of 74~\sneii. We first converted the bolometric luminosities (Sect.~\ref{sec:blc_calc}) into bolometric magnitudes using
\begin{equation}
    M_{\rm bol} = M_{\odot,\rm bol} - 2.5\,\mathrm{log}_{10}\,
    \left( \frac{L_{\rm bol}}{L_{\odot,\rm bol}}  \right) \,,
\end{equation}
where $L_{\odot,\rm bol}$~=~3.845~$\times$~10$^{33}$~erg~s$^{-1}$ and $M_{\odot,\rm bol}$~=~4.74 are the luminosity and the absolute bolometric magnitude of the Sun \citep{allen2000}, and then we calculated the BC from Eq.~\ref{eq:bc}.

We then looked for correlations between the BC and three colour indices. Figure~\ref{fig:bc} shows the BC relative to the $V$ band (BC$_{V}$) as a function of $B-V$ colour (top panel), and the BC relative to the $g$ band (BC$_{g}$) as a function of $g-r$ and $g-i$ colours (middle and bottom panels, respectively).
For all cases, we find strong relations. Particularly, BC$_{g}$ versus $g-r$ and $g-i$ present the tightest correlations.
In Fig.~\ref{fig:bc}, we also include polynomial fits to the data.
The order of the polynomials has been chosen to give the smallest dispersions.
For each colour, we decide to separate the data into three groups determined by the phase of evolution in which the \snii\ is: cooling phase (i.e. at times earlier than \ttrans), plateau phase (for times between \ttrans\ and the end of the plateau), and radioactive tail phase.
The coefficients of the polynomial fits are shown in Table~\ref{table:bc}.
The cooling phase needs a separate examination for the following reason. 
The lack of near-UV data forces an extrapolation of the flux bluewards of the $u$ band. As already pointed out in Sect.~\ref{sec:uvcorr}, the UV correction utilised in the present study may not predict the actual flux in the UV at times earlier than $\sim$20~days from explosion epoch, when the flux at wavelengths shorter than the $u$ band contains a significant fraction of the bolometric flux. Therefore, at early times, the BC can also be underestimated. For this reason, we produced fits to early data (i.e. during the cooling phase) separately.
However, care has to be taken when using our fits to predict early BCs, as these may be underestimated. At early times, it may be more appropriate to use the calibrations from \citet{pejcha+15b} as they include \emph{Swift} observations \citep[see also][]{pritchard+14}.

The scatter of the BC$_{V}$ against $B-V$ is relatively large during the plateau and radioactive tail phases (Fig.~\ref{fig:bc}, top panel). In fact, the rms dispersion between the BC and the fits is 0.19~mag and 0.21~mag respectively, which is about a factor of two larger than previous works.
We also calculated the BC relative to the $B$ band for comparison, but similar scatter is found.
BC$_{g}$ versus Sloan optical colours display smaller dispersions.
The calibration using $g-r$ colours displays a dispersion of 0.14~mag in the plateau phase. While this value is smaller than what we found for $B-V$, it is still a factor of three larger than previous calibrations \citep{lyman+14}.
We note a tight relation at blue colours ($g-r$~$\lesssim$~0.5~mag), however the scatter increases at redder colours (later times). The increase in strength over time of the \halpha\ profile, which falls in the $r$ band at low redshifts like those from the CSP-I \snii\ sample, and the diversity in the equivalent widths of the absorption and emission components of \halpha\ \citep{gutierrez+17I} may explain such behaviour.
The calibration of BC$_{g}$ versus $g-i$ colours exhibits the tightest relation of the three. The scatter is 0.07~mag during the plateau phase, and 0.08~mag during the tail phase.
For this reason, we suggest future works to use our BC$_{g}$ versus $g-i$ calibration to construct bolometric LCs when NIR data are not sufficient for reliable BB fits.

The bottom panel of Fig.~\ref{fig:bc} also illustrates the effect of extinction in the BC$_{g}$ versus $g-i$ colour diagram. For this purpose, we re-analysed SN~2008ag but with a simulated host-galaxy extinction of 0.6~mag (yellow solid line). The original curve for SN~2008ag (filled cyan diamonds) and that with extra host-galaxy extinction are offset from the data for better visualisation.
Additional extinction makes the SN intrinsically bluer, while the BC$_{g}$ takes larger values. Thus, both contributions move the \sneii\ in the same direction as the fits.
This means that host-galaxy extinction uncertainties do not significantly affect the BC calibration producing the low dispersion found \citep[see also][for discussion]{lyman+14}.

\subsection{Comparison to previous calibrations}

\input{bc_comp.tab}

\begin{figure*}
\centering
\includegraphics[width=1\textwidth]{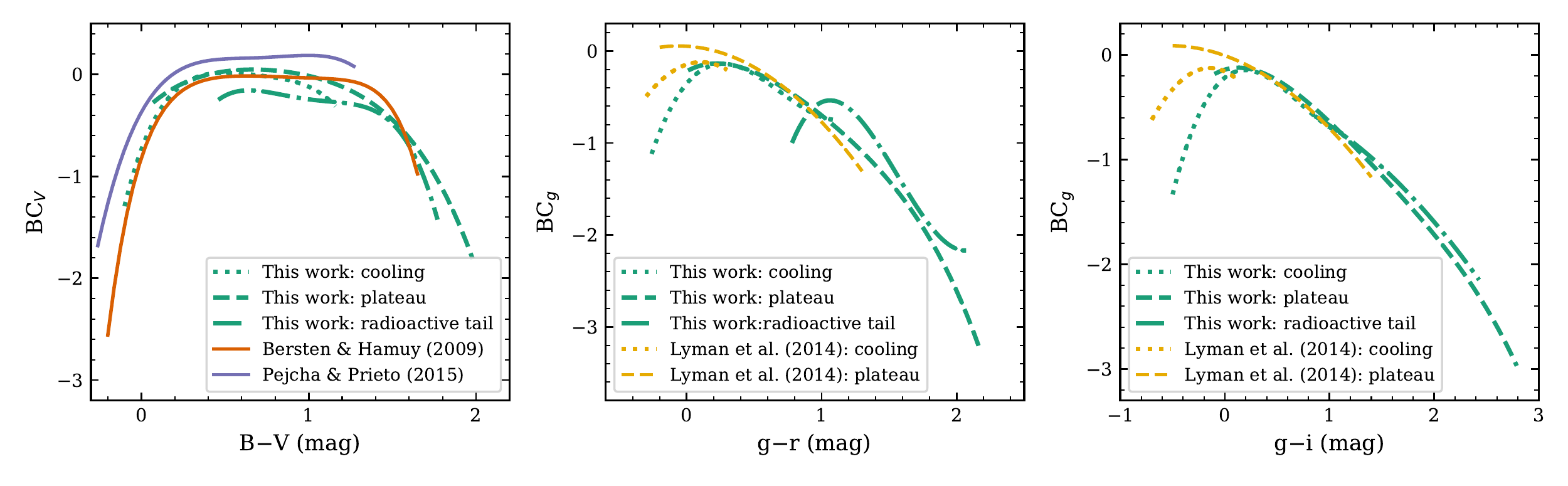}
\caption{Comparison of the bolometric corrections estimated in this study with other calibrations found in the literature. Left panel: BC$_{V}$ as a function of $B-V$ colour. Middle panel: BC$_{g}$ as a function of $g-r$ colour. Right panel: BC$_{g}$ as a function of $g-i$ colour. The BC calibrations from the literature extend bluer, possibly because those studies correct for host-galaxy reddening.}
\label{fig:bc_comp}
\end{figure*}

\begin{figure*}
\centering
\includegraphics[width=0.33\textwidth]{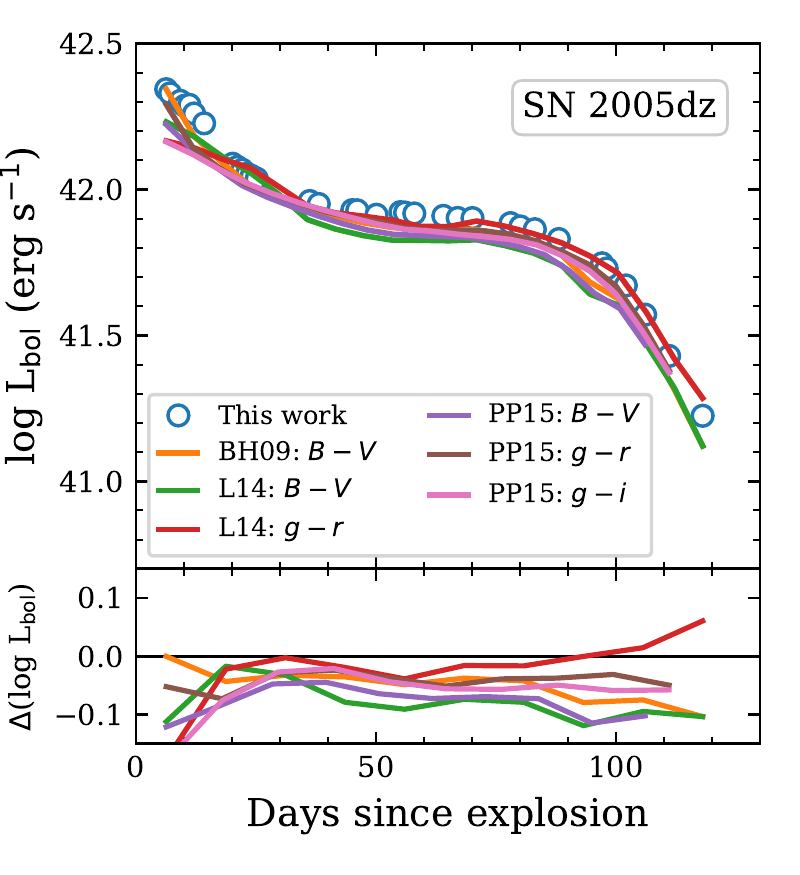}
\includegraphics[width=0.33\textwidth]{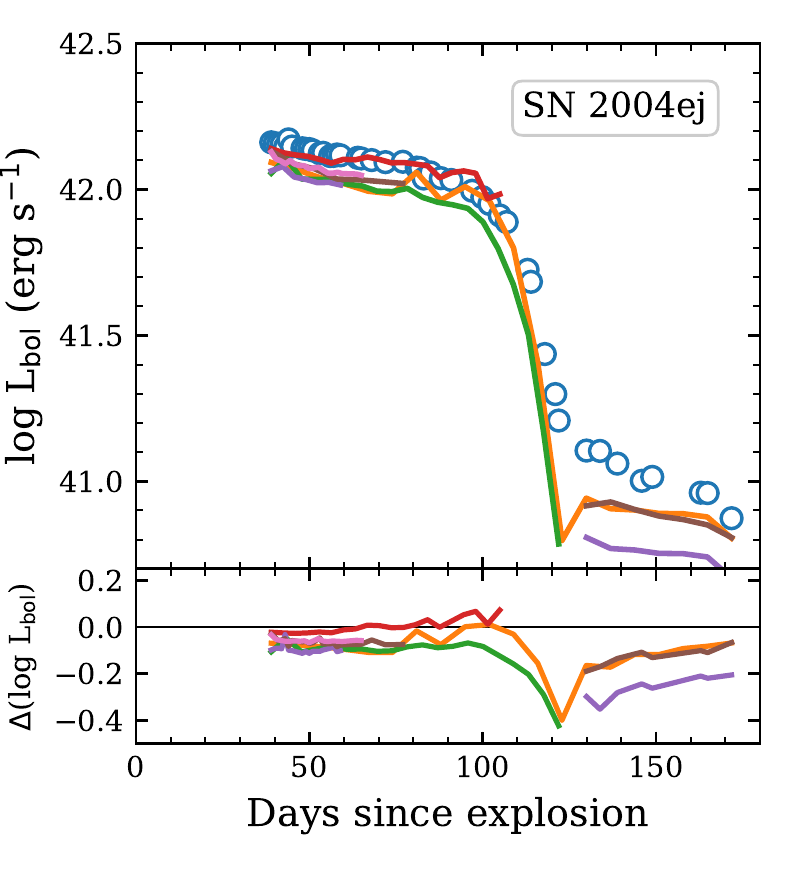}
\includegraphics[width=0.33\textwidth]{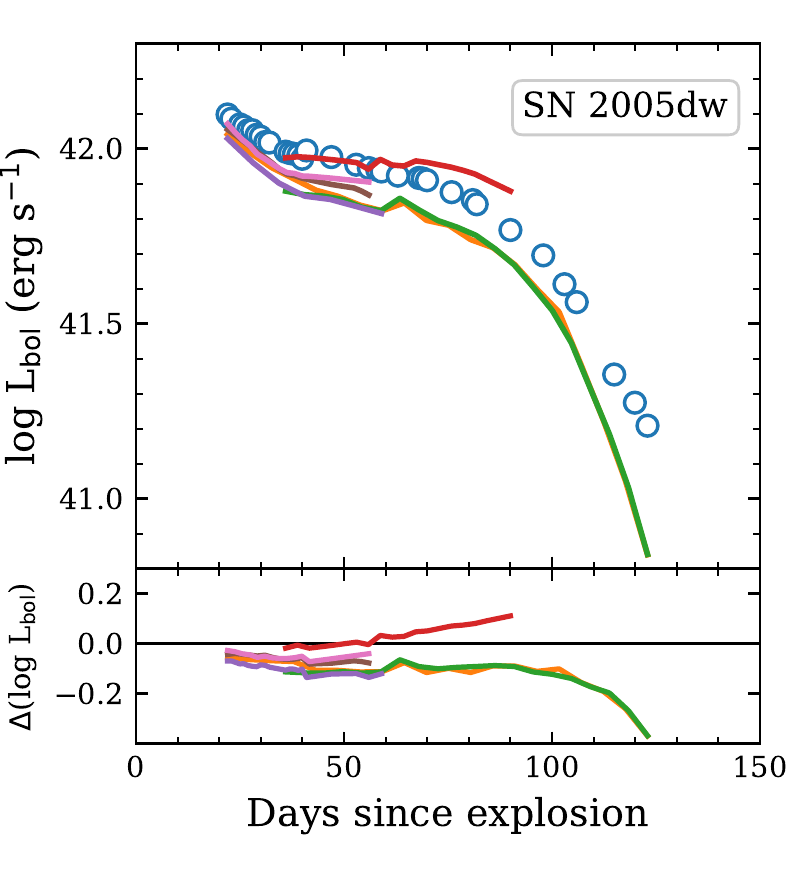}
\caption{Bolometric LCs for SN~2005dz (left panel), SN~2004ej (middle panel), and SN~2005dw (right panel) in comparison with those calculated from calibrations of BC versus colour found in the literature. BH09, L14, and PP15 refer to \citet{bersten+09}, \citet{lyman+14}, and \citet{pejcha+15b}, respectively.}
\label{fig:bc_comp_blcs}
\end{figure*}

\begin{figure*}
\centering
\includegraphics[width=0.33\textwidth]{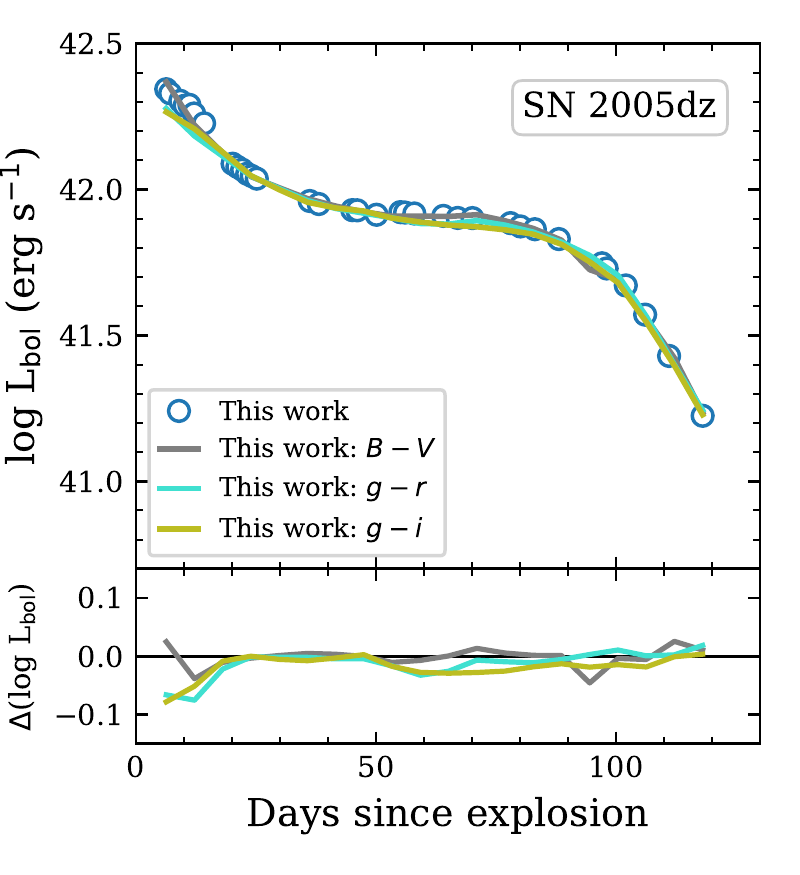}
\includegraphics[width=0.33\textwidth]{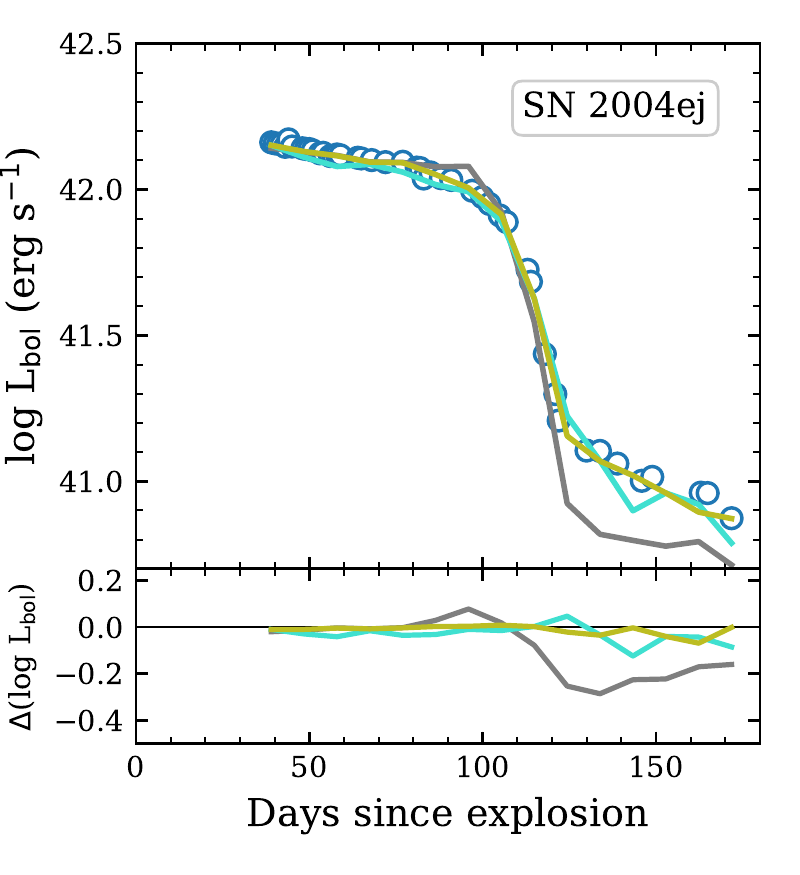}
\includegraphics[width=0.33\textwidth]{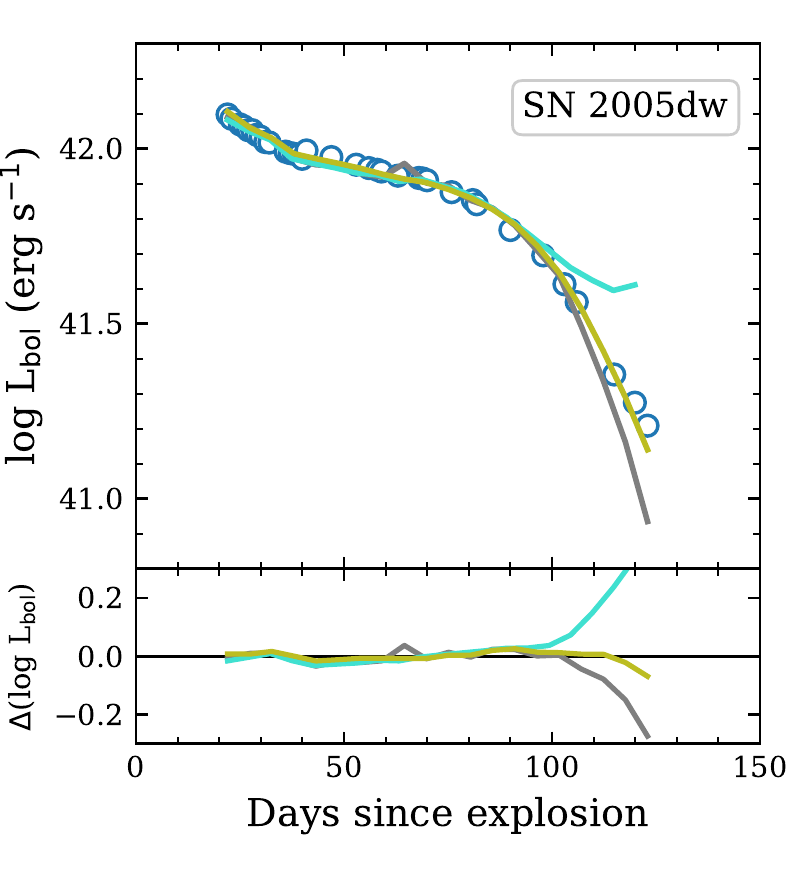}
\caption{Bolometric LCs for SN~2005dz (left panel), SN~2004ej (middle panel), and SN~2005dw (right panel) in comparison with those calculated from the calibrations of BC versus colour presented in the current study.}
\label{fig:bc_comp_blcs_M21}
\end{figure*}

In this section, we compare our results with the calibrations of the BC versus colour found in the literature. We begin by comparing the calibrations directly, and then we compare our bolometric LCs (Sect.~\ref{sec:csp_snii_blcs}) with those calculated using the different calibrations in the literature for the same sample of 74~\sneii, whenever possible.

The left panel of Fig.~\ref{fig:bc_comp} shows our fits to the BC$_{V}$ against $B-V$. We also include the results from \citet{bersten+09} and \citet{pejcha+15b}. We do not include the fits by \citet{lyman+14} as they calculate the BC relative to the $B$ band and a direct comparison is not possible. We find a very good agreement with the BCs of \citet{bersten+09}. The difference between the latter and that from \citet{pejcha+15b} may be due to a difference in the zero point \citep{pejcha+15b}.
Fits to the BC$_{g}$ versus $g-r$ and $g-i$ are shown in the middle and right panels of Fig.~\ref{fig:bc_comp}, respectively. In these cases, we can only compare with the study of \citet{lyman+14} since \citet{pejcha+15b} present BCs relative to the $r$ band for $g-r$ colour and relative to the $i$ band for $g-i$, and \citet{bersten+09} uses photometry in Johnson filters. In both cases, we find good agreements with the BCs of \citet{lyman+14}.
We are not able to compare with the BCs of \citet{pritchard+14}, as they use colours from \emph{Swift}-UVOT bands.

In the three cases analysed above we note that the BC calibrations from the literature extend bluer than ours. 
The calibration of \citet{bersten+09} results in lower BC values than our fits at blue colours (early times), probably because \citet{bersten+09} used spectral models to extrapolate the missing UV flux at these early times which suggest larger UV corrections than observations.
Moreover, the analysis of the BC$_{g}$ versus $g-r$ and $g-i$ shows that our fits yield lower BCs for $g-r$~<~0.5~mag and $g-i$~<~0.2~mag, respectively.
The bottom panel of Fig.~\ref{fig:bc} shows the effect of additional extinction in the BC$-$colour plane, finding that the BCs are larger and the SN moves to bluer colours \citep[see also][their Fig.~8]{lyman+14}.
Therefore, the differences found with previous calibrations may point to uncorrected host-galaxy extinction in our sample. 
However, while the data from \citet{bersten+09}, \citet{lyman+14}, and \citet{pejcha+15b} are corrected for host-galaxy extinction, the reader should again note the issues present in the calculation of these values (Sect.~\ref{sec:sample}).
Another important point to mention is with respect to the different ranges of validity of the calibrations. Our calibrations of BC$_{V}$ and BC$_{g}$ can be used for a larger range of colours which means that a larger number of bolometric LCs of \sneii\ can be predicted and to later times, as the SN evolves to redder colours.

We also estimated the bolometric luminosity for the 74~\sneii\ in the CSP-I sample by means of the BC calibrations reported in the literature and those presented in the current work, and compare them with our bolometric LCs presented in Sect.~\ref{sec:csp_snii_blcs}.
The calibrations of BC versus colour used in this analysis are shown in Table~\ref{table:bc_comp}. It also shows the average of the differences for the entire sample and the standard deviation.
We find major agreements for most BC calibrations. However, we note some peculiarities that are important to mention.
First, we compare the bolometric LCs presented in this study with those calculated using the BC calibrations from the literature (Fig.~\ref{fig:bc_comp_blcs}) and then with those using the new calibrations from the current study (Fig.~\ref{fig:bc_comp_blcs_M21}).

The bolometric LCs presented in Fig.~\ref{fig:bc_comp_blcs} represent the behaviour of the vast majority of the \sneii\ in the sample.
The left panel of Fig.~\ref{fig:bc_comp_blcs} shows that all the bolometric LCs using previous BC calibrations are similar to each other, although they are dimmer than our bolometric LC. This behaviour is also seen for a number of other CSP-I \sneii.
In several cases, there is a significant offset between the bolometric LCs from previous BC calibrations and our bolometric LCs. This is shown in the middle panel of Fig.~\ref{fig:bc_comp_blcs} taking SN~2004ej as an example. The difference is obvious and tends to increase with time, being even more significant in the radioactive tail.
For this SN, the $g-r$ calibration from \citet{lyman+14} is the only one showing good agreement in the photospheric phase.
The large differences found during the tail phase are seen for almost all the \sneii\ in the sample. Sometimes the discrepancy is even larger than that for SN~2004ej, which can significantly bias the \Ni\ mass estimates (see Sect.~\ref{sec:comp_blcs}).
This may be due to the constant BC adopted by \citet{bersten+09} and \citet{pejcha+15b} during the exponential decay phase. Although the BC at late times may be constant for individual \sneii, it does not mean that the same value should be adopted for all \sneii.
In Fig.~\ref{fig:bc} it is clearly shown that the BC during the radioactive phase is not constant when considering all \sneii\ together.
\citet{lyman+14} do not present BC calibrations during the radioactive phase.

In the right panel of Fig.~\ref{fig:bc_comp_blcs}, the bolometric LCs for SN~2005dw are shown as an example. The bolometric LCs constructed using the calibration of BC versus $B-V$ from \citet{bersten+09} and \citet{lyman+14} (orange and green lines, respectively) systematically underestimate the luminosity of our LC, while the opposite is found for the BC calibration against $g-r$ colours from \citet{lyman+14} (red line). However, we are most interested in how the distinct validity ranges of the BC calibrations affect the luminosity estimation during the \snii\ evolution.
In Fig.~\ref{fig:bc_comp}, we already indicated that the BC calibrations from \citet{lyman+14} are valid over shorter ranges of colours than the colours from our \snii\ sample. 
This means that the bolometric luminosity cannot be estimated when the \snii\ colours fall outside the validity range. We found that for most \sneii\ in our sample, the luminosity estimates using some of the calibrations from the literature do not cover the entire evolution in the photospheric phase.
Figure~\ref{fig:bc_comp_blcs} (right panel) shows that the bolometric LC using the BC calibration versus $g-r$ colours from \citet{lyman+14} only covers 90 days of evolution of SN~2005dw.
This is even worse when using the BC calibrations from \citet{pejcha+15b} since these calibrations are valid over a much shorter range of colours. The bolometric LCs using the BC calibrations from \citet{pejcha+15b} only cover the first 55~days of evolution. This behaviour occurs for the majority of the \sneii\ in our sample.
While the bolometric LCs using the relations from \citet{pejcha+15b} show small differences with our LCs (see Table~\ref{table:bc_comp}), we note that most of the times the BC calibrations from \citet{pejcha+15b} estimate the luminosity only up to phases around the middle of the plateau phase.
In principle, this issue could be connected with the fact that we do not correct for host-galaxy extinction, resulting in redder \snii\ events that fall outside the validity ranges. We tested this by randomly sampling 74 host-galaxy extinction values from an uniform distribution with values between 0 and 0.6~mag. CSP-I photometry was then corrected for host-galaxy extinction (i.e. one value for each \snii) and we constructed the bolometric LCs using the BC calibrations from the literature. Some of the bolometric LCs now cover the plateau phase up to later epochs, but only for $\sim$10 additional days. The luminosity estimates still do not cover the entire photospheric phase of the \sneii.

Finally, Fig.~\ref{fig:bc_comp_blcs_M21} compares the bolometric LCs for the same three \sneii\ (SNe~2004ej, 2005dw, 2005dz) calculated using the method presented in Sect.~\ref{sec:blc_calc} with those constructed from the BC calibrations provided in the current work (Fig.~\ref{fig:bc}).
We find excellent agreement between both methods, except for the radioactive tail of SN~2004ej which is systematically dimmer when the BC calibration against $B-V$ is used.
We note that the calibration of BC$_{g}$ against $g-i$ colours presents the smallest differences when all the bolometric LCs from CSP-I \snii\ sample are used (Table~\ref{table:bc}). Such behaviour is expected since that calibration shows the lowest dispersion.

In summary, we have constructed BC calibrations against optical colours from 74~\sneii\ that are valid over a wide range of colours. Therefore, we are confident that our BC calibrations are an appropriate method to estimate bolometric luminosities, particularly the calibration of BC$_{g}$ versus $g-i$ as it shows the tightest correlation.

\section{Summary and conclusions}
\label{sec:conclusions}

In this paper, we have presented bolometric LCs for 74~\sneii\ using $uBgVriYJH$ photometry from the CSP-I. Our calculation method is based on the integration of the observed SED and extrapolations for the flux that falls bluewards and redwards of the wavelength range. The contribution of the IR regime was derived from BB fits. At early times, the UV flux was assumed to be that of a BB. Once the observed SED departs from a BB model a linear extrapolation was used to estimate the flux in the UV. Thus, we have built the most homogeneous sample of \snii\ bolometric LCs so far in the literature.

Several studies calculate bolometric LCs from BB fits to optical data only. We compared the latter calculation method with ours and found that a BB fitted only with optical data systematically overestimate the SN luminosity for the vast majority of the \sneii\ in our sample.
These differences cause variations in the \snii\ physical parameters derived from LC modelling. Specifically, we found significant discrepancies in the explosion energy and \Ni\ mass estimates.
This argues that NIR observations are crucial to obtain reliable BB fits, and therefore, accurate bolometric luminosities.
For this reason, we recommend the following options to calculate bolometric luminosities when NIR data is missing: 
(a) using CSP-I \snii\ NIR photometry as a template. In Table~\ref{table:colour_calibration}, we provide relationships between the $g-i$ colour and colours involving the $i$ magnitude and a NIR magnitude ($YJH$). Using these relationships, NIR magnitudes can be estimated using optical colours;
(b) the bolometric luminosity can be determined by means of BCs. In Sect.~\ref{sec:bc}, we present calibrations of BC against three optical colours (Table~\ref{table:bc}) using the 74~\sneii\ in the CSP-I sample. Specifically, the calibration of BC$_{g}$ versus $g-i$ colour presents the tightest relation.
Previous BC calibrations from the literature agree quite well with those from the current study. However, the previous calibrations are often valid only over a small range of colours and most of the times the luminosity estimates do not cover the entire optically-thick phase of the \snii.
Our calibrations were constructed from 74~\sneii\ which cover a larger range of colours.

A full characterisation of the bolometric LCs based on several observed parameters was provided. Magnitudes were measured at different epochs, as well as durations and decline rates of different phases of the evolution.
The analyses of the parameter distributions give a continuous sequence of observed LC properties which is consistent with previous analyses using optical LCs.
Bolometric LCs cover a wide range of observed parameters in terms of their luminosities and decline rates. 
During the radioactive phase, most of the \sneii\ decline more quickly than expected if full trapping of gamma-ray photons is assumed. This is further discussed in \citetalias{paper3_submitted}.
The \optd\ distribution shows a continuous sequence between 42~days and 146~days. 

The bolometric LCs from this work are used in \citetalias{paper2_submitted} to determine the progenitor and explosion properties of \sneii\ through the hydrodynamical modelling of LCs and expansion velocities.
In \citetalias{paper3_submitted}, the physical parameters are correlated with the observed LC properties measured in the current paper and several other spectroscopic properties to understand the observed \snii\ diversity from the variety of progenitor systems and explosion properties.

Finally, we remind the reader that our bolometric fluxes are available for widespread use. Such data will be useful for independent modelling to test the consistency between different approaches and further refine how \snii\ observations can be understood in terms of massive star explosions.

\begin{acknowledgements}
The work of the Carnegie Supernova Project was supported by the National Science Foundation under grants AST-0306969, AST-0607438, AST-1008343, AST-1613426, AST-1613472, and AST-1613455.
L.M. acknowledges support from a CONICET fellowship.
L.M. and M.O. acknowledge support from UNRN~PI2018~40B885 grant.
M.H. acknowledges support from the Hagler Institute of Advanced Study at Texas A\&M University.
S.G.G. acknowledges support by FCT under Project CRISP PTDC/FIS-AST-31546/2017 and~Project~No.~UIDB/00099/2020.
M.S. is supported by grants from the VILLUM FONDEN (grant number 28021) and the Independent Research Fund Denmark (IRFD; 8021-00170B).
F.F. acknowledges support from the National Agency for Research and Development (ANID) grants: BASAL Center of Mathematical Modelling AFB-170001, Ministry of Economy, Development, and Tourism’s Millennium Science Initiative through grant IC12009, awarded to the Millennium Institute of Astrophysics, and FONDECYT Regular \#1200710.
L.G. acknowledges financial support from the Spanish Ministry of Science, Innovation and Universities (MICIU) under the 2019 Ram\'on y Cajal program RYC2019-027683 and from the Spanish MICIU project PID2020-115253GA-I00.
P.H. acknowledges the support by National Science Foundation (NSF) grant AST-1715133.
\\
\textit{Software:} \texttt{NumPy} \citep{numpyguide2006,numpy2011}, \texttt{matplotlib} \citep{matplotlib}, \texttt{scikit-learn} \citep{scikit-learn}, \texttt{SciPy} \citep{scipy2020}, \texttt{emcee} \citep{emcee}, \texttt{Pandas} \citep{pandas}, \texttt{ipython/jupyter} \citep{jupyter}.
\end{acknowledgements}

\bibliographystyle{aa}
\bibliography{biblio.bib}

\onecolumn
\begin{appendix}
\begin{multicols}{2}
\section{\snii\ sample}

\input{sn_info.tab}
\input{blc_parameters.tab}

\sneii\ from the CSP-I sample are presented in 
\columnbreak
Table~\ref{table:sn_info}, together with their host galaxies, distance moduli, explosion epochs, and Milky Way reddenings.
Table~\ref{table:blc_parameters} presents the measured bolometric LC parameters as defined in Sect.~\ref{sec:distributions}.
\end{multicols}

\section{Additional plots}
\begin{multicols}{2}
\subsection{Colour-colour diagrams}
\begin{figure*}[h]
\centering
\includegraphics[width=0.31\textwidth]{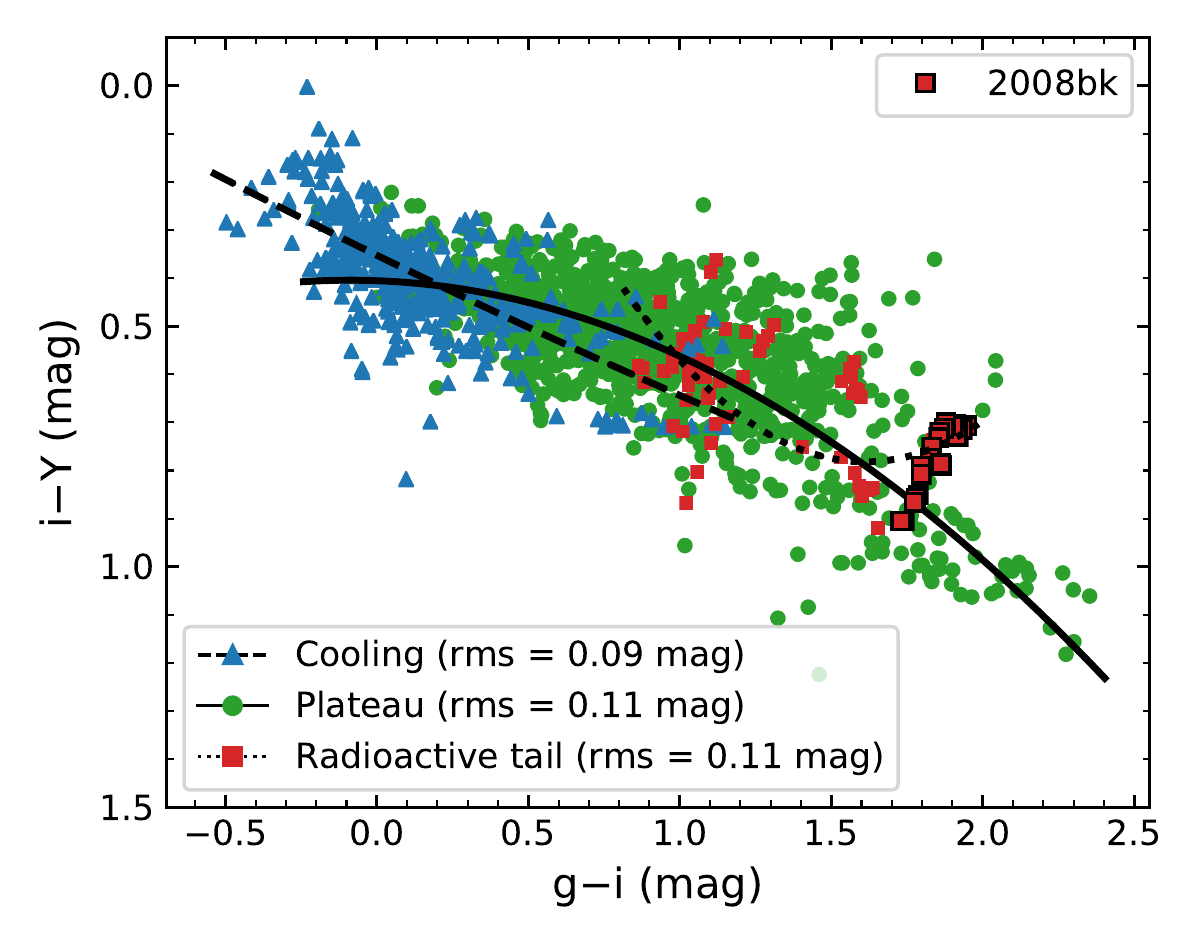}
\includegraphics[width=0.31\textwidth]{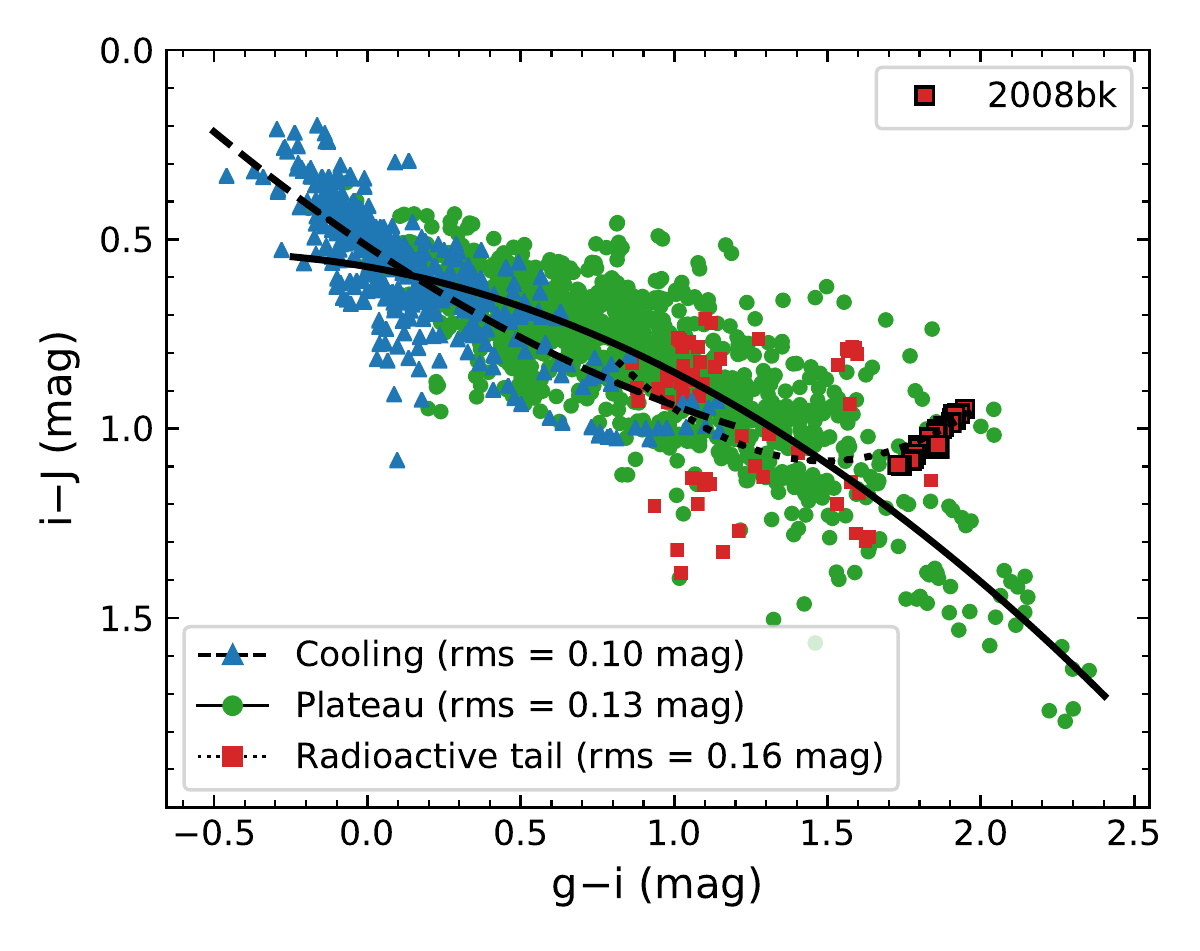}
\includegraphics[width=0.31\textwidth]{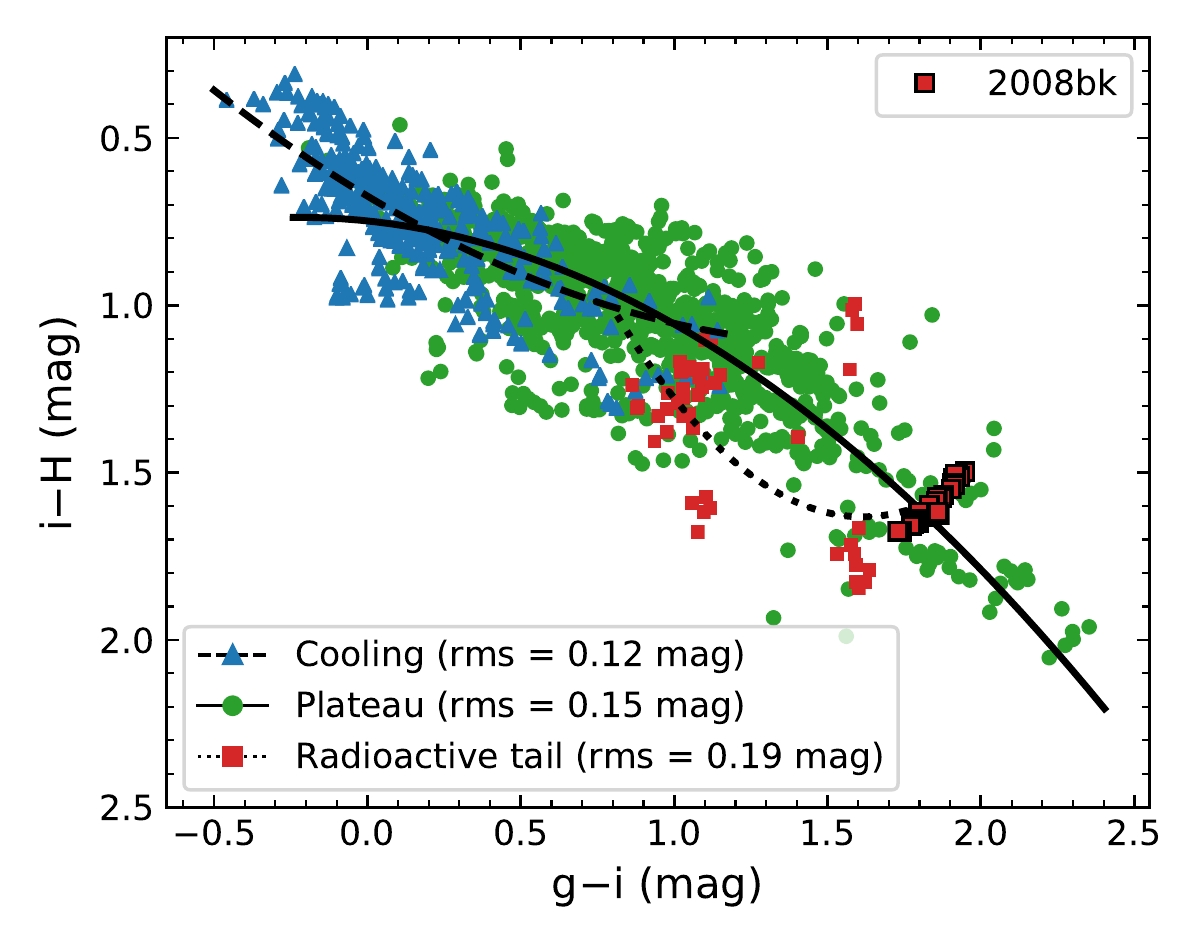}
\caption{Colour-colour diagrams when separating the sample into three epochs: cooling (blue triangles), plateau (green dots), and radioactive tail phase (red squares). All measurements were first corrected for Milky Way extinction. The solid line indicates the polynomial fit to the plateau phase data, while dashed and dotted lines to the cooling and radioactive tail phases, respectively.}
\label{fig:ccd_phases}
\end{figure*}

Additional plots of $g-i$ against $i-Y$, $i-J$, and $i-H$ colours are presented in Fig.~\ref{fig:ccd_phases}. 
\columnbreak
Here, the data were separated into three epochs: cooling phase (blue triangles), plateau phase (green dots), and radioactive tail phase (red squares). Additionally, polynomial fits are presented to each phase of SN evolution.
\end{multicols}

\FloatBarrier

\begin{multicols}{2}
\subsection{Light-curve parameters}
\begin{figure*}[h]
\centering
\includegraphics[width=0.415\textwidth]{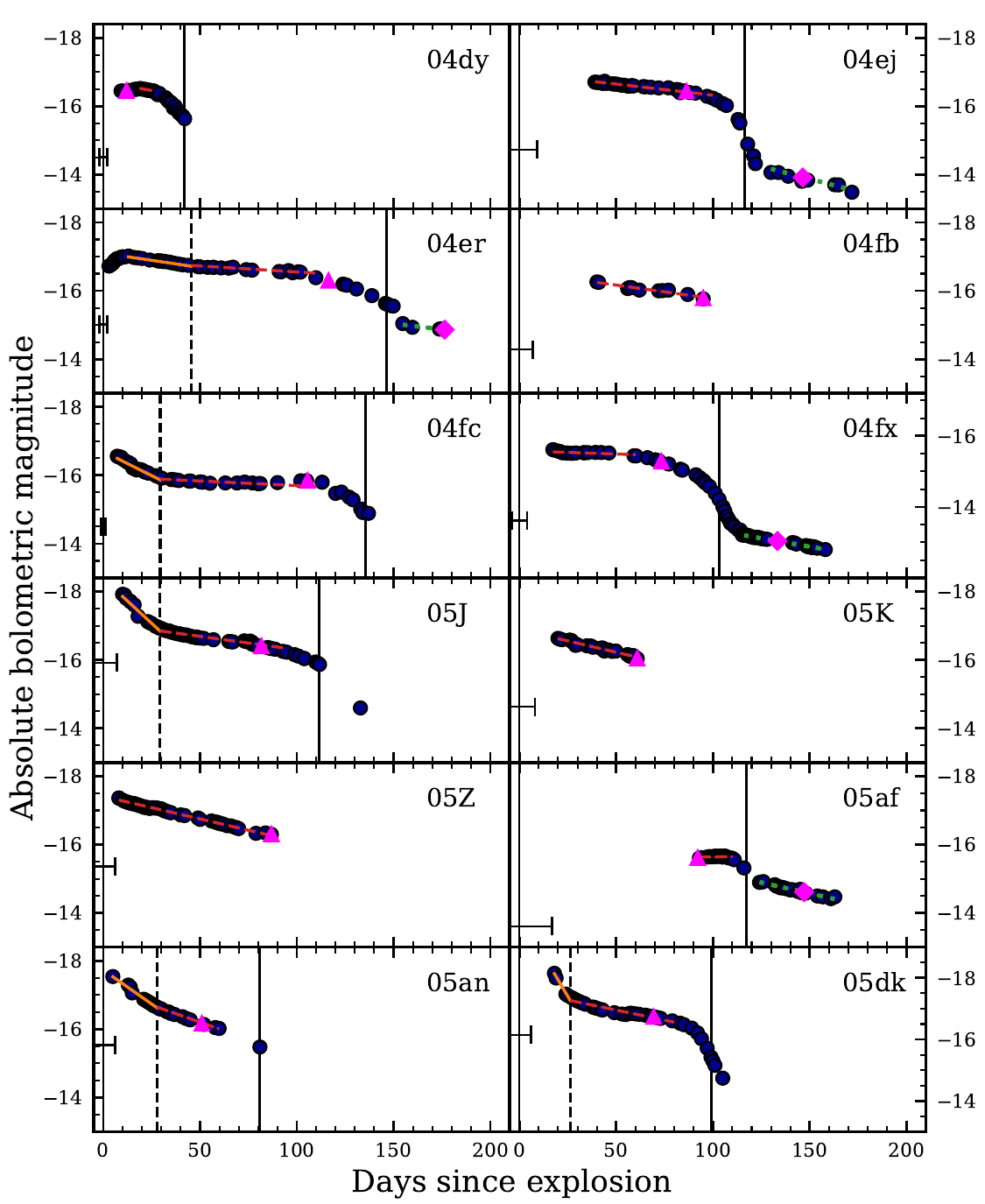} \includegraphics[width=0.415\textwidth]{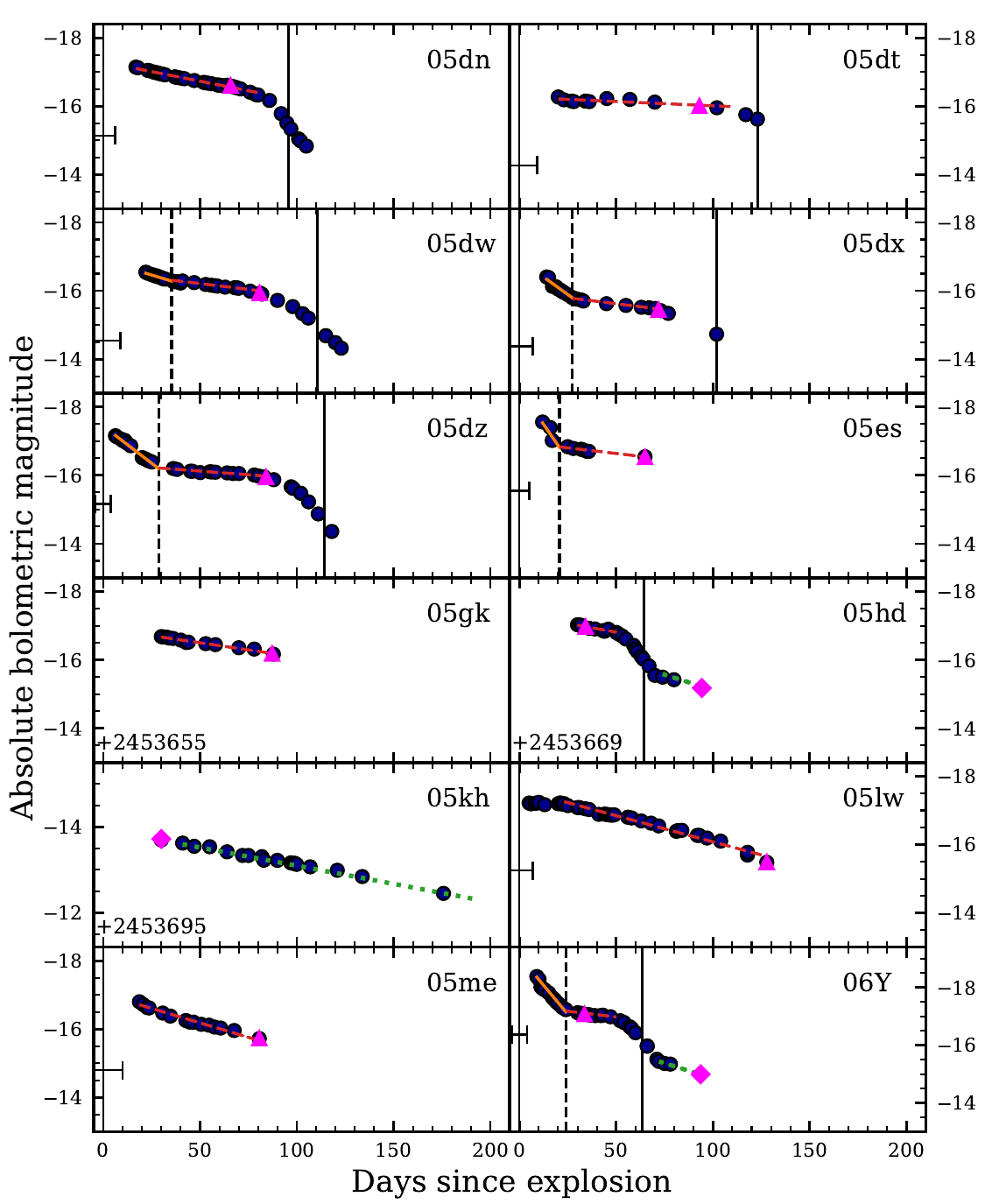}
\caption{Absolute bolometric LCs of \sneii\ in our sample. The panels present SNe in order of their discovery dates, from SN~2004dy to SN~2006Y. Bolometric luminosities are presented as blue dots. The errors are not plotted for better visualisation, but in general they are small. Measured bolometric LC parameters are also presented: \mbolend\ as magenta triangles, \mboltail\ as magenta diamonds, \sone\ as solid orange lines, \stwo\ as dashed red lines, \sthr\ as dotted green lines, \tpt\ as thick vertical black lines, and \ttrans\ as dashed vertical black lines. Thin vertical black lines show the explosion epoch and its uncertainty. SNe~2005gk, 2005hd, and 2005kh do not show this line because they have no explosion epoch estimation.}
\label{fig:fits1}
\end{figure*}

\begin{figure*}
\centering
\includegraphics[width=0.415\textwidth]{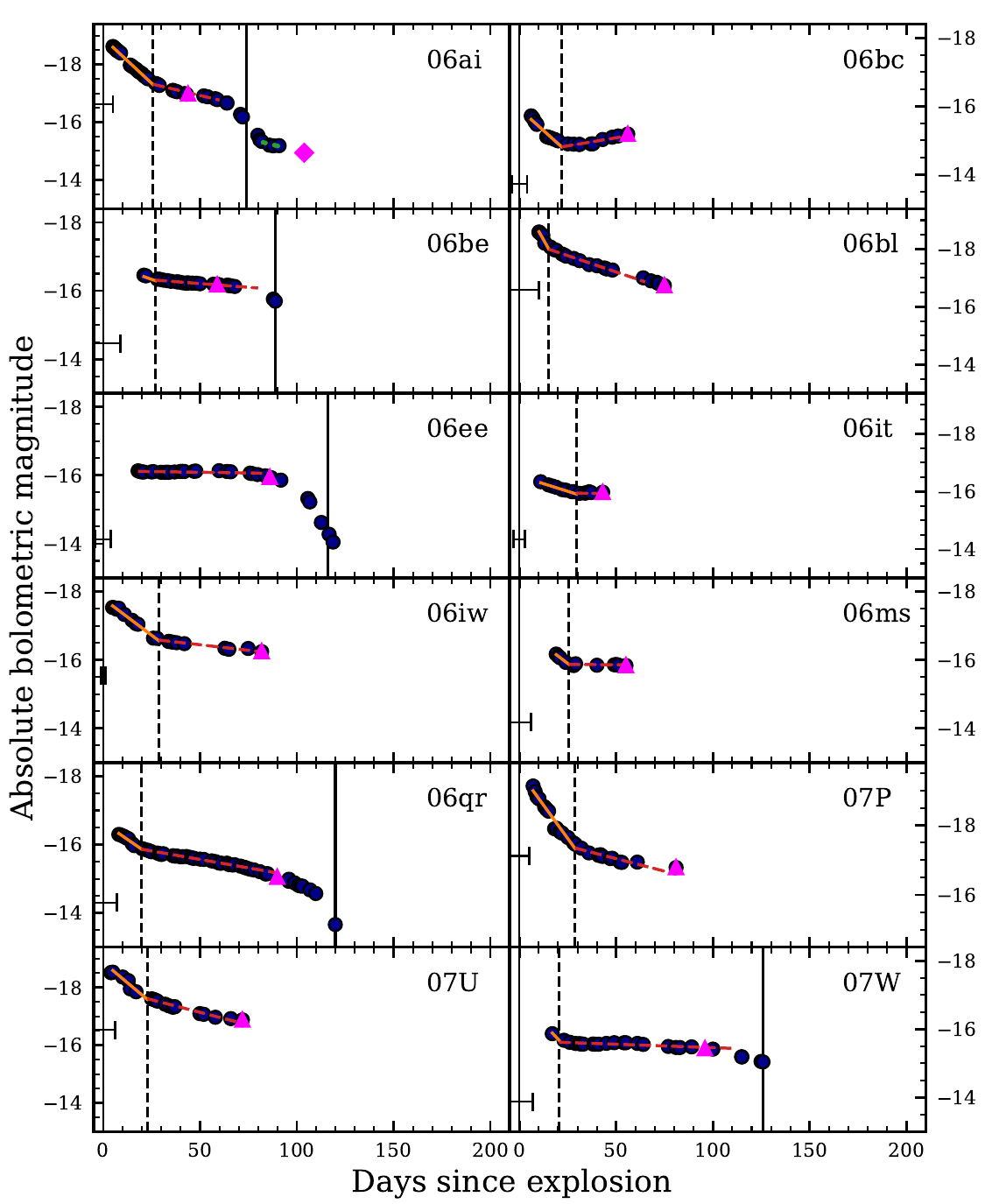}
\includegraphics[width=0.415\textwidth]{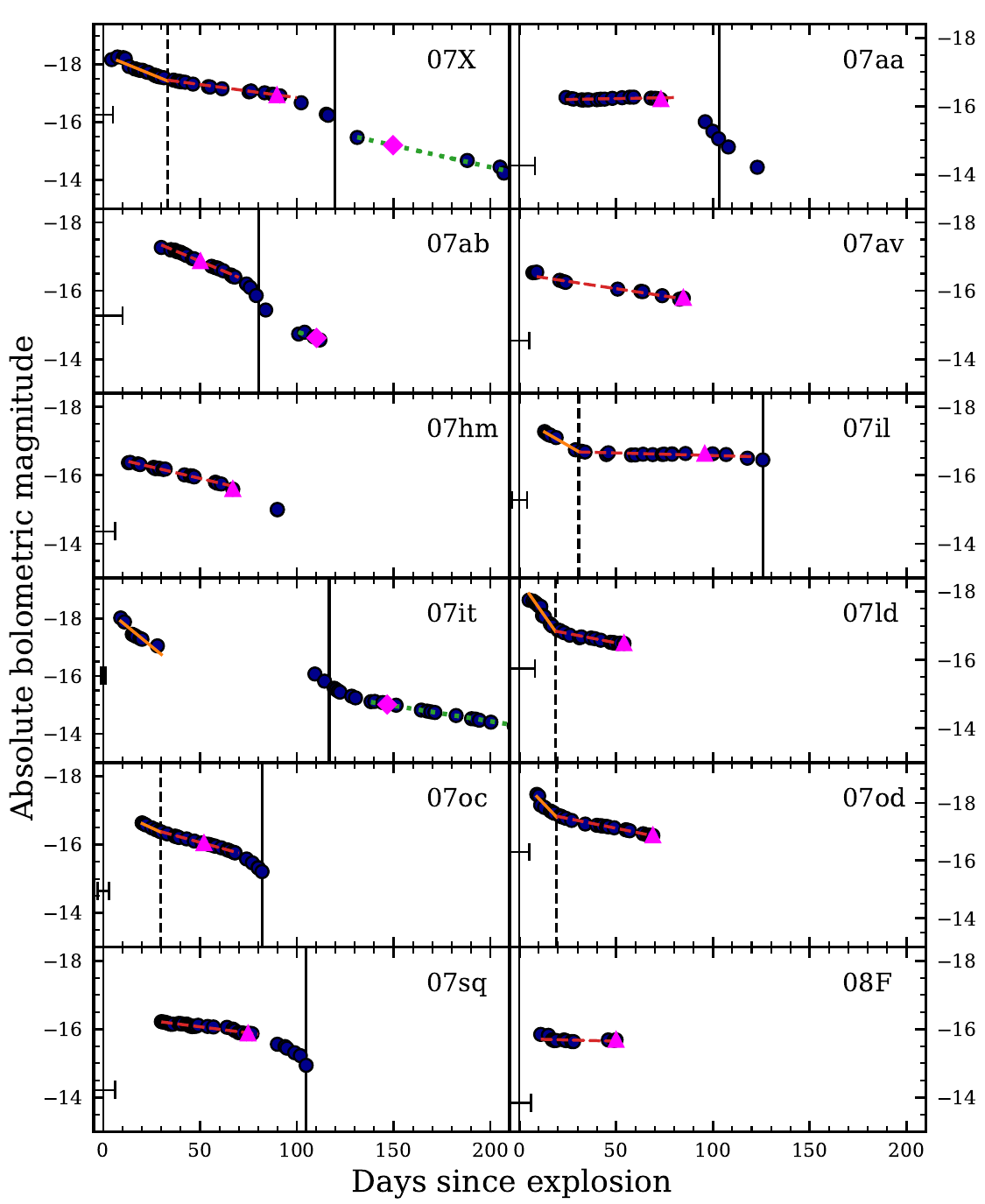}
\includegraphics[width=0.415\textwidth]{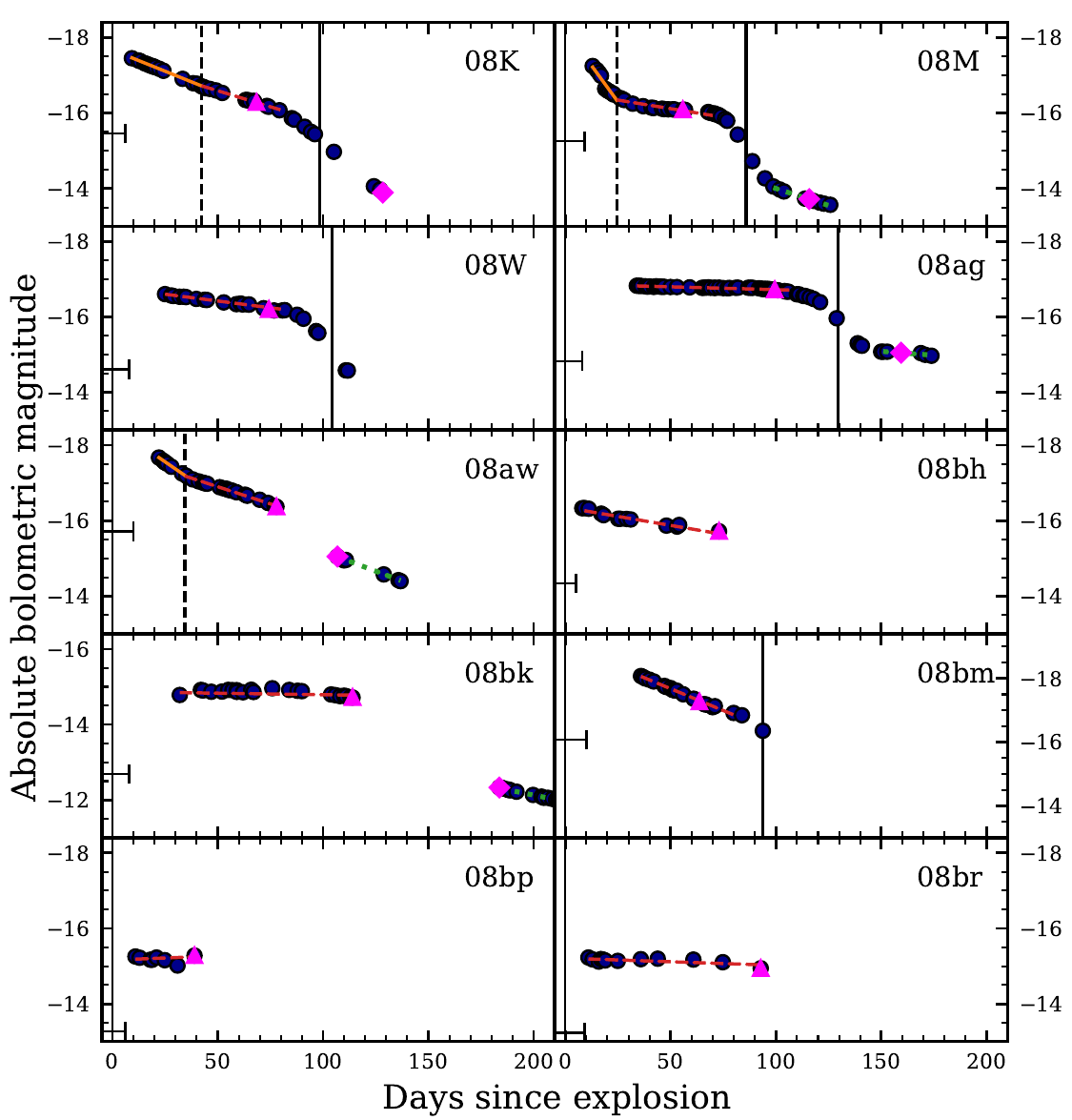} \includegraphics[width=0.415\textwidth]{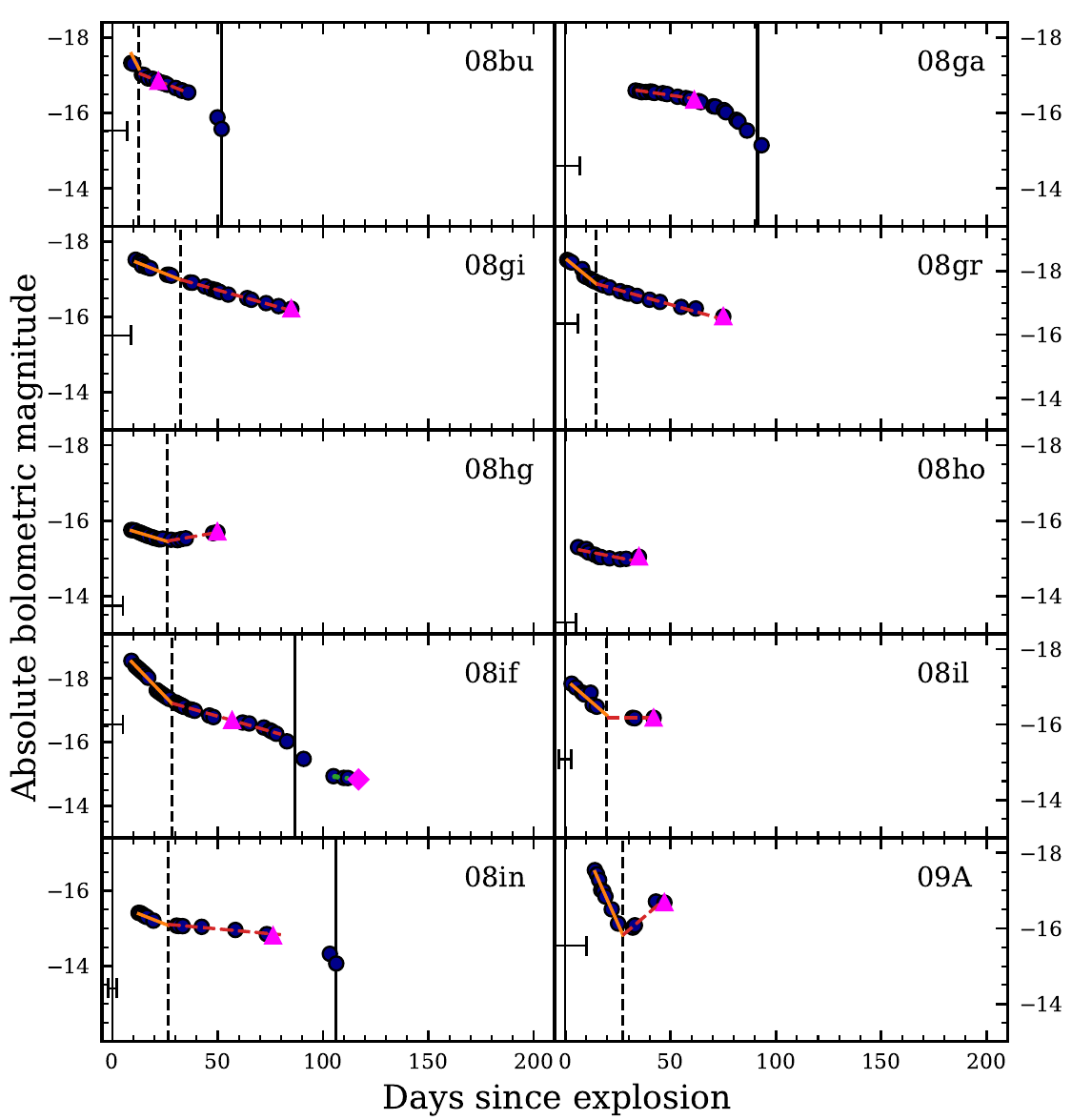}
\caption{Same as in Fig.~\ref{fig:fits1}, but from SN~2006ai to SN~2009A.}
\label{fig:fits2}
\end{figure*}

\begin{figure}
\centering
\includegraphics[width=0.415\textwidth]{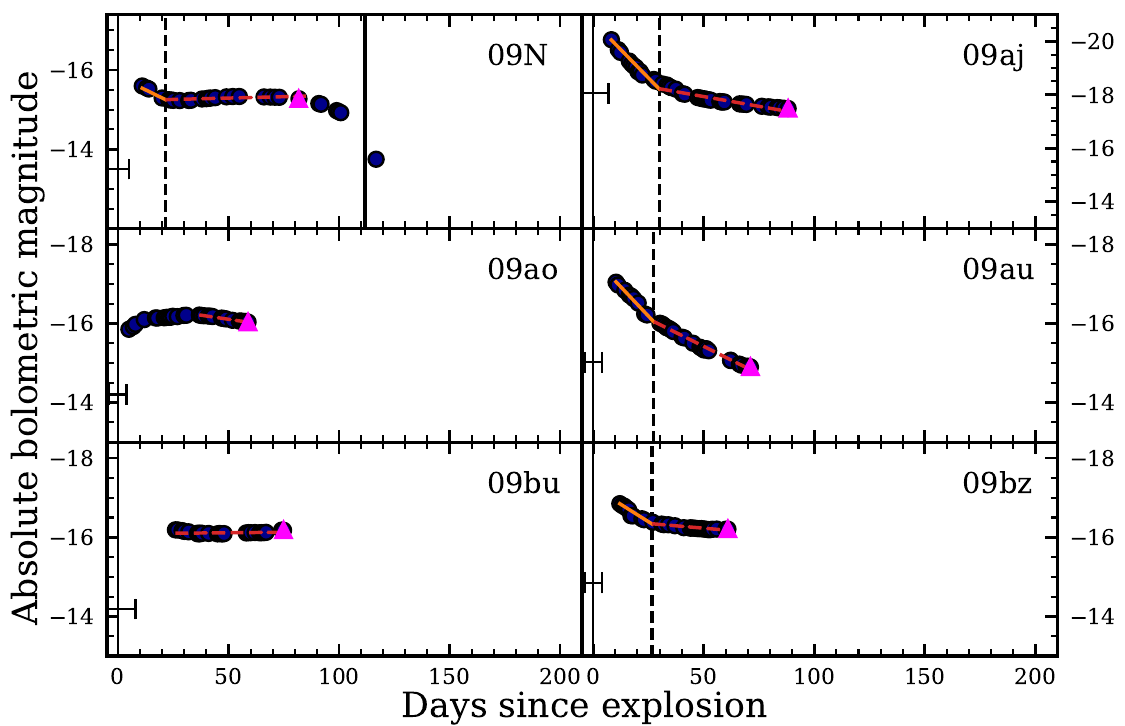} 
\caption{Same as in Fig.~\ref{fig:fits1}, but from SN~2009N to SN~2009bz.}
\label{fig:fits3}
\end{figure}

Figures \ref{fig:fits1}, \ref{fig:fits2}, and \ref{fig:fits3} present absolute bolometric  
\columnbreak
magnitude LCs for the 74~\sneii\ within the CSP-I sample, together with their measured parameters listed in Table~\ref{table:blc_parameters}.
\end{multicols}

\FloatBarrier

\twocolumn
\section{IR and UV corrections}
\label{app:extrapolations}

In Sect.~\ref{sec:bb}, we mentioned that BB fits are not suitable to extrapolate IR fluxes during the radioactive tail phase since at these times the SN radiation is dominated by line emission.
However, Fig.~\ref{fig:ircorr} shows that the IR correction takes a value of about $\sim$15\% by the end of the photospheric phase, meaning that it is certainly important to correct for the missing flux in the IR during the radioactive tail phase or the bolometric luminosities will be considerably underestimated.
In the following, we tested if BB extrapolations can account for the flux beyond the $H$ band during the tail phase, while keeping in mind the strong caveat that BB fits are not physically justified at these epochs.
We took two \sneii\ with mid-infrared (MIR) observations in the literature: SN~1987A and SN~2004et. SN~1987A presents extensive NIR and MIR observations, while SN~2004et has five observations in the MIR (four of them at nebular times). It is not expected that the flux longwards of 5~microns significantly contributes to the total flux.
For SN~1987A, we took optical and IR ($JHKLM$ bands) data from \citet{hamuy+90} and \citet{bouchet+89}, respectively. Optical and NIR data for SN~2004et was taken from \citet{maguire+10}, and MIR data from \citet{kotak+09}. The latter observations were obtained in the 3.6~$\mu$m and 4.5~$\mu$m bands (I1 and I2, respectively) using the InfraRed Array Camera (IRAC) on the \emph{Spitzer} Space Telescope.

\begin{figure}
\centering
\includegraphics[width=0.47\textwidth]{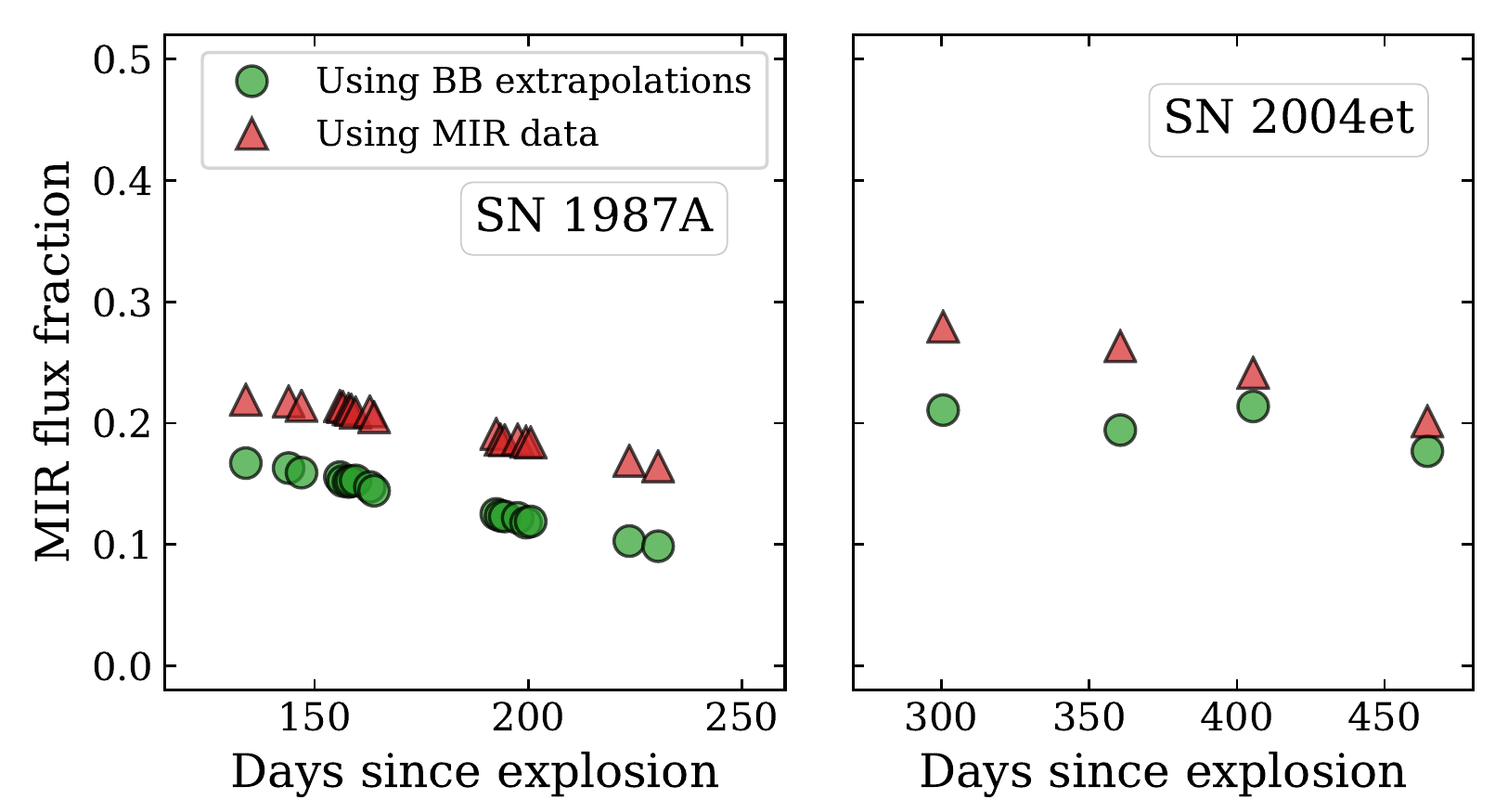} 
\caption{MIR flux fraction as a function of time for SN~1987A and SN~2004et. Green dots show MIR fluxes extrapolated using BB models. Red triangles represent the observed MIR flux.}
\label{fig:mir_comp}
\end{figure}

The goal of this test is to compare the observed flux between the $H$ and $M$ ($I2$) bands for SN~1987A (SN~2004et) with the flux from BB extrapolations in the same regime. BB fits to the observed data were performed using the same procedure described above, but only considering fluxes at shorter wavelengths than the $H$ band to be consistent with the reddest band in CSP-I data.
Figure~\ref{fig:mir_comp} shows the MIR flux normalised by the bolometric flux for SN~1987A (left panel) and SN~2004et (right panel). Red triangles correspond to the observed flux in the MIR, while green dots represent the extrapolated flux in that regime.
We find that BB extrapolations can account for most of the flux between the $H$ and $M$ (I2) bands, from 60\% to almost 90\%, depending on the epoch and the SN. 
In neither case, IR corrections overpredict the actual flux in the MIR.
We could use SN~1987A and SN~2004et as representative of all \sneii\ and derive an IR correction from these two SNe that will then be used for the \sneii\ in the CSP-I sample.
However, this correction would be dominated by possible systematics from these two \sneii.
Thus, while BB fits are not appropriate to model the SN radiation during the radioactive tail from the physical point of view, we used them to estimate IR corrections given that we have shown that they can represent most of the flux at long wavelengths.

In order to test our UV correction, we used \emph{Swift}-UV data for seven \sneii\ in the CSP-I sample. These observations were already analysed by \citet{pritchard+14}. 
Before using \emph{Swift}-UV observations we first need to consider the effect of the red tail of the \emph{Swift} $uvw1$ and $uvw2$ filters \citep{breeveld+11} since it produces a red leak that might bias the flux at these bands.
We used BB fits to model the red-leak contribution. BB fits were performed using CSP-I and \emph{Swift}-$uvm2$ data, since the latter is the only \emph{Swift}-UV filter that is not affected by a red leak. 
We assumed that the red-leak contribution represents the wavelengths more than half the equivalent width redwards of the central wavelength of the filter \citep[similar to][]{ergon+14}.
Once the red-leak contribution was determined for each observation, it was removed from the observed magnitudes. Finally, CSP-I data together with red-leak corrected \emph{Swift}-UV observations were used to calculate bolometric luminosities.

\begin{figure*}
\centering
\includegraphics[width=0.93\textwidth]{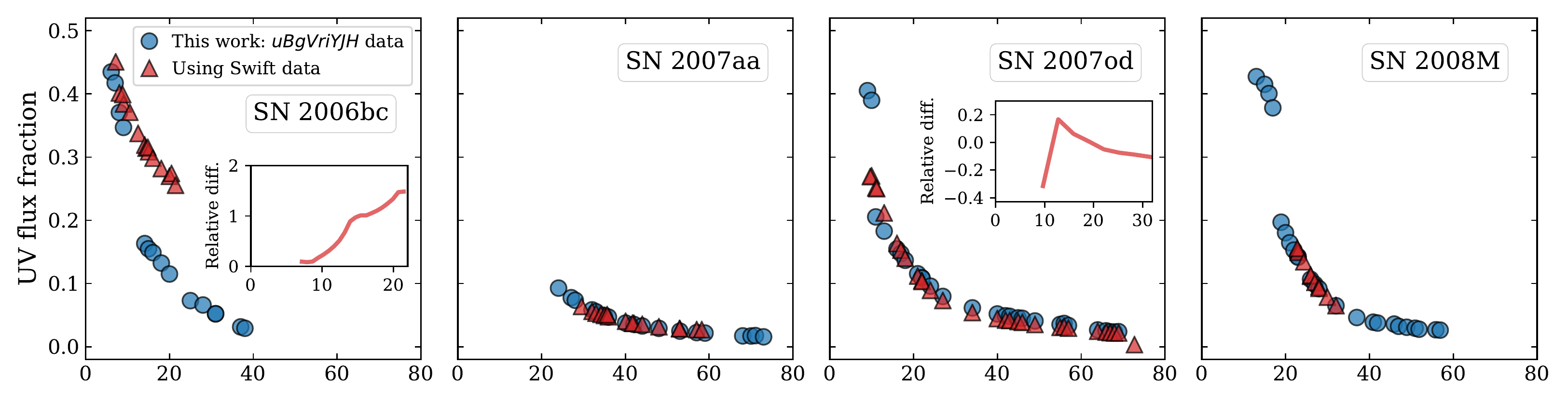} \\
\includegraphics[width=0.93\textwidth]{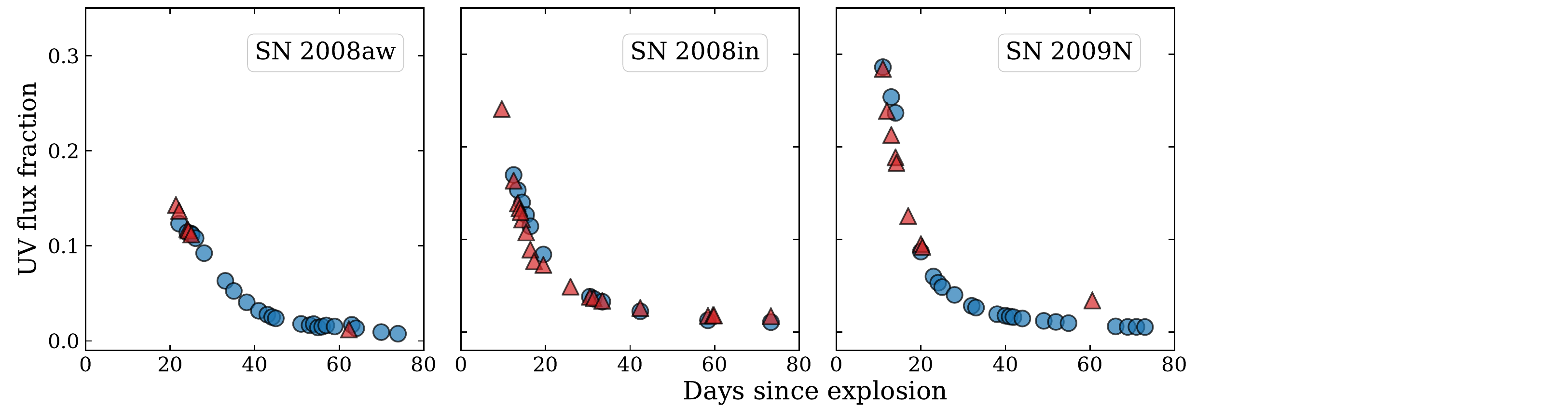}
\caption{UV flux fraction as a function of time for the \sneii\ in the CSP-I sample with available \emph{Swift}-UV observations. Blue dots represent our UV extrapolation method to account for the unobserved flux at shorter wavelengths than $u$ band, while red triangles consider red-leak corrected \emph{Swift}-UV data. For the cases of SN~2006bc and SN~2007od the relative differences between both UV fractions are shown since these two \sneii\ present the largest differences.}
\label{fig:uvcorr_comp}
\end{figure*}

Figure~\ref{fig:uvcorr_comp} compares the extrapolated fluxes using our method with observed UV fluxes for seven \sneii\ in the CSP-I sample with \emph{Swift}-UV data. Blue dots represent the UV correction using our technique as described above, while red triangles represent the integrated flux within the bluest-\emph{Swift} band and $u$ band. Both quantities are normalised by the bolometric flux.
In addition, we considered BB extrapolations to account for the missing flux bluewards of the observed \emph{Swift} filters, but only as long as the bluest-\emph{Swift} band does not drop below the BB model.
We note excellent agreement between both techniques at times $\gtrsim$\,20~days from explosion, while sometimes we find discrepancies at earlier times. 
Unfortunately, \emph{Swift}-UV observations earlier than $\sim$20~days (when a considerable part of the bolometric flux lies in the UV) are available only for four out of the seven \sneii\ we analyse here and, therefore, we cannot achieve any general conclusion.
SN~2008in and SN~2009N show good agreements at times $\lesssim$\,20~days, however, the comparison for SN~2006bc and SN~2007od displays some discrepancy between both methods.
In the case of SN~2006bc, both methods agree well until 10~days after explosion, but then, the observed UV fraction using Swift data is considerably larger ($\sim$50$-$150\%) than the UV fraction using our extrapolation method.
For SN~2007od, we find the opposite, that is our UV correction is 20\% larger than the observed flux in the first 10~days of evolution.

The different results found at times earlier than 20~days may imply that there is no unique method to correct the missing flux in the UV at such early epochs. 
This may be reasonable if we consider that most \sneii\ may be affected by an additional energy source at early times, such as the interaction of the ejecta with CSM. 
Such an effect produces an increment of the flux in the UV (as well as in optical wavelengths), which might not be fully represented by BB fits to optical and NIR observations.
CSM interaction affects the early evolution of \sneii\ and its effect depends on the density of the CSM and its extent, among other CSM properties. Therefore, the diversity of CSM configurations that could be present in Nature may explain the different results we found when comparing our extrapolated and observed UV fluxes.
With this test we conclude that (a) we trust our bolometric calculations for times $\gtrsim$\,20~days from explosion, (b) results involving bolometric luminosities at times $\lesssim$\,20~days should be taken with caution as our calculation method may not predict the actual UV contribution at those epochs, and (c) early-time UV observations are necessary for reliable estimations of the UV flux, and therefore, of the bolometric flux.

\end{appendix}
\end{document}

%% file: colour_calibration.tab
\begin{table}
    \caption{Coefficients of the polynomial fits to the intrinsic colour relations.}
    \label{table:colour_calibration}      
    \centering          
    \begin{tabular}{ccccccc}
    \hline\hline\noalign{\smallskip}
    Colour & $g-i$ range & $c_{0}$ & $c_{1}$ & $c_{2}$ & $c_{3}$ & rms \\
    \hline\noalign{\smallskip}
    $i-Y$ & ($-$0.5, 2.4) & 0.35 & 0.27 & $-$0.11 & 0.06 & 0.10 \\
    $i-J$ & ($-$0.5, 2.4) & 0.52 & 0.44 & $-$0.22 & 0.12 & 0.13 \\
    $i-H$ & ($-$0.5, 2.4) & 0.67 & 0.43 & $-$0.16 & 0.11 & 0.15 \\
    \hline
    \end{tabular}
\tablefoot{colour = $\sum_{k=0}^{3} c_{k}(g-i)^{k}$, where colour is chosen from column 1. The coefficients are the median of the marginal distributions.}
\end{table}

%% file: physical_pars.tab
\begin{table*}
   \caption{Physical parameters derived from the hydrodynamic modelling of bolometric LCs and expansion velocities for three SNe II in our sample when their bolometric LCs are constructed using different methods.}             
   \label{table:physical_pars}      
   \centering          
   \begin{tabular}{lcccccc}
   \hline\hline\noalign{\smallskip}
   SN & BLCs & $M_{\rm ZAMS}$ & $M_{\rm ej}$ & $E$ & $M_{\rm Ni}$ & $^{56}$Ni mixing \\
   & & ($M_{\odot}$) & ($M_{\odot}$) & (foe) & ($M_{\odot}$) & (\%) \\
   \hline\noalign{\smallskip}
   \multirow{2}*{2005dn} & This work & 10.7$^{+1.7}_{-0.6}$ & 8.5$^{+1.2}_{-0.3}$ & 0.94$^{+0.10}_{-0.12}$ & 0.041$^{+0.009}_{-0.010}$ & 39$^{+7}_{-9}$ \\
   \noalign{\smallskip}
   & Method 1 & 11.8$^{+1.9}_{-1.4}$ & 9.2$^{+1.8}_{-0.9}$ & 1.12$^{+0.16}_{-0.15}$ & 0.049$^{+0.014}_{-0.015}$ & 45$^{+19}_{-16}$ \\
   \noalign{\smallskip}   
   \hline\noalign{\smallskip}
   \multirow{2}*{2007ab} & This work & 11.2$^{+0.3}_{-0.2}$ & 8.8$^{+0.2}_{-0.2}$ & 1.15$^{+0.09}_{-0.04}$ & 0.033$^{+0.003}_{-0.002}$ & 22$^{+2}_{-1}$ \\
   \noalign{\smallskip}
   & Method 1 & 10.9$^{+0.5}_{-0.5}$ & 8.6$^{+0.3}_{-0.2}$ & 1.42$^{+0.05}_{-0.07}$ & 0.051$^{+0.006}_{-0.006}$ & 26$^{+6}_{-4}$ \\
   \noalign{\smallskip}   
   \hline\noalign{\smallskip}
   \multirow{2}*{2008if} & This work & 10.1$^{+0.2}_{-0.1}$ & 8.2$^{+0.1}_{-0.1}$ & 1.00$^{+0.06}_{-0.06}$ & 0.042$^{+0.003}_{-0.002}$ & 21$^{+2}_{-1}$ \\
   \noalign{\smallskip}
   & Method 1 & 10.2$^{+1.0}_{-0.4}$ & 8.3$^{+0.5}_{-0.2}$ & 1.33$^{+0.11}_{-0.13}$ & 0.065$^{+0.010}_{-0.013}$ & 33$^{+11}_{-8}$ \\
   \noalign{\smallskip}   
   \hline
   \end{tabular}
   \tablefoot{BLCs indicates the method used to calculate the bolometric luminosity. $M_{\rm ZAMS}$, $M_{\rm ej}$, $E$, and $M_{\rm Ni}$ refer to the progenitor initial mass, ejecta mass, explosion energy, and $^{56}$Ni mass, respectively. $^{56}$Ni mixing describes the spatial distribution of $^{56}$Ni within the ejecta and is presented in percentage of the pre-SN mass in mass coordinate. The reader is referred to Paper~II for a full description of our modelling and fitting procedures.}
\end{table*}

%% file: temperature.tab
\begin{table}
    \caption{Mean BB temperatures and the standard deviation for our SNe~II sample as a function of time.}
    \label{table:temp}      
    \centering          
    \begin{tabular}{cc}
    \hline\hline\noalign{\smallskip}
    Epoch\tablefootmark{a} (days) & T$_{\rm BB}$ (K) \\
    \hline\noalign{\smallskip}
    5 & 9328~$\pm$~2164 \\
    15 & 8307~$\pm$~1711 \\
    25 & 6766~$\pm$~1138 \\
    35 & 6211~$\pm$~967 \\
    45 & 5948~$\pm$~882 \\
    55 & 5786~$\pm$~774 \\
    65 & 5692~$\pm$~741 \\
    75 & 5673~$\pm$~881 \\
    85 & 5692~$\pm$~965 \\
    95 & 5595~$\pm$~1211 \\
    105 & 5270~$\pm$~1007 \\
    115 & 5342~$\pm$~1011 \\
    125 & 4994~$\pm$~995 \\
    135 & 4379~$\pm$~657 \\
    \hline
    \end{tabular}
\tablefoot{
\tablefoottext{a}{Days since explosion.}}
\end{table}

%% file: bc.tab
\begin{table*}
   \caption{Coefficients of the polynomial fits to the BC versus different colours.}             
   \label{table:bc}
   \centering          
   \begin{tabular}{lcccccccc}
   \hline\hline\noalign{\smallskip}
   Colour & Phase & Range & $c_{0}$ & $c_{1}$ & $c_{2}$ & $c_{3}$ & $c_{4}$ & $\sigma$ \\
   \hline\noalign{\smallskip}
   \multirow{3}*{$B-V$} & Cooling & ($-$0.10, 1.16) & $-$0.740 & 4.472 & $-$9.637 & 9.075 & $-$3.290 & 0.12 \\
   & Plateau & (0.07, 1.98) & $-$0.384 & 1.692 & $-$2.370 & 1.524 & $-$0.476 & 0.19 \\
   & Tail & (0.46, 1.78) & $-$2.696 & 11.532 & $-$18.805 & 13.040 & $-$3.315 & 0.21 \\
   \hline\noalign{\smallskip}
   \multirow{3}*{$g-r$} & Cooling & ($-$0.26, 1.09) & $-$0.352 & 1.753 & $-$4.078 & 1.961 & --- & 0.11 \\
   & Plateau & (0.01, 2.17) & $-$0.219 & 0.813 & $-$2.194 & 1.205 & $-$0.305 & 0.14 \\
   & Tail & (0.78, 2.07) & $-$9.994 & 21.507 & $-$15.343 & 3.273 & --- & 0.22 \\
   \hline\noalign{\smallskip}
   \multirow{3}*{$g-i$} & Cooling & ($-$0.50, 1.15) & $-$0.214 &  0.789 & $-$2.357 & 1.097 & --- & 0.11 \\
   & Plateau & ($-$0.10, 2.79) & $-$0.140 & 0.292 & $-$1.224 & 0.522 & $-$0.090 & 0.07 \\
   & Tail & (0.86, 2.43) & $-$0.263 & $-$0.154 & $-$0.256 & --- & --- & 0.08 \\
   \hline
   \end{tabular}
   \tablefoot{BC = $\sum_{k=0}^{n} c_{k}(colour)^{k}$, where colour is taken from Column 1. $\sigma$ is the standard deviation about the fit.}
\end{table*}

%% file: bc_comp.tab
\begin{table}
    \caption{Average difference and standard deviation between our bolometric LCs and those calculated using different BC calibrations from the literature.}
    \label{table:bc_comp}
    \centering   
    \resizebox{0.49\textwidth}{!}{
    \begin{tabular}{ccl}
    \hline\hline\noalign{\smallskip}
    Colour & Average difference (dex) & BC calibration \\
    \hline\noalign{\smallskip}
    $B-V$ & 0.05~$\pm$~0.04 & This work \\
    $B-V$ & 0.08~$\pm$~0.05 & \citet{bersten+09} \\
    $B-V$ & 0.08~$\pm$~0.04 & \citet{lyman+14} \\
    $B-V$ & 0.07~$\pm$~0.04 & \citet{pejcha+15b} \\
    $g-r$ & 0.05~$\pm$~0.04 & This work \\
    $g-r$ & 0.05~$\pm$~0.02 & \citet{lyman+14} \\
    $g-r$ & 0.05~$\pm$~0.03 & \citet{pejcha+15b} \\
    $g-i$ & 0.03~$\pm$~0.01 & This work \\
    $g-i$ & 0.04~$\pm$~0.01 & \citet{pejcha+15b} \\
    \hline
    \end{tabular}}
\end{table}

%% file: sn_info.tab
\longtab[1]{
\begin{longtable}{llccc}
\caption{SN~II sample.}
\label{table:sn_info}
\\ \hline\hline\noalign{\smallskip}
SN & Host galaxy & Distance modulus & Explosion epoch & $E(B-V)_{\rm MW}$ \\
   & & (mag) & (MJD) & (mag) \\
\hline\noalign{\smallskip}
\endfirsthead
\caption{continued.}\\
\hline\hline\noalign{\smallskip}
SN & Host galaxy & Distance modulus & Explosion epoch & $E(B-V)_{\rm MW}$ \\    
   & & (mag) & (MJD) & (mag) \\
\hline\noalign{\smallskip}
\endhead
\hline\noalign{\smallskip}
\endfoot
2004dy & IC 5090 & 35.46(0.07) & 53240.5(2) & 0.045 \\
2004ej & NGC 3095 & 33.10(0.21) & 53223.9(9) & 0.061 \\
2004er & MCG -01-7-24 & 33.79(0.16) & 53271.8(2) & 0.023 \\
2004fb & ESO 340-G7 & 34.54(0.11) & 53258.6(7) & 0.056 \\
2004fc & IC 701 & 31.68(0.31) & 53293.5(1) & 0.023 \\
2004fx & MCG -02-14-3  & 32.82(0.24) & 53303.5(4) & 0.090 \\
2005J & NGC 4012 & 33.96(0.14) & 53379.8(7) & 0.025 \\
2005K & NGC 2923 & 35.33(0.08) & 53369.8(8) & 0.035 \\
2005Z & NGC 3363 & 34.61(0.11) & 53396.7(6) & 0.025 \\
2005af & NGC 4945 & 27.75(0.36) & 53320.8(17) & 0.156 \\
2005an & ESO 506-G11 & 33.43(0.18) & 53431.8(6) & 0.083 \\
2005dk & IC 4882 & 34.01(0.14) & 53601.5(6) & 0.043 \\
2005dn & NGC 6861 & 32.83(0.24) & 53602.6(6) & 0.048 \\
2005dt & MCG -03-59-6 & 35.02(0.09) & 53605.6(9) & 0.025 \\
2005dw & MCG -05-52-49 & 34.17(0.13) & 53603.6(9) & 0.020 \\
2005dx & MCG -03-11-9 & 35.18(0.08) & 53611.8(7) & 0.021 \\
2005dz & UGC 12717 & 34.32(0.12) & 53619.5(4) & 0.072 \\
2005es & MCG +01-59-79  & 35.87(0.06) & 53638.7(5) & 0.076 \\
2005gk & 2MASX J03081572 & 35.36(0.08) & --- & 0.050 \\
2005hd & anon & 35.38(0.08) & --- & 0.054 \\
2005kh & NGC 3094 & 32.94(0.04) & --- & 0.025 \\
2005lw & IC 672 & 35.22(0.08) & 53713.8(7) & 0.043 \\
2005me & ESO 244-31  & 34.76(0.10) & 53717.9(10) & 0.022 \\
2006Y & anon & 35.73(0.06) & 53766.5(4) & 0.115 \\
2006ai & ESO 005- G009 & 34.01(0.14) & 53781.6(5) & 0.113 \\
2006bc & NGC 2397 & 31.97(0.26) & 53815.5(4) & 0.181 \\
2006be & IC 4582 & 32.44(0.29) & 53802.8(9) & 0.026 \\
2006bl & MCG +02-40-9 & 35.65(0.07) & 53822.7(10) & 0.045 \\
2006ee & NGC 774 & 33.87(0.15) & 53961.9(4) & 0.054 \\
2006it & NGC 6956 & 33.88(0.15) & 54006.5(3) & 0.087 \\
2006iw & 2MASX J23211915  & 35.42(0.07) & 54010.7(1) & 0.044 \\
2006ms & NGC 6935 & 33.90(0.15) & 54028.5(6) & 0.031 \\
2006qr & MCG -02-22-023 & 34.02(0.14) & 54062.8(7) & 0.040 \\
2007P & ESO 566-G36  & 36.18(0.05) & 54118.7(5) & 0.036 \\
2007U & ESO 552-65 & 35.14(0.08) & 54133.6(6) & 0.046 \\
2007W & NGC 5105 & 33.22(0.20) & 54130.8(7) & 0.045 \\
2007X & ESO 385-G32 & 33.11(0.21) & 54143.5(5) & 0.060 \\
2007aa & NGC 4030 & 31.95(0.27) & 54126.7(8) & 0.023 \\
2007ab & MCG -01-43-2  & 34.94(0.09) & 54123.9(10) & 0.235 \\
2007av & NGC 3279 & 32.56(0.22) & 54173.8(5) & 0.032 \\
2007hm & SDSS J205755.65 & 34.98(0.09) & 54336.6(6) & 0.059 \\
2007il & IC 1704 & 34.63(0.11) & 54349.8(4) & 0.042 \\
2007it & NGC 5530 & 30.34(0.50) & 54348.5(1) & 0.103 \\
2007ld & anon & 35.00(0.09) & 54376.5(8) & 0.081 \\
2007oc & NGC 7418 & 31.29(0.15) & 54388.5(3) & 0.014 \\
2007od & UGC 12846 & 31.91(0.80) & 54400.6(5) & 0.032 \\
2007sq & MCG -02-23-5 & 34.12(0.13) & 54422.8(6) & 0.183 \\
2008F & MCG -01-8-15 & 34.31(0.12) & 54469.6(6) & 0.044 \\
2008K & ESO 504-G5 & 35.29(0.08) & 54475.5(6) & 0.035 \\
2008M & ESO 121-26 & 32.55(0.28) & 54471.7(9) & 0.040 \\
2008W & MCG -03-22-7  & 34.59(0.11) & 54483.8(8) & 0.086 \\
2008ag & IC 4729 & 33.91(0.15) & 54477.9(8) & 0.074 \\
2008aw & NGC 4939 & 33.36(0.19) & 54517.8(10) & 0.036 \\
2008bh & NGC 2642 & 34.02(0.14) & 54543.5(5) & 0.020 \\
2008bk & NGC 7793 & 27.68(0.05) & 54540.9(8) & 0.017 \\
2008bm & CGCG 071-101 & 35.66(0.07) & 54486.5(10) & 0.023 \\
2008bp & NGC 3095 & 33.10(0.21) & 54551.7(6) & 0.061 \\
2008br & IC 2522 & 33.30(0.20) & 54555.7(9) & 0.083 \\
2008bu & ESO 586-G2 & 34.81(0.10) & 54566.8(7) & 0.376 \\
2008ga & LCSB L0250N & 33.99(0.14) & 54711.5(7) & 0.582 \\
2008gi & CGCG 415-004 & 34.94(0.09) & 54742.7(9) & 0.060 \\
2008gr & IC 1579 & 34.76(0.10) & 54769.6(6) & 0.012 \\
2008hg & IC 1720 & 34.36(0.12) & 54779.8(5) & 0.016 \\
2008ho & NGC 922 & 32.98(0.23) & 54792.7(5) & 0.017 \\
2008if & MCG -01-24-10 & 33.56(0.17) & 54807.8(5) & 0.029 \\
2008il & ESO 355-G4 & 34.61(0.11) & 54825.6(3) & 0.015 \\
2008in & NGC 4303 & 30.38(0.47) & 54825.4(2) & 0.020 \\
2009A & KUG 0150-036B & 34.40(0.03) & 54823.0(10) & 0.025 \\
2009N & NGC 4487 & 31.49(0.40) & 54846.8(5) & 0.019 \\
2009aj & ESO 221-G018 & 33.89(0.20) & 54880.5(7) & 0.130 \\
2009ao & NGC 2939 & 33.33(0.20) & 54890.7(4) & 0.034 \\
2009au & ESO 443-21 & 33.16(0.21) & 54897.5(4) & 0.081 \\
2009bu & NGC 7408 & 33.32(0.19) & 54901.9(8) & 0.022 \\
2009bz & UGC 9814 & 33.34(0.19) & 54915.8(4) & 0.035
\end{longtable}
\tablefoot{Columns 1 and 2 list the SN name and its host-galaxy, respectively. In Column 3 the distance modulus is presented, followed by explosion epochs in Column 4, and Milky-Way reddening values in Column 5. Errors are indicated in parenthesis.}
}

%% file: blc_parameters.tab
\longtab[2]{
\begin{landscape}
\begin{longtable}{lcccccccc}
\caption{SN~II bolometric light$-$curve parameters.}
\label{table:blc_parameters}
\\ \hline\hline\noalign{\smallskip}
SN & $M_{\rm bol,end}$ & $M_{\rm bol,tail}$ & $s_{\rm 1}$ & $s_{\rm 2}$ & $s_{\rm 3}$ & $Cd$ & $pd$ & $optd$ \\
 & (mag) & (mag) & (mag 100 day$^{-1}$) & (mag 100 day$^{-1}$) & (mag 100 day$^{-1}$) & (days) & (days) & (days) \\
\hline\noalign{\smallskip}
\endfirsthead
\caption{continued.}\\
\hline\hline\noalign{\smallskip}
SN & $M_{\rm bol,end}$ & $M_{\rm bol,tail}$ & $s_{\rm 1}$ & $s_{\rm 2}$ & $s_{\rm 3}$ & $Cd$ & $pd$ & $optd$ \\
 & (mag) & (mag) & (mag 100 day$^{-1}$) & (mag 100 day$^{-1}$) & (mag 100 day$^{-1}$) & (days) & (days) & (days) \\
\hline\noalign{\smallskip}
\endhead
\hline\noalign{\smallskip}
\endfoot
2004dy & $-$16.4(0.1) & --- & --- & 1.01(2.68) & --- & --- & --- & 42.0(2.0) \\
2004ej & $-$16.4(0.2) & $-$13.9(0.2) & --- & 0.65(0.12) & 1.48(0.04) & --- & --- & 116.4(9.0) \\
2004er & $-$16.3(0.2) & $-$14.9(0.2) & 0.84(0.70) & 0.35(0.09) & 0.68(0.03) & 45.5(10.8) & 100.9(8.8) & 146.4(2.0) \\
2004fb & $-$15.8(0.1) & --- & --- & 0.79(0.22) & --- & --- & --- & --- \\
2004fc & $-$15.8(0.4) & --- & 2.83(0.37) & 0.27(0.13) & --- & 29.4(5.3) & 106.2(4.3) & 135.6(1.0) \\
2004fx & $-$15.3(0.3) & $-$13.0(0.2) & --- & 0.18(0.13) & 0.94(0.01) & --- & --- & 103.3(4.0) \\
2005J  & $-$16.4(0.2) & --- & 5.23(0.36) & 0.78(0.06) & --- & 29.2(8.5) & 82.5(1.5) & 111.7(7.0) \\
2005K  & $-$16.0(0.1) & --- & --- & 1.33(0.20) & --- & --- & --- & --- \\
2005Z  & $-$16.3(0.1) & --- & --- & 1.32(0.03) & --- & --- & --- & --- \\
2005af & $-$15.6(0.4) & $-$14.6(0.4) & --- & $-$0.08(0.68) & 1.28(0.02) & --- & --- & 117.1(17.0) \\
2005an & $-$16.2(0.2) & --- & 3.97(0.59) & 1.99(0.47) & --- & 27.8(11.9) & 53.1(5.9) & 80.9(6.0) \\
2005dk & $-$16.7(0.1) & --- & 10.68(0.87) & 1.31(0.05) & --- & 26.2(6.7) & 73.1(0.7) & 99.3(6.0) \\
2005dn & $-$16.6(0.2) & --- & --- & 1.12(0.04) & --- & --- & --- & 95.8(6.0) \\
2005dt & $-$16.0(0.1) & --- & --- & 0.25(0.14) & --- & --- & --- & 123.0(9.0) \\
2005dw & $-$15.9(0.1) & --- & 1.72(0.74) & 0.68(0.13) & --- & 35.3(16.5) & 75.4(7.5) & 110.7(9.0) \\
2005dx & $-$15.4(0.1) & --- & 4.08(1.16) & 0.70(0.25) & --- & 27.4(12.2) & 74.5(5.2) & 101.9(7.0) \\
2005dz & $-$15.9(0.1) & --- & 4.34(0.27) & 0.43(0.08) & --- & 28.6(6.1) & 85.5(2.1) & 114.1(4.0) \\
2005es & $-$16.5(0.1) & --- & 8.23(4.81) & 0.66(0.49) & --- & 20.7(20.6) & --- & --- \\
2005gk & $-$16.2(0.1) & --- & --- & 0.83(0.17) & --- & --- & --- & --- \\
2005hd & $-$17.0(0.1) & $-$15.2(0.1) & --- & 1.02(0.17) & 1.80(0.04) & --- & --- & --- \\
2005kh & ---          & $-$13.7(0.1) & --- & --- & 0.85(0.01) & --- & --- & --- \\
2005lw & $-$15.5(0.1) & --- & --- & 1.53(0.08) & --- & --- & --- & --- \\
2005me & $-$15.7(0.1) & --- & --- & 1.69(0.11) & --- & --- & --- & --- \\
2006Y  & $-$17.1(0.1) & $-$15.0(0.1) & 8.02(0.54) & 0.80(0.41) & 2.11(0.17) & 23.9(5.7) & 39.6(1.7) & 63.6(4.0) \\
2006ai & $-$17.0(0.2) & $-$14.9(0.2) & 6.25(0.30) & 1.61(0.18) & 1.66(0.14) & 25.7(6.5) & 48.2(1.5) & 73.8(5.0) \\
2006bc & $-$15.2(0.3) & --- & 5.09(0.66) & $-$0.95(0.30) & --- & 21.8(5.5) & --- & --- \\
2006be & $-$16.2(0.3) & --- & 2.03(5.06) & 0.44(0.11) & --- & 26.9(20.3) & 62.0(11.3) & 88.9(9.0) \\
2006bl & $-$16.7(0.1) & --- & 11.93(2.75) & 2.27(0.22) & --- & 15.1(11.7) & --- & --- \\
2006ee & $-$16.0(0.2) & --- & --- & 0.08(0.07) & --- & --- & --- & 116.0(4.0) \\
2006it & $-$16.0(0.2) & --- & 2.11(0.79) & 0.07(0.81) & --- & 29.4(8.3) & --- & --- \\
2006iw & $-$16.3(0.1) & --- & 4.27(0.41) & 0.65(0.14) & --- & 28.6(4.1) & --- & --- \\
2006ms & $-$15.8(0.2) & --- & 4.33(4.11) & 0.04(0.50) & --- & 25.5(13.1) & --- & --- \\
2006qr & $-$15.1(0.2) & --- & 3.91(0.88) & 0.99(0.07) & --- & 19.6(11.8) & 100.3(4.8) & 119.9(7.0) \\
2007P  & $-$16.8(0.1) & --- & 7.65(0.60) & 1.44(0.37) & --- & 28.7(7.3) & --- & --- \\
2007U  & $-$16.9(0.1) & --- & 5.63(1.18) & 1.74(0.26) & --- & 22.7(13.1) & --- & --- \\
2007W  & $-$15.4(0.2) & --- & 6.07(18.98) & 0.20(0.16) & --- & 20.5(17.2) & 105.3(10.2) & 125.8(7.0) \\
2007X  & $-$16.9(0.2) & $-$15.2(0.2) & 2.74(0.15) & 0.91(0.05) & 1.51(0.01) & 33.2(7.9) & 86.5(2.9) & 119.7(5.0) \\
2007aa & $-$16.2(0.3) & --- & --- & $-$0.10(0.07) & --- & --- & --- & 103.1(8.0) \\
2007ab & $-$16.9(0.1) & $-$14.6(0.1) & --- & 2.37(0.14) & 1.79(0.21) & --- & --- & 80.2(10.0) \\
2007av & $-$15.8(0.2) & --- & --- & 0.86(0.04) & --- & --- & --- & --- \\
2007hm & $-$15.6(0.1) & --- & --- & 1.34(0.10) & --- & --- & --- & --- \\
2007il & $-$16.6(0.1) & --- & 3.25(1.21) & 0.15(0.10) & --- & 30.7(9.1) & 95.1(5.1) & 125.7(4.0) \\
2007it & ---          & $-$15.0(0.5) & 5.51(0.38) & --- & 1.08(0.01) & --- & --- & 116.7(1.0) \\
2007ld & $-$16.5(0.1) & --- & 8.08(0.60) & 1.10(0.14) & --- & 18.6(9.1) & --- & --- \\
2007oc & $-$16.0(0.2) & --- & 2.61(0.86) & 1.55(0.09) & --- & 29.6(12.3) & 52.4(9.3) & 82.0(3.0) \\
2007od & $-$16.9(0.8) & --- & 6.98(0.48) & 1.36(0.05) & --- & 19.2(6.1) & --- & --- \\
2007sq & $-$15.9(0.1) & --- & --- & 0.72(0.07) & --- & --- & --- & 104.8(6.0) \\
2008F  & $-$15.7(0.1) & --- & --- & 0.14(0.13) & --- & --- & --- & --- \\
2008K  & $-$16.3(0.1) & $-$13.9(0.1) & 2.25(0.33) & 1.70(0.31) & --- & 42.4(21.0) & 56.1(15.0) & 98.4(6.0) \\
2008M  & $-$16.1(0.3) & $-$13.7(0.3) & 7.66(0.52) & 0.91(0.08) & 1.80(0.01) & 24.4(10.1) & 61.4(1.1) & 85.8(9.0) \\
2008W  & $-$16.2(0.1) & --- & --- & 0.72(0.11) & --- & --- & --- & 104.3(8.0) \\
2008ag & $-$16.7(0.2) & $-$15.1(0.2) & --- & 0.15(0.04) & 0.41(0.05) & --- & --- & 129.4(8.0) \\
2008aw & $-$16.4(0.2) & $-$15.1(0.2) & 3.84(1.23) & 1.82(0.11) & 2.11(0.03) & 34.4(15.0) & --- & --- \\
2008bh & $-$15.7(0.2) & --- & --- & 0.95(0.15) & --- & --- & --- & --- \\
2008bk & $-$14.7(0.1) & $-$12.3(0.1) & --- & 0.08(0.08) & 1.13(0.02) & --- & --- & --- \\
2008bm & $-$17.3(0.1) & --- & --- & 2.74(0.12) & --- & --- & --- & 93.7(10.0) \\
2008bp & $-$15.3(0.2) & --- & --- & $-$0.22(0.41) & --- & --- & --- & --- \\
2008br & $-$14.9(0.2) & --- & --- & 0.19(0.12) & --- & --- & --- & --- \\
2008bu & $-$16.8(0.1) & --- & 10.92(6.38) & 2.21(0.19) & --- & 12.7(8.8) & 39.2(1.8) & 51.9(7.0) \\
2008ga & $-$16.3(0.2) & --- & --- & 0.82(0.26) & --- & --- & --- & 91.2(7.0) \\
2008gi & $-$16.2(0.1) & --- & 2.22(1.42) & 1.56(0.38) & --- & 32.3(33.1) & --- & --- \\
2008gr & $-$16.6(0.1) & --- & 5.43(0.36) & 1.86(0.11) & --- & 14.7(7.4) & --- & --- \\
2008hg & $-$15.7(0.1) & --- & 1.68(0.46) & $-$0.96(0.35) & --- & 26.2(8.8) & --- & --- \\
2008ho & $-$15.0(0.2) & --- & --- & 1.11(0.25) & --- & --- & --- & --- \\
2008if & $-$16.7(0.2) & $-$14.8(0.2) & 6.84(0.25) & 1.89(0.08) & 0.88(0.19) & 28.3(6.1) & 58.5(1.1) & 86.8(5.0) \\
2008il & $-$16.1(0.1) & --- & 4.84(3.24) & 0.00(3.23) & --- & 19.7(17.2) & --- & --- \\
2008in & $-$14.8(0.5) & --- & 2.19(2.05) & 0.51(0.45) & --- & 26.4(24.5) & 79.8(22.5) & 106.3(2.0) \\
2009A  & $-$16.7(0.0) & --- & 12.60(0.68) & $-$4.79(0.47) & --- & 27.35(10.7) & --- & --- \\
2009N  & $-$15.3(0.4) & --- & 2.74(1.62) & $-$0.14(0.07) & --- & 21.4(9.1) & 90.3(4.1) & 111.8(5.0) \\
2009aj & $-$17.5(0.2) & --- & 8.46(0.14) & 1.46(0.05) & --- & 29.8(8.0) & --- & --- \\
2009ao & $-$16.0(0.2) & --- & --- & 0.83(0.21) & --- & --- & --- & --- \\
2009au & $-$14.9(0.2) & --- & 5.91(0.77) & 2.78(0.17) & --- & 27.4(7.5) & --- & --- \\
2009bu & $-$16.2(0.2) & --- & --- & $-$0.05(0.08) & --- & --- & --- & --- \\
2009bz & $-$16.2(0.2) & --- & 3.51(0.74) & 0.48(0.14) & --- & 26.6(8.0) & --- & --- \\

\end{longtable}
\tablefoot{Column 1 lists the SN names. Columns 2 and 3 show the absolute bolometric magnitudes of $M_{\rm bol,end}$ and and $M_{\rm bol,tail}$. In columns 4, 5, and 6, we list the decline rates $s_{\rm 1}$, $s_{\rm 2}$, and $s_{\rm 3}$, respectively. Columns 7, 8, and 9 present the \emph{Cd}, \emph{pd}, and \emph{optd}, respectively. Errors are indicated in parenthesis.}
\end{landscape}
}